\documentclass[preprint,12pt,nonatbib]{elsarticle}

\usepackage[english]{babel}
\usepackage[utf8x]{inputenc}
\usepackage[tbtags]{amsmath}
\usepackage{mathrsfs}
\usepackage{graphicx}
\usepackage[colorinlistoftodos]{todonotes}
\usepackage{setspace}
\usepackage{fixltx2e}
\usepackage{amssymb}
\usepackage[paper=a4paper]{geometry}
\usepackage{color,soul}
\usepackage{amsthm}
\usepackage[section]{placeins}
\usepackage{mathtools}
\usepackage{caption}
\usepackage{subcaption}
\usepackage{appendix}
\usepackage{float}
\usepackage{enumerate}
\usepackage{titlesec}
\usepackage{setspace}
\usepackage{bm}
\usepackage[lined,ruled]{algorithm2e}
\usepackage{hyperref}
\usepackage{natbib}

\SetKwBlock{RRepeat}{repeat}{}

\theoremstyle{plain}
\newtheorem*{thm*}{Theorem}
\newtheorem{thm}{Theorem}[section]

\theoremstyle{plain}
\newtheorem*{defn*}{Definition}
\newtheorem{defn}[thm]{Definition} 
\newtheorem{lemma}[thm]{Lemma} 
\newtheorem{prop}[thm]{Proposition}
\newtheorem{notate}[thm]{Notation}
\newtheorem{problem}[thm]{Problem}
\newtheorem{ass}[thm]{Assumption}
\newtheorem{cond}[thm]{Condition}
\newtheorem{corr}[thm]{Corollary}
\newtheorem{alg}[thm]{Algorithm}

\journal{ }

\begin{document}
	
	\begin{frontmatter}
		
		\title{Bayesian inference from time series of allele frequency data using exact simulation techniques}
		
		\author{Jaromir Sant\corref{Jaro}$^{1}$}
		\ead{jaromir.sant@stats.ox.ac.uk}
		\cortext[Jaro]{Corresponding author}
		\author{Paul A.~Jenkins$^{2,3}$}
		\author{Jere Koskela$^{2,4}$}
		\author{Dario Span{\`o}$^{2}$}
		
		\address{Department of Statistics$^{1}$, University of Oxford, Oxford, OX1 3LB, United Kingdom}
		\address{Department of Statistics$^{2}$ \& Department of Computer Science$^{3}$ \\ University of Warwick, Coventry CV4 7AL, United Kingdom}
		\address{School of Mathematics, Statistics and Physics$^{4}$, Newcastle University, Newcastle upon Tyne NE1 7RU, United Kingdom}
		
		\begin{abstract}
			A central statistical problem in population genetics is to infer evolutionary and biological parameters such as the strength of natural selection and allele age from DNA samples extracted from a contemporary population. That all samples come only from the present-day has long been known to limit statistical inference; there is potentially more information available if one also has access to ancient DNA so that inference is based on a time-series of historical changes in allele frequencies. We introduce a Markov Chain Monte Carlo (MCMC) method for Bayesian inference from allele frequency time-series data based on an underlying Wright--Fisher diffusion model of evolution, through which one can infer the parameters of essentially any selection model including those with frequency-dependent effects. The chief novelty is that we show this method to be exact in the sense that it is possible to augment the state space explored by MCMC with the unobserved diffusion trajectory, even though the transition function of this diffusion is intractable. Through careful design of a proposal distribution, we describe an efficient method in which updates to the trajectory and accept/reject decisions are calculated without error. We illustrate the method on data capturing changes in coat colour over the past 20,000 years, and find evidence to support previous findings that the mutant alleles ASIP and MC1R responsible for changes in coat color have experienced very strong, possibly overdominant, selection and further provide estimates for the ages of these genes.
		\end{abstract}
		
		
	\end{frontmatter}
	
\section{Introduction}
Recent advances in DNA sequencing as well as improvements in ancient DNA (aDNA) retrieval technologies have given rise to a growing number of genetic datasets spanning several centuries (see for instance \citet{Fages2019,Hummel,Ludwig,SandovalCastellanos,Mathieson2020}, but also the Allen Ancient DNA Resource in the context of human aDNA \citep{AADR}). Such time-series data hold a wealth of information on various evolutionary factors which have shaped and are still shaping the population upon which they act, such as the strength and time of onset of natural selection. Detecting and quantifying such phenomena has generated a considerable interest in the literature (see for instance \citet{Ludwig15, Wutke2016, Fages2019, SandovalCastellanos, DerSarkissian15, FangEtAl09, Beaumont1, Beaumont2, LiBarton, Sohail1, Sohail2}, whilst the recently released \href{https://reich-ages.rc.hms.harvard.edu/}{AGES} resource provides genome-wide information on selection for humans), offering empirical evidence for well-documented shifts in selective regimes such as lactase persistence \citep{Tishkoff} and the HLA region \citep{Klein} in humans, as well as the  exogenous effect of humans on plants and animals through domestication \citep{Purugganan, Frantz}. Since most traditional population genetic methods developed to estimate such quantities rely on a static snapshot of the population being considered (as captured for instance by the contemporaneously sampled genetic information present in the myriad biobanks that now exist), they possess a limited amount of statistical power \citep{Dehasque} even when leveraging whole-genome information. More sophisticated statistical techniques are therefore needed to elicit the additional information gained by suitably accounting for the temporal dimension available from time-series data. \\ \newline
A standard way to do this is to assume that allele frequency changes are modelled by the \emph{Wright--Fisher diffusion} \citep[Chapter 15, Section 2]{KarlinTaylor}. This diffusion is mathematically well understood, allows for several important genetic features such as mutation and selection to be incorporated rather easily, and offers a realistic approximation to patterns of genetic diversity observed in nature, particularly when genetic drift is acting on the same timescale as mutation and selection. Here we shall focus on conducting inference for the selection parameters as well as the allele age at a single locus. In this context, one pitfall in adopting the Wright--Fisher diffusion is the fact that the resulting expression for the likelihood consists of a product of intractable diffusion transition densities. This has led to a number of methods in the literature which in some way or another rely on a combination of discretisation and approximation to conduct inference, be it numerical approximation of the likelihood \citep{Bollback, Beaumont1, Zivkovic, Steinrucken, Schraiber, LiBarton, Sohail1, Sohail2, Rito}, or approximation of the diffusion itself through the use of simpler alternative models \citep{Malaspinas, FerrerAdmetlla, MathiesonMcVean, Mathieson2020, Paris, Beaumont2}. Although the discretisations and approximations employed vary from one method to another, all introduce a bias into the inference which is hard to quantify and distinguish from other sources of error. For instance, in \citet{Malaspinas} the authors approximate the Wright--Fisher diffusion through the use of a birth-and-death process; \citet{Steinrucken} employ truncation to be able to evaluate a spectral expansion of the transition density for the Wright--Fisher diffusion; whereas in \citet{Schraiber} the Lebesgue integrals present in the acceptance probabilities used with their path updating scheme need to be approximated through Riemann sums. In contrast, there is a growing body of literature that has attempted to address this problem by avoiding discretisation or approximation entirely \citep{KonKamKing, Goncalves, GarciaPareja2022}, however none gives direct access to the true posterior distribution of the selection parameters and the allele age. A review on inference for natural selection in the presence of time-series genetic data can be found in \citet{Dehasque}. \\ \newline
In this paper we present an exact and practicable inferential framework which allows for Bayesian inference of the allele age and the selection parameters associated with a diallelic locus under essentially any frequency-dependent selective regime (including apostatic selection, under- and over-dominance, balancing selection, and so on), based on discrete noisy observations of a Wright--Fisher diffusion. By identifying a suitable state space augmentation, we are able to construct a Markov Chain Monte Carlo (MCMC) method which updates prior beliefs on the parameters of interest by incorporating the observed likelihoods and returns exact draws from the posteriors of interest. The method is exact in the sense that we avoid any form of approximation or discretisation in tackling the underlying model's intractability, and thus avoid introducing approximation error into the inferential output. The approach adopted here builds upon work done in the context of Brownian motion \citep{Beskos2006, Sermaidis} and exploits the availability of an exact algorithm for the Wright--Fisher class of diffusions \citep{JenkinsSpano, GriffithsJenkinsSpano}.\\ \newline
The main challenge in our setup is that the likelihood involves a finite product of intractable transition densities, which we circumvent through a suitable state space augmentation. Thus we are able to construct a Metropolis-within-Gibbs sampler targeting the joint posterior distribution of the selection parameters and the age of the allele $t_0$. We henceforth assume that $t_0$ will be the first time a \emph{de novo} allele appears in the population, however we point out that our method can easily be modified for the case when the allele arises from standing variation (i.e.\ when the initial frequency of the diffusion $X_0$ is sampled from the stationary distribution, and where $t_0$ would now be the time of onset of selection). \\ \newline
We implemented our method in \texttt{C++} callable from within Python, and tested it by simulating draws from the Wright--Fisher diffusion through the use of the simulation software \texttt{EWF} \citep{Sant2} for different choices of selection and allele age parameters. We show that the method performs well and is able to recover the true parameters even when the prior is poorly specified. Subsequently we apply our method to the horse coat colouration aDNA dataset found in \citet{Ludwig}, where the allele frequencies for two genes (ASIP and MC1R) at two distinct loci (Agouti and Extension respectively) involved in the production of pigmentation are made available over a timespan of approximately 20,000 years. We find strong evidence for overdominance with a large selection coefficient for both genes (mirroring previous studies conducted on the same dataset), whilst the ASIP allele is estimated to be slightly older than the MC1R allele. \\ \newline  
The remainder of this article is organised as follows: Section \ref{Methods} introduces the inferential setup together with the mathematical framework within which we can embed an exact algorithm into a Markov chain Monte Carlo (MCMC) sampler allowing us to target the posterior of interest. The results when considering both simulated and real data, together with details pertaining to the implementation and computational considerations are presented in Section \ref{Results}. We conclude with a brief discussion in Section \ref{Discussion}. Details on the exact algorithm for the Wright--Fisher diffusion are found in \ref{AppendixExactSimulation}, with the calculations leading to a joint tractable likelihood present in \ref{AppendixTractableLikelihood}. \ref{AppendixUpdatingProcedure} provides the exact details pertaining to the updating mechanism used in the algorithm, whilst a `Poisson estimator' required for the analysis is elaborated upon in \ref{AppendixPoissonEstimator}. Further plots for the various simulation and real data scenarios can be found in \ref{AppendixExtraPlots}.
    
\section{Methods}\label{Methods}
We are interested in conducting Bayesian inference on the selection parameters and allele age of an allele at a diallelic locus given that the underlying allele frequency dynamics are driven by a non-neutral Wright--Fisher diffusion. We shall assume that we have $K$ observation times $\boldsymbol{t}=\{t_i\}_{i=1}^{K}$, and at each time $t_i$ we have an observation $y_{i}$ counting the number of individuals in the sample carrying the allele of interest (where we shall make use of the convention $y_i := y_{t_i}$ to make notation less cumbersome). We shall further assume that the observations $\boldsymbol{Y} = \{ Y_{t_i} \}_{i=1}^{K} = \{ Y_{i} \}_{i=1}^{K}$ are each binomially distributed with sample sizes $\boldsymbol{n} := \{n_{i}\}_{i=1}^{K}$, and respective success probabilities $\boldsymbol{X} = \{ X_{t_i} \}_{i=1}^{K} = \{X_i\}_{i=1}^{K}$, where $(X_t)_{t \geq t_0}$ is a non-neutral Wright--Fisher diffusion with mutation parameter $\boldsymbol{\theta} = (\theta_1, \theta_2)$, selection coefficient $\sigma$ and polynomial selection function $\eta(x) = \sum_{i=0}^{d}\eta_i x^i$ for $\boldsymbol{\eta} = \{\eta_i\}_{i=0}^{d} \in \mathbb{R}^{d+1}$ with $d$ fixed and finite. We specify that when referring to $\sigma$ and $\boldsymbol{\eta}$ \emph{jointly}, we shall make use of the term ``selection parameters". \\ \newline
Thus we have that
\[
(Y_{i}\mid X_i=x_{i}) \overset{\textnormal{ind}}{\sim} \text{Binomial}(n_{i},x_{i}), \qquad i=1,\dots,K,
\]
and the diffusion $(X_t)_{t \geq t_0}$ satisfies the following stochastic differential equation, where $t_0$ denotes the time of birth of the allele:
\begin{align}\label{WFDiff}
	dX_{t} &= \frac{1}{2}\Big(\sigma X_{t}(1-X_{t})\eta(X_t) - \theta_{2}X_{t} + \theta_{1}(1-X_{t})\Big)dt + \sqrt{X_{t}(1-X_{t})}dW_{t},
\end{align}
and $X_{s}\equiv0$ for all $s\leq t_0$. Here $\boldsymbol{\theta}$ parametrises ongoing recurrent mutation taking place after time $t_0$. We assume $\boldsymbol{\theta}$ to be fixed and known throughout (since when inferring both the mutation and selection parameters jointly, $\boldsymbol{\theta}$ can be inferred at a faster rate \citep{Jenkins24}), and that recurrent mutation does not take place earlier than time $t_0$. Here $\eta(x)$ is a function accounting for any frequency-dependent effects of selection. For example, genic selection is modelled by taking $\eta(x) = 1$ while diploid selection is obtained when $\eta(x) = x + h(1-2x)$ with $h$ the dominance parameter. \\ \newline
The main object of interest for inference is the posterior of the selection parameters and allele age, whose density we denote by $p(\sigma, \boldsymbol{\eta}, t_0 | \boldsymbol{y})$, and which is given by   
\begin{align*}
    p(\sigma, \boldsymbol{\eta}, t_0 | \boldsymbol{y} ) \propto q_{1}(\sigma) q_{2}(\boldsymbol{\eta}) q_{3}(t_0) \ell\left( \boldsymbol{y} | \sigma, \boldsymbol{\eta}, t_0 \right)
\end{align*}
where we place independent prior densities $q_{1}(\cdot), q_{2}(\cdot), q_{3}(\cdot)$ on the selection parameters and the allele age, while $\ell( \boldsymbol{y} | \sigma, \boldsymbol{\eta}, t_0)$ denotes the likelihood of the observations $\boldsymbol{y}$ given that the selection parameters are given by $\sigma$ and $\boldsymbol{\eta}$, and the allele age is equal to $t_0$. \\ \newline
An expression for the likelihood can be obtained by marginalising out the values of the latent diffusion at the sampling times:
\begin{align*}
	\ell(\boldsymbol{y}|\sigma, \boldsymbol{\eta}, t_0) = \int_{[0,1]^{K}} \prod_{i=1}^{K}\mathcal{B}_{n_{i},x_i}\left(y_{i}\right) p_{\sigma, \boldsymbol{\eta}}^{\boldsymbol{\theta}}(t_{i}-t_{i-1},x_{i-1},x_i)dx_i
\end{align*}
where $\mathcal{B}_{n,x}(\cdot)$ denotes the probability mass function of a Binomial random variable with parameters $n, x$, whilst $p_{\sigma, \boldsymbol{\eta}}^{\boldsymbol{\theta}}(\cdot,\cdot,\cdot)$ denotes the transition density of a non-neutral Wright--Fisher diffusion solving \eqref{WFDiff}, with selection parameters $\sigma$ and $\boldsymbol{\eta}$, and mutation parameter $\boldsymbol{\theta} = (\theta_1, \theta_2)$. That is, $p_{\sigma, \boldsymbol{\eta}}^{\boldsymbol{\theta}}(t,x,u)$ denotes the probability density of moving to allele frequency $u$ given that the frequency was $x$ a time $t$ ago. However, the transition densities $p_{\sigma, \boldsymbol{\eta}}^{\boldsymbol{\theta}}(t_i-t_{i-1},x_{i-1},x_i)$ for $i=1, \dots, K$ are intractable (see \ref{WFNonNeutralTransitionDecomposition} in \ref{AppendixExactSimulation}), and thus so is the likelihood. 
\subsection{A tractable joint density}
In spite of the above mentioned intractability, \citet{JenkinsSpano} showed how to simulate exact draws from the non-neutral Wright--Fisher diffusion and its corresponding bridge diffusion (i.e.\ the diffusion conditioned on both start and end points). Augmenting the state space with the values of the latent diffusion at the sampling times, $\boldsymbol{X} = \{X_i\}_{i=1}^{K}$, leads to the following augmented likelihood
\begin{align}\label{AugmentedLikelihood}
	\ell(\boldsymbol{y}|\sigma, \boldsymbol{\eta}, t_0,\boldsymbol{X}) = \prod_{i=1}^{K}\mathcal{B}_{n_{i},X_i}\left(y_{i}\right) p_{\sigma, \boldsymbol{\eta}}^{\boldsymbol{\theta}}(t_{i}-t_{i-1},X_{i-1},X_i).
\end{align}
Having access to an exact algorithm, one might consider applying a standard Metropolis-within-Gibbs scheme which alternates between performing Gibbs updates of the underlying path and Metropolis updates of the parameters. The main difficulty in such a setting would be computing the acceptance probabilities for the updates involving the selection parameters, as these would involve evaluating ratios of intractable transition densities. \\ \newline
The key idea is that by an appropriate further augmentation of the state space, the joint density of the data $\boldsymbol{y} = \{ y_{i} \}_{i=1}^{K}$, the selection parameters $\sigma$ and $\boldsymbol{\eta}$, the allele age $t_0$, the values of the diffusion at the sampling times $\boldsymbol{X} = \{ X_i \}_{i=1}^{K}$, and an additional collection of latent variables termed the \emph{skeleton points} $\boldsymbol{\Phi} = \{ \Phi_i \}_{i=1}^{K}$, is tractable and given by
\begin{align}\label{AugmentedLikelihoodGlobal}
	p(\sigma, \boldsymbol{\eta}, t_0, \boldsymbol{y}, \boldsymbol{X}, \boldsymbol{\Phi}) &= q_{1}(\sigma)q_{2}(\boldsymbol{\eta})q_{3}(t_0)e^{A_{\sigma, \boldsymbol{\eta}}(X_K) - \left(t_{K}-t_{0}\right)\varphi_{\sigma, \boldsymbol{\eta}}^{-}}\nonumber \\
    &\quad{}\times\prod_{i=1}^{K}\mathcal{B}_{n_{i},X_i}\left(y_{i}\right) f(\sigma, \boldsymbol{\eta}, \Phi_i, X_i)
\end{align}
for a known function $f$ which can be easily evaluated (see \eqref{AugmentedLikelihoodGlobalExplicit} for its exact form), $\varphi_{\sigma, \boldsymbol{\eta}}^{-}$ a known constant, and $A_{\sigma, \boldsymbol{\eta}}(x)$ a known polynomial in $x \in [0,1]$. 
Through this augmentation the intractability is subsumed within the exact simulation routine, resulting in a tractable expression for the augmented likelihood. The latter in turn leads to an augmented posterior, from which the original posterior can be recovered by marginalising out the auxiliary variables. The price for this tractability is that we have introduced additional latent variables, the skeleton points, that are a by-product of the state space augmentation and thus need to be tracked in order to run an MCMC chain. Additionally, we observe that \eqref{AugmentedLikelihoodGlobal} is a density with respect to a dominating measure which is independent of any of the quantities we wish to infer or marginalise over (see the display beneath \eqref{AugmentedLikelihoodGlobalExplicit}), thereby ensuring that our MCMC scheme mixes appropriately \citep{RobertsStramer01}. For full details relating to the derivation and precise form of \eqref{AugmentedLikelihoodGlobal} as well as the corresponding dominating measure, please see \ref{AppendixTractableLikelihood}. 

\subsection{Updating procedure}\label{Updating}
Our proposed MCMC scheme sequentially iterates between the following two steps:
\begin{enumerate}
    \item Update the selection parameters conditional on the allele age and auxiliary variables.
    \item Update the allele age and auxiliary variables conditional on the selection parameters.
\end{enumerate}
Step 1 can be performed through the use of a simple Metropolis--Hastings step. If proposals for the selection parameters are generated according to some proposal kernel $g_{\textnormal{sel}}((\cdot, \cdot)|(\sigma, \boldsymbol{\eta}))$, then using \eqref{AugmentedLikelihoodGlobal} the acceptance probability for a move from a configuration $(\sigma_1, \boldsymbol{\eta}_1)$ to $(\sigma_2, \boldsymbol{\eta}_2)$ has acceptance probability given by
\begin{align}\label{AcceptanceSigma}
    \alpha_{\textnormal{sel}} &= \alpha((\sigma_2, \boldsymbol{\eta}_2)|(\sigma_1, \boldsymbol{\eta}_1)) \nonumber \\
    &= \min\Bigg\{ 1, \frac{g_{\textnormal{sel}}((\sigma_1, \boldsymbol{\eta}_1)|(\sigma_2, \boldsymbol{\eta}_2))}{g_{\textnormal{sel}}((\sigma_2, \boldsymbol{\eta}_2)|(\sigma_1, \boldsymbol{\eta}_1))}\frac{q_1(\sigma_2)}{q_1(\sigma_1)}\frac{q_{2}(\boldsymbol{\eta}_2)}{q_{2}(\boldsymbol{\eta}_{1})}e^{A_{\sigma_2, \boldsymbol{\eta}_2}(X_K) - A_{\sigma_1, \boldsymbol{\eta}_1}(X_K)} \nonumber \\
    &{}\qquad{}\qquad{}\quad{}\times e^{-\left(t_K-t_0\right)\left(\varphi_{\sigma_2, \boldsymbol{\eta}_2}^{-}-\varphi_{\sigma_1, \boldsymbol{\eta}_1}^{-}\right)} \frac{\displaystyle\prod_{i=1}^{K}f(\sigma_2, \boldsymbol{\eta}_2, \Phi_i, X_i)}{\displaystyle\prod_{i=1}^{K}f(\sigma_1, \boldsymbol{\eta}_1, \Phi_i, X_i)} \Bigg\},
\end{align}
and as noted previously, all expressions on the right-hand side can be computed precisely. \\ \newline
For Step 2 on the other hand we employ multiple smaller updates over the observation intervals $[t_0, t_2]$, $[t_{i-1}, t_{i+1}]$ for $i = 2, \dots, K-2$, and $[t_{K-1}, t_K]$ (which we term initial, interior, and end segment updates respectively), which helps boost the algorithm's performance. Whilst the whole path could be updated in a single pass, the resulting acceptance probability would be very low, resulting in a poorly mixing chain. By splitting the update into smaller overlapping steps (inspired by the approach in \citet{GolightlyWilkinson06}), we are able to retain exactness whilst also achieving a reasonable acceptance probability.
\subsubsection{Initial path segment update}\label{InitialPathSegmentUpdate}
We briefly summarise how to update the auxiliary variables, starting with an initial segment update. 
\begin{enumerate}
    \item At iteration $m > 0$, propose a new allele age $\widetilde{t}_0$ conditional on the allele age at iteration $m-1$, $t_0^{(m-1)}$, according to some proposal kernel $g_{\textnormal{age}}(\cdot | t_{0}^{(m-1)})$.
    \item Conditional on the proposal $\widetilde{t}_0$ and the value of the latent diffusion at time $t_2$ at iteration $m-1$, $X_2^{(m-1)}$, propose a draw $\widetilde{X}_{t_1}$ from the law of a non-neutral Wright--Fisher bridge going from frequency $0$ to $X_2^{(m-1)}$ in time $t_2 - \widetilde{t}_0$, sampled at time $t_1 - \widetilde{t_0}$. We denote the corresponding proposal kernel by $g_{\textnormal{WF}}(\cdot|0, X_2^{(m-1)}, \widetilde{t}_0, t_1, t_2)$.
    \item Conditional on $\widetilde{t}_0, \widetilde{X}_{t_1}, X_2^{(m-1)}$, simulate skeleton points $\widetilde{{\Phi}}_1, \widetilde{{\Phi}}_2$ over the time intervals $[\widetilde{t}_0, t_1]$ and $[t_1, t_2]$ respectively. We denote the corresponding proposal kernels by $g_{\textnormal{skel}}(\cdot|0, \widetilde{X}_{1}, \widetilde{t}_{0}, t_1)$ and $g_{\textnormal{skel}}(\cdot|\widetilde{X}_{1}, X_{2}^{(m-1)}, t_1, t_2)$. In particular, following \citet{Sermaidis}, we shall only ever propose skeleton points from their true conditional distribution, and thus these terms shall cancel out in the below acceptance probability (see \eqref{SkeletonDensity} and the paragraph preceding it for further details).
\end{enumerate}
The resulting proposal $(\widetilde{t}_0, \widetilde{X}_{t_1}, \widetilde{{\Phi}}_1, \widetilde{{\Phi}}_2)$ has acceptance probability given by 
\begin{align}\label{AcceptanceT0}
    \alpha_{\textnormal{init}} &:= \min\Bigg\{1, \frac{q_3(\widetilde{t}_0)}{q_3(t_0^{(m-1)})}\frac{g_{\textnormal{age}}(t_0^{(m-1)} | \widetilde{t}_0)}{g_{\textnormal{age}}(\widetilde{t}_0 | t_0^{(m-1)})}e^{-\varphi_{\sigma, \boldsymbol{\eta}}^{-}(t_0^{(m-1)} - \widetilde{t}_0)}\frac{p_{0, \boldsymbol{0}}^{\boldsymbol{\theta}}(t_2 - t_0^{(m-1)}, 0, X_2^{(m-1)})}{p_{0, \boldsymbol{0}}^{\boldsymbol{\theta}}(t_2 - \widetilde{t}_0, 0, X_2^{(m-1)})} \\ \nonumber
    &{}\qquad{}\qquad{}\quad{}\times\frac{a(t_2 - \widetilde{t}_0, 0, X_2^{(m-1)}, \sigma, \boldsymbol{\eta})}{a(t_2 - t_0^{(m-1)}, 0, X_2^{(m-1)}, \sigma, \boldsymbol{\eta})} \Bigg\}
\end{align}
where $a(\cdot, \cdot, \cdot, \cdot, \cdot)$ is defined in \eqref{WFNonNeutralTransitionDecomposition}. Despite the fact that the last ratio in \eqref{AcceptanceT0} cannot be evaluated exactly, one can obtain an unbiased estimate through the \emph{Poisson estimator} (see \citet[Section 6]{Beskos2006} and \citet{Wagner88}; full details on how this estimator is used in our algorithm can be found in \ref{AppendixPoissonEstimator}). The resulting unbiased estimator can then be used within a Metropolis--Hastings update, leading to a pseudo-marginal scheme which is less efficient than one which has access to the ratio without error but which is guaranteed to target the true posterior \citep{Andrieu, Beaumont2003}. We further mention that although the neutral transition densities appearing in \eqref{AcceptanceT0} are available only as infinite sums, one can nonetheless make the accept/reject decision without having to approximate these infinite sums via an `alternating series' trick \citep{JenkinsSpano}. 
\subsubsection{Interior path segment updates}\label{InteriorEndPathSegmentUpdate}
Following an initial segment update, the algorithm moves to update the path over the time interval $\{[t_{i-1}, t_i] \cup [t_i, t_{i+1}]\}_{i=2}^{K-2}$ sequentially. The values of the latent diffusion at the sampling times $t_{i-1}$ and $t_{i+1}$, $X_{i-1}^{(m)}$ and $X_{i+1}^{(m-1)}$, are fixed (we note that $X_{i-1}^{(m)}$ would have been updated in the previous step), and conditional on these values a proposal is generated as follows:
\begin{enumerate}
    \item Conditional on $X_{i-1}^{(m)}, X_{i+1}^{(m-1)}$, draw $\widetilde{X}_{i}$ from the law of a Wright--Fisher diffusion bridge going from $X_{i-1}^{(m)}$ to $X_{i+1}^{(m-1)}$ over the time interval $t_{i+1} - t_{i-1}$ sampled at time $t_i - t_{i-1}$. Denote the corresponding proposal kernel by $g_{\textnormal{WF}}(\cdot|X_{i-1}^{(m)}, X_{i+1}^{(m-1)}, t_{i-1}, t_i, t_{i+1})$. 
    \item Conditional on $\widetilde{X}_{i}$, simulate skeleton points $\widetilde{{\Phi}}_{i-1}$ over the time interval $[t_{i-1}, t_i]$, and $\widetilde{{\Phi}}_i$ over the time interval $[t_i, t_{i+1}]$. We denote the corresponding proposal kernels by $g_{\textnormal{skel}}(\cdot|X_{i-1}^{(m)}, \widetilde{X}_{i}, t_{i-1}, t_i)$ and $g_{\textnormal{skel}}(\cdot|\widetilde{X}_{i}, X_{i+1}^{(m-1)}, t_i, t_{i+1})$.
\end{enumerate}
The resulting acceptance probability for the proposal $(\widetilde{X}_{i}, \widetilde{\Phi}_{i-1}, \widetilde{\Phi}_{i})$, is given by
\begin{align}\label{AcceptanceXti}
    \alpha_{\textnormal{int}, i} := \min\left\{ 1, \left(\frac{X_i^{(m-1)}}{\widetilde{X}_{i}}\right)^{y_i} \left(\frac{1-X_i^{(m-1)}}{1-\widetilde{X}_{i}}\right)^{n_i - y_i} \right\},
\end{align}
which can be evaluated exactly. All the relevant details to derive \eqref{AcceptanceXti} can be found in \ref{AppendixTractableLikelihood}. The above interior path segment updating routine is repeated over all time intervals $[t_{i-1}, t_i] \cup [t_i, t_{i+1}]$ for $i \in \{2, \dots, K-1\}$. 
\subsubsection{End path segment updates}\label{EndPathSegmentUpdates}
\begin{figure}[h]
    \centering \includegraphics[width=0.99\textwidth]{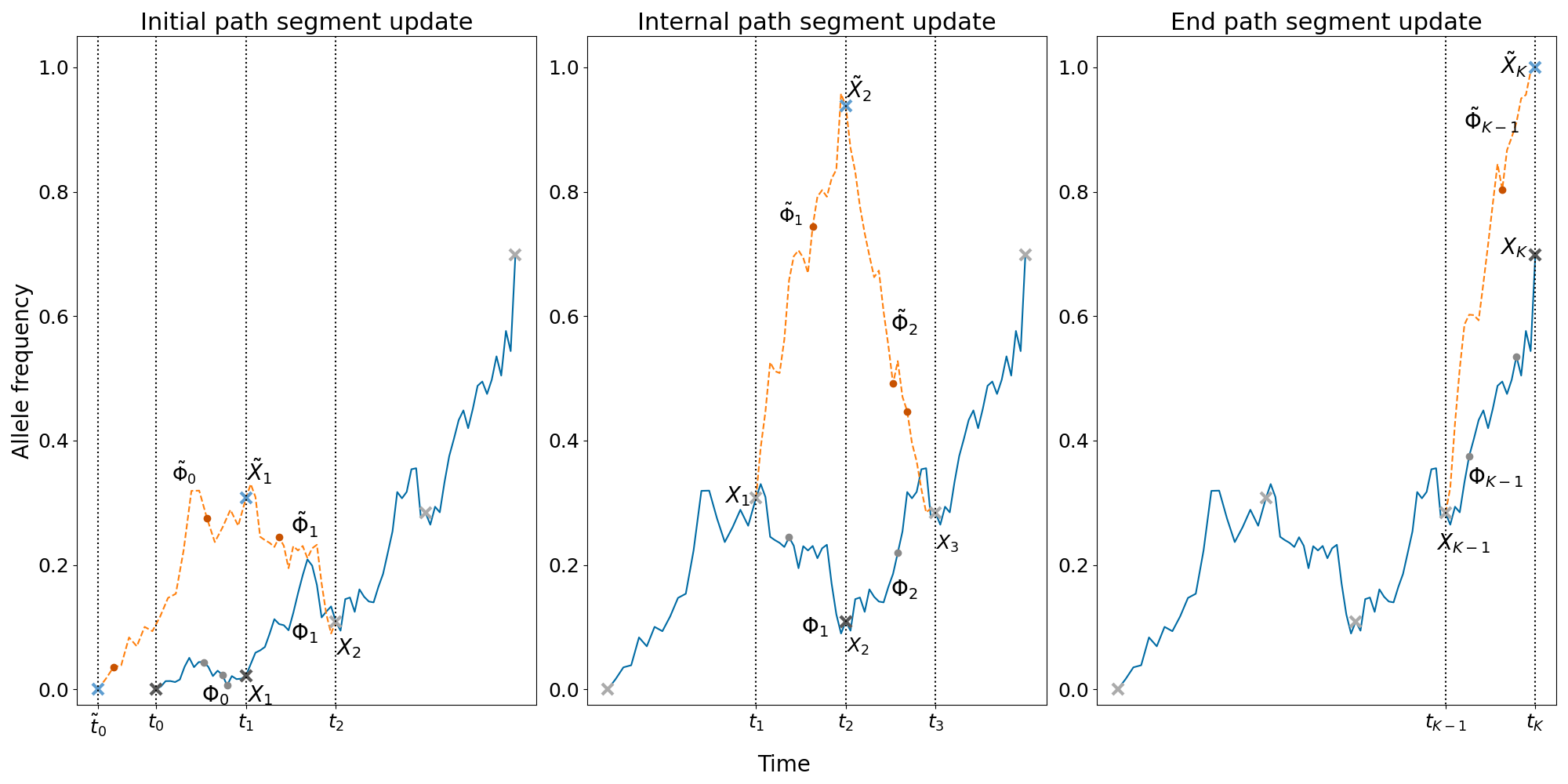}
    \caption[Schematic of the initial, internal and end path updates]{Diagram illustrating the initial, internal, and end path segment update procedures. In this particular instance, the proposal for the initial path segment is accepted, whilst the internal path segment is rejected. \textbf{Initial path segment update (left):} At the current iteration, the latent path is given by the full blue line, the values of the diffusion at the observation times are denoted by the dark grey crosses, whilst the skeleton points are the light grey circles. Fixing the latent diffusion path and skeleton points beyond sampling time $t_2$, we generate a proposal as per Subsection \ref{InitialPathSegmentUpdate}. The proposed initial path segment is given by the dashed orange line, with proposed new allele age $\widetilde{t_0}$, latent diffusion value $\widetilde{X}_{1}$ given by the blue cross, and corresponding skeleton points $\widetilde{\Phi}_0, \widetilde{\Phi}_1$ given by the red circles. By comparing this panel to the middle one we deduce that this initial path segment proposal was accepted. \textbf{Internal path segment update (middle):} For the internal path segment over the time interval $[t_{i-1}, t_{i+1}]$, we fix the latent diffusion and skeleton points outside of said interval, and generate a proposal as per Subsection \ref{InteriorEndPathSegmentUpdate}. The proposal generated in this particular instance is rejected as can be observed by comparing the middle panel to the right-most one. \textbf{End path segment update (right):} Fixing the current path until time $t_{K-1}$, we propose a new diffusion end path segment as per Subsection \ref{EndPathSegmentUpdates}.} 
    \label{UpdatingMechanism}
\end{figure}

For the final segment $[t_{K-1}, t_K]$ a similar updating technique is applied, where however a Wright--Fisher diffusion (rather than diffusion bridge) sampled at time $t_K - t_{K-1}$ is used to generate a proposal:
\begin{enumerate}
    \item Conditional on $X_{{K-1}}^{(m)}$, draw $\widetilde{X}_{K}$ from the law of a Wright--Fisher diffusion started at $X_{{K-1}}^{(m)}$ and sampled at time $t_K - t_{K-1}$. The corresponding proposal kernel is denoted by $g_{\textnormal{WF}}(\cdot|X_{K-1}^{(m)}, t_{K-1}, t_K)$.
    \item Conditional on $\widetilde{X}_{K}$, simulate skeleton points $\widetilde{\Phi}_{K}$ over the time interval $[t_{K-1}, t_K]$. Denote the corresponding proposal kernel by $g_{\textnormal{skel}}(\cdot|X_{K-1}^{(m)}, \widetilde{X}_{K}, t_{K-1}, t_K)$
\end{enumerate}
The resulting proposal $(\widetilde{X}_{K}, \widetilde{\Phi}_{K})$ has acceptance probability given by
\begin{align}\label{AcceptanceXtn}
    \alpha_{\textnormal{end}} = \min\left\{ 1, \left(\frac{X_{K}^{(m-1)}}{\widetilde{X}_{K}}\right)^{y_K} \left(\frac{1-X_{K}^{(m-1)}}{1-\widetilde{X}_{K}}\right)^{n_K - y_K} \right\}.
\end{align}
Figure \ref{UpdatingMechanism} illustrates the above-described update mechanism, which we further summarise in Algorithm \ref{Algorithm1}.  
	
\begin{algorithm}
    \SetAlgoLined
    \DontPrintSemicolon
    Initialise $\sigma^{(0)}, \boldsymbol{\eta}^{(0)},  t_{0}^{(0)}, \boldsymbol{X}^{(0)}, \boldsymbol{\Phi}^{(0)}$, and set $k = 0$\;
    \Repeat{\textnormal{convergence}}{
	Draw $\widetilde{\sigma}, \widetilde{\boldsymbol{\eta}} \sim g_{\textnormal{sel}}((\cdot, \cdot)|(\sigma^{(k)}, \boldsymbol{\eta}^{(k)}))$\;
	Draw $\widetilde{t}_0 \sim g_{\textnormal{age}}(\cdot|t_{0}^{(k)})$\;
        Draw $\widetilde{X}_{t_1} \sim g_{\textnormal{WF}}(\cdot|0, X_{2}^{(k)}, \widetilde{t}_0, t_1, t_2)$\; 
        Draw $\widetilde{\Phi}_{1} \sim g_{\textnormal{skel}}(\cdot|0, \widetilde{X}_{1}, \widetilde{t}_0, t_1)$, $\widetilde{\Phi}_{2} \sim g_{\textnormal{skel}}(\cdot|\widetilde{X}_{t_1}, X_{2}^{(k)}, t_1, t_2)$\;
	   Compute $\alpha_{\textnormal{init}}$ as in \eqref{AcceptanceT0}, draw $U_1 \sim \textnormal{Unif}([0,1])$\;
	   \eIf{$\alpha_{\textnormal{init}}<U_1$}{$(t_0^{(k+1)}, X_1^{(k+1)}, \Phi_{1}^{(k+1)}, \Phi_{2}^{(k+1)}) \leftarrow (\widetilde{t}_0, \widetilde{X}_{t_1}, \widetilde{\Phi}_{1}, \widetilde{\Phi}_{2})$}{$(t_0^{(k+1)}, X_1^{(k+1)}, \Phi_{1}^{(k+1)}, \Phi_{2}^{(k+1)})  \leftarrow$ $(t_0^{(k)}, X_1^{(k)}, \Phi_{1}^{(k)}, \Phi_{1}^{(k)})$}
	    \For{$i=2,\dots,K-1$}{Draw $\widetilde{X}_{i} \sim g_{\textnormal{WF}}(\cdot|X_{i-1}^{(k+1)}, X_{i+1}^{(k)}, t_{i-1}, t_i, t_{i+1})$\;
        Draw $\widetilde{\Phi}_{i} \sim g_{\textnormal{skel}}(\cdot|X_{i-1}^{(k+1)}, \widetilde{X}_{i}, t_{i-1}, t_{i})$, $\widetilde{\Phi}_{i+1} \sim g_{\textnormal{skel}}(\cdot|\widetilde{X}_{i}, X_{i+1}^{(k)}, t_{i}, t_{i+1})$\;
		      Compute $\alpha_{\textnormal{int},i}$ as in \eqref{AcceptanceXti}, draw $U_2 \sim \textnormal{Unif}([0,1])$\;
			\eIf{$\alpha_{\textnormal{int},i}<U_2$}{$(X_i^{(k+1)}, \Phi_{i}^{(k+1)}, \Phi_{i+1}^{(k+1)}) \leftarrow (\widetilde{X}_{i}, \widetilde{\Phi}_{i}), \widetilde{\Phi}_{i+1})$}{$(X_i^{(k+1)}, \Phi_{i}^{(k+1)}, \Phi_{i+1}^{(k+1)}  \leftarrow$ $(X_i^{(k)}, \Phi_{i}^{(k)}, \Phi_{i+1}^{(k)})$}}
			Draw $\widetilde{X}_{K} \sim g_{\textnormal{WF}}(\cdot|X_{K-1}^{(k+1)}, t_{K-1}, t_K)$\;
                Draw $\widetilde{\Phi}_{K} \sim g_{\textnormal{skel}}(\cdot|X_{K-1}^{(k+1)}, \widetilde{X}_K, t_{K-1}, t_K)$\;
			     Compute $\alpha_{\textnormal{end}}$ as in \eqref{AcceptanceXtn}, draw $U_3 \sim \textnormal{Unif}([0,1])$\;
				\eIf{$\alpha_{\textnormal{end}}<U_3$}{$(X_{K}^{(k+1)}, \Phi_{K}^{(k+1)}) \leftarrow (\widetilde{X}_{K}, \widetilde{\Phi}_{K})$}{$(X_{K}^{(k+1)}, \Phi_{K}^{(k+1)})  \leftarrow$ $(X_{K}^{(k)}, \Phi_{K}^{(k)})$}
	Compute $\alpha_{\textnormal{sel}}$ as in \eqref{AcceptanceSigma}, draw $U_4 \sim \textnormal{Unif}([0,1])$\;
	\eIf{$\alpha_{\textnormal{sel}}<U_4$}{$\sigma^{(k+1)}, \boldsymbol{\eta}^{(k+1)} \leftarrow \widetilde{\sigma}, \widetilde{\boldsymbol{\eta}}$}{$\sigma^{(k+1)}, \boldsymbol{\eta}^{(k+1)} \leftarrow \sigma^{(k)}, \boldsymbol{\eta}^{(k)}$}
    Increment $k \leftarrow k + 1$\;
}
\caption[Metropolis-within-Gibbs scheme for exact inference of allele age and selection parameters]{Metropolis-within-Gibbs scheme for exact inference on selection parameters $\sigma, \boldsymbol{\eta}$, and allele age $t_0$, given noisy observations $\boldsymbol{y}$}
\label{Algorithm1}
\end{algorithm}

\section{Results}\label{Results}
The above-described inferential framework was implemented in C++, and can be called from within Python through the package \texttt{MCMC4WF}, which is available for download at \url{https://github.com/JaroSant/MCMC4WF}. The method was tested on both simulated data and real data, as we now describe.
\subsection{Simulated data}
Through the use of the \texttt{EWF} software package \citep{Sant2}, a non-neutral Wright--Fisher diffusion trajectory was simulated, and subsequently the values of said diffusion at each sampling time was used as the success probability within a binomial sampling framework to generate a synthetic dataset. We tested our method in the case of genic and diploid selection (i.e.\ when $\eta(x) = 1$ and $\eta(x) = x + h(1-2x)$ for $h$ the dominance parameter), but we point out that it can be used to infer the parameters of essentially any frequency-dependent selection function. Below we present the output obtained for four particular simulation scenarios with parameters given in Table \ref{SetupTable}. In the first two experiments, A and B, our method was set to infer solely genic selection, whilst in the latter two experiments, C and D, the method performed inference on both $\sigma$ and $h$. We further observe that the lack of statistical power in inferring allele ages that extend far back in time motivates the choice of true allele ages below. 
\begin{center}
\begin{tabular}{c|c|c|c|c|c|c|c|c|c|c|c}
	Experiment & Datapoints & $\sigma$ & $h$ & $\boldsymbol{\theta}$ & $t_0$ & $t_1$ & $t_2$ & $t_3$ & $t_4$ & $t_5$ & $n_{i}$ \\ \hline
	A & 5 & 10 & 0.5 & (0.1, 0.1) & 0.2 & 0.25 & 0.35 & 0.36 & 0.4 & 0.48 & 20 \\
	B & 5 & 0 & 0.5 & (0.1, 0.1) & 0.2 & 0.25 & 0.35 & 0.36 & 0.4 & 0.48 & 20 \\
	C & 5 & 10 & 0.0 & (0.1, 0.1) & 0.2 & 0.25 & 0.35 & 0.36 & 0.4 & 0.48 & 20 \\
	D & 5 & 10 & 1.0 & (0.1, 0.1) & 0.2 & 0.25 & 0.35 & 0.36 & 0.4 & 0.48 & 20 
\end{tabular}
\captionof{table}[Parameter configuration for the simulated data]{List of the parameter configurations for the simulated dataset.\label{SetupTable}}
\end{center}
The number of datapoints, sample sizes, and order of magnitude of the mutation parameters chosen was inspired by similar datasets in the literature (see for instance \citet{Ludwig,Bollback,Wutke2016,Fages2019}). In choosing the observation time spacing, we chose a maximum spacing between successive observations of 0.1 such that the resulting observations are not spaced too far apart, nor too closely together. As for the choice of $\sigma$ and $h$, we wanted to emulate both neutral scenarios as well as cases where the selection parameter $\sigma$ is significantly different from 0, but where it is not so large as to lead to a regime where the Wright--Fisher diffusion is no longer an appropriate model for the underlying dynamics. \\ \newline
In order to gauge how well our method deals with mis-specified and flat priors, a Gaussian prior of $\textnormal{N}(-10,10)$ as well as a $\textnormal{Unif}([-30,30])$ were used for $\sigma$, whereas a $\textnormal{Unif}([-1.5, 1.5])$ prior was adopted for $h$ (which covers the most common range of values for the dominance parameters observed in nature \citep{ChenluLohmueller24}). For $t_0$, a suitably transformed exponential distribution was used such that $t_0 = t_c - Z$, for $Z \sim \textnormal{Exp}(\beta)$, $t_c := \min\{i \geq 1 : y_{i} \neq 0\}$ and some user-defined $\beta > 0$. In particular, due to the fact that the data was found to be quite uninformative for the distribution of allele age (as evidence through the flat likelihood evaluations in Figure \ref{T0KSDensityGenicSigmaT0ExpA}), $\beta$ was chosen such that the prior assigned 0.99 of its mass to values within a user-defined distance $\gamma$ of the first non-zero observation time $t_c$, namely $\beta$ was chosen such that $\int_{0}^{\gamma}\beta e^{-\beta z}dz = 0.99$. This ensures that the method spends less time in parameter configurations which are uninformative for the allele age but which drive up the computational cost through a larger initial observation time interval. Longer time intervals imply longer run-times in view of the increased number of skeleton points that need to be simulated (see \ref{AppendixExactSimulation} as well as \citet[Section 5]{JenkinsSpano} for further details). In what follows, we took $\gamma = 0.25$ which from Table \ref{SetupTable} can be seen to be of the same order of magnitude as the whole observation interval over which the observations were generated.  \\ \newline
After some initial tuning, Gaussian proposals centred at the previous value and with standard deviation equal to 10 were adopted for $\sigma$, whilst for $h$ proposals came from a conditional Gaussian centred at the previous value and with standard deviation 0.25. Thus any proposals for $h$ falling outside of the interval $[-1.5, 1.5]$ are redrawn. In the case of $t_0$, a conditional Gaussian proposal (again centred at the previous value), with standard deviation 0.25 and with truncation threshold equal to $t_c$ was used. Once again, any proposed values of $t_0$ exceeding $t_c$ are automatically redrawn. Convergence of the MCMC chain was assessed by first monitoring the mean and standard deviation of the inferred parameters, and subsequently consulting the traceplots and autocorrelation functions for the parameters of interest to ensure appropriate mixing of the corresponding chains (see \ref{AppendixExtraPlots}).
\subsubsection{Inferring genic selection}\label{GenicSelection}
\begin{figure}[h]
    \centering
    \begin{subfigure}[b]{0.495\textwidth}
    \centering
    \includegraphics[width=\textwidth]{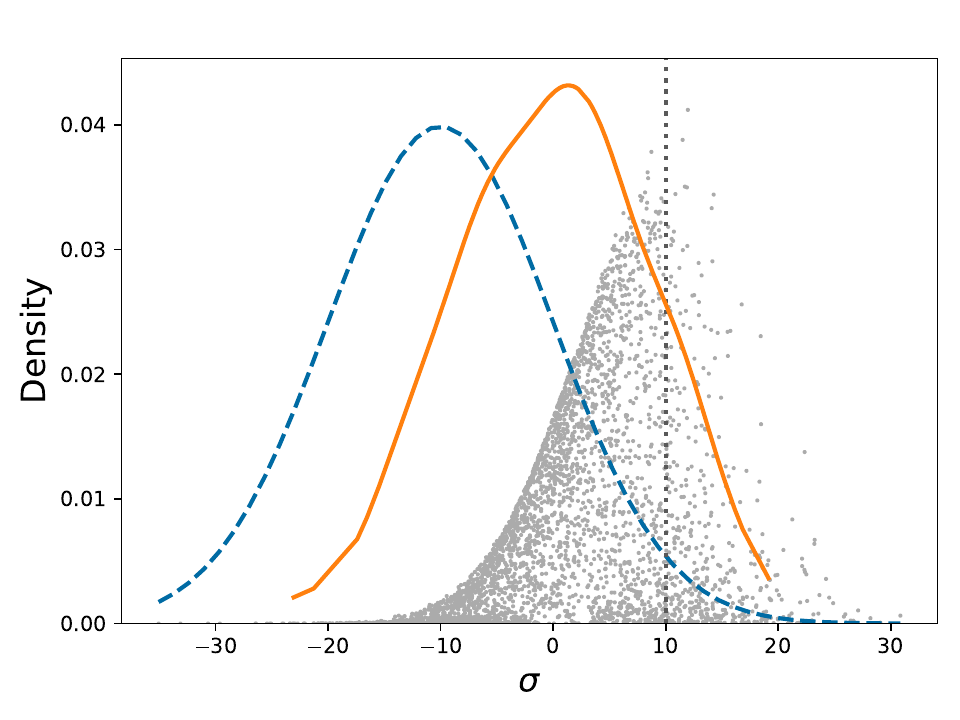}\\
    \caption{}\label{SigmaKSDensityGenicSigmaT0ExpA}
    \end{subfigure}
    \hfill
    \begin{subfigure}[b]{0.495\textwidth}  
    \centering 
    \includegraphics[width=\textwidth]{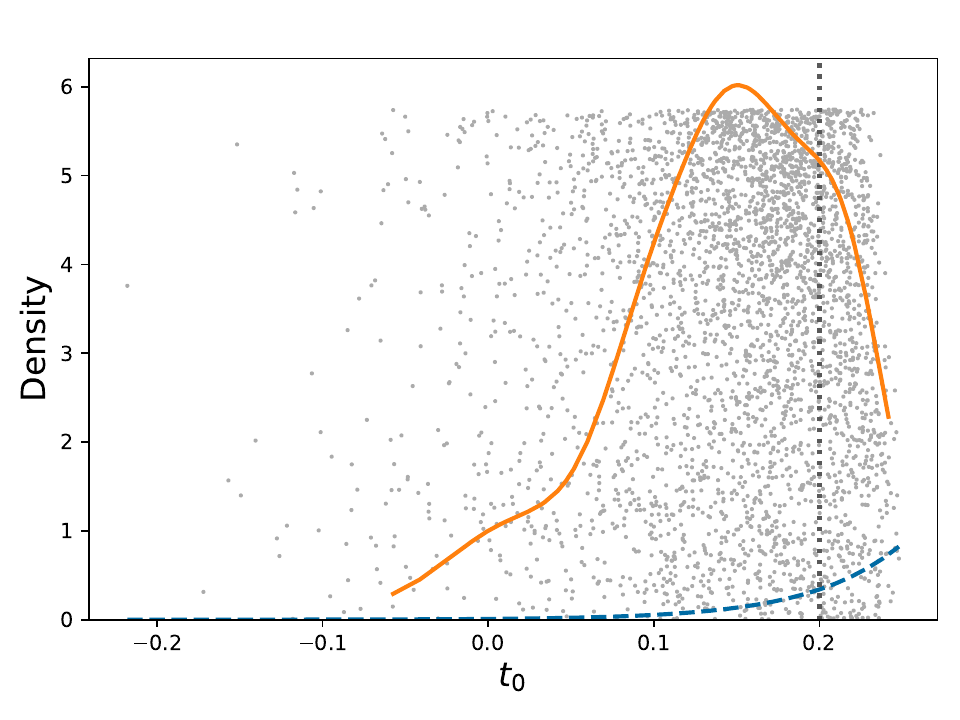}\\
    \caption{}\label{T0KSDensityGenicSigmaT0ExpA}
    \end{subfigure}\\
    \caption[Plots of the prior, likelihood, posterior and true value of the selection coefficient]{Plots of the prior (dashed blue line), likelihood (grey dots), kernel smoothed posterior (solid orange line) and the truth (vertical dotted black line) for (a) the selection coefficient $\sigma$ and (b) the allele age $t_0$ for Experiment A. We point out that the likelihood evaluations are unnormalised.} 
    \label{SigmaT0KSDensityGenicSigmaT0ExpA}
\end{figure}

Using the configurations given in the first two rows of Table \ref{SetupTable}, the generated sequence of true allele frequencies together with the observed binomial samples are plotted in Figure \ref{ObsA} in blue and orange respectively. We present the output obtained for Experiment A below, whilst the corresponding output for Experiment B can be found in \ref{AppendixExtraPlots}, specifically Figures \ref{ObsB}, \ref{KSDensityGenicSigmaT0ExpB}, \ref{TraceplotSigmaB}, \ref{AutoCorrSigmaB}, \ref{SigmaMeanB}, \ref{TraceplotT0B}, \ref{AutoCorrT0B}, \ref{T0MeanB}, \ref{MixingXB}. Despite using two poorly specified Gaussian priors centred at $-10$, the method correctly detects signatures of selection, with the posterior concentrating around the true value of $\sigma$ as observed in Figure \ref{SigmaT0KSDensityGenicSigmaT0ExpA}. Here, posterior and prior densities are plotted against the left axis, whilst the likelihood evaluations (i.e.\ the contributions to the likelihood which depend on $\sigma$) are plotted against the right axis. Comparing all three on the same axis is infeasible due to the intractable normalising constant that would need to be computed for the likelihood evaluations. Furthermore, this normalising constant is very noisy and hard to estimate via Riemann-sum-type approximations owing to its dependence on $\boldsymbol{X}, \boldsymbol{\Phi}$, and $t_0$. \\ \newline
The posterior plots for the allele age $t_0$ in Figure \ref{T0KSDensityGenicSigmaT0ExpA} reveal a relatively flat likelihood, with the smoothed posterior seemingly retaining much of its shape from the prior. This comes as no surprise, as one would expect there to be relatively little signal in the data about the allele age. Indeed, as the initial sampling time interval $[t_0, t_1]$ increases, inferring $t_0$ becomes harder as the transition density of the diffusion converges to the invariant density (which is independent of $t_0$) rather quickly. This implies that there is little benefit to be gained (in terms of statistical power) by adopting more diffuse priors which allow for deeper values in time for $t_0$, especially given the non-negligible increase in computational cost incurred. Further plots showing the mixing of the inferred parameters and latent diffusion values at sampling times, as well as convergence diagnostics, can be found in \ref{AppendixExtraPlots}. 

\subsubsection{Inferring diploid selection}\label{DiploidSelection}
\begin{figure}[h]
	\centering
	\begin{subfigure}[b]{0.49\textwidth}
		\centering
		\includegraphics[width=\textwidth]{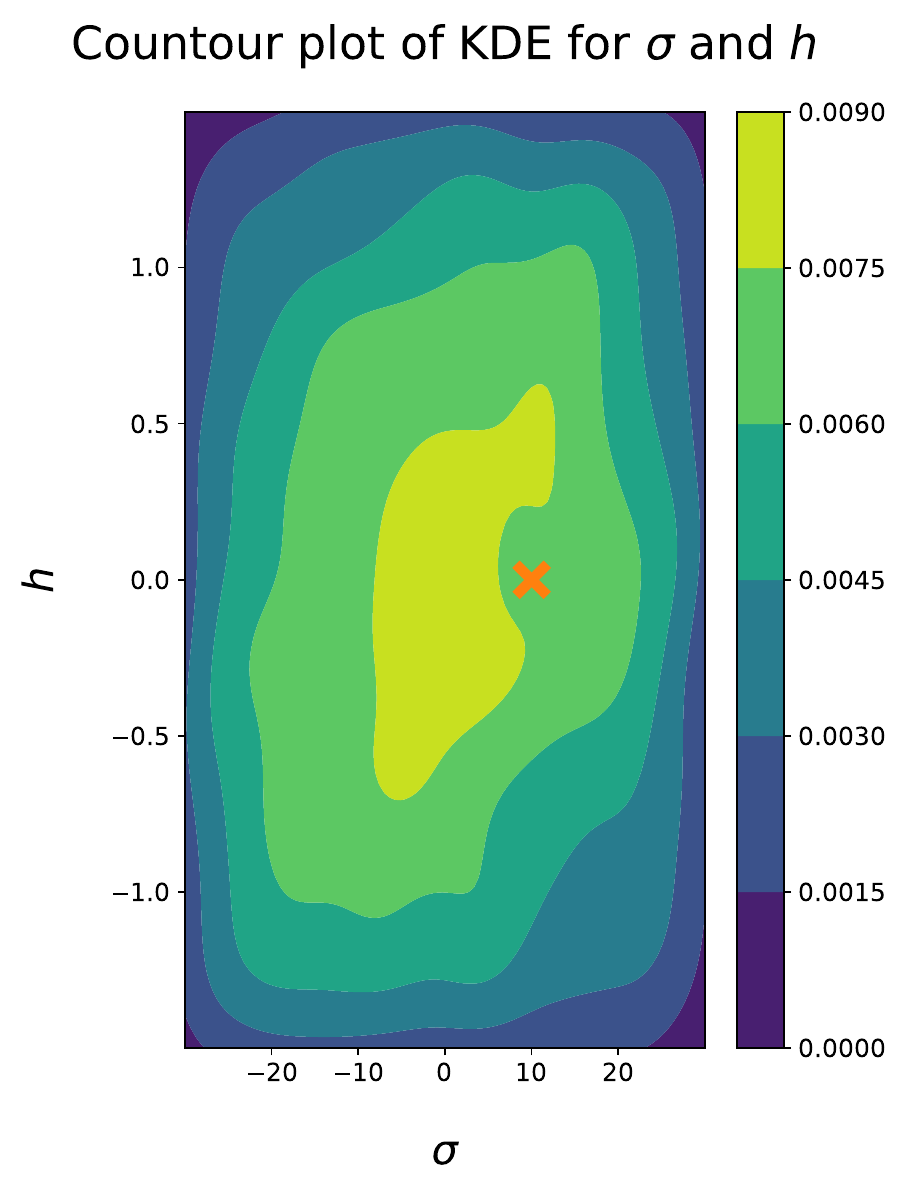}\\
		\caption{}  
		\label{SigmaHKSDensityC}
	\end{subfigure}
	\begin{subfigure}[b]{0.49\textwidth}  
		\centering 
		\includegraphics[width=\textwidth]{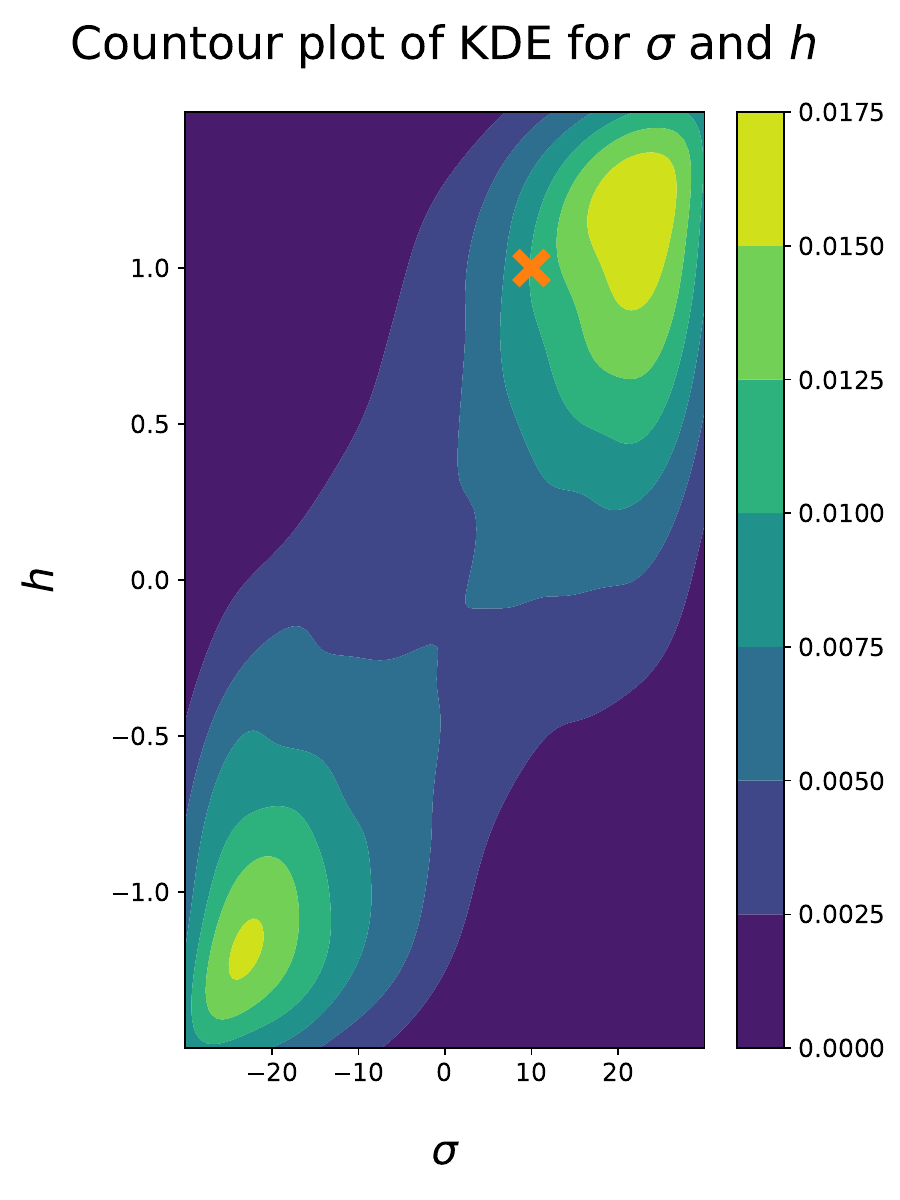}\\
		\caption{}  
		\label{SigmaHKSDensityD}
	\end{subfigure}\\
	\caption[KDE and contour plots of the joint posterior of $\sigma$ and $h$]{Smoothed KDE surface plot for the joint posterior of $\sigma$ and $h$, with the true value indicated by an orange cross for (a) Experiment C and (b) Experiment D.} 
	\label{SigmaHKSDensitySigma_diploid}
\end{figure}

Using the configurations in the last two rows of Table \ref{SetupTable}, the generated sequence of true and observed allele frequencies are plotted in Figure \ref{Observations_diploid}. The same prior distribution for the allele age as in the genic selection case was used, however uniform priors were imposed on both the selection coefficient $\sigma$ and the dominance parameter $h$, namely $\textnormal{Unif}([-30, 30])$ for $\sigma$ and $\textnormal{Unif}([-1.5, 1.5])$ for $h$. This was done primarily to ensure that the inferred parameter values remained within realistic regimes (in most biological situations $h \in [-1, 1]$ \citep{ChenluLohmueller24}), and ones for which the Wright--Fisher diffusion is a suitable model. Additionally we observed that the larger parameter space induced by considering diploid selection resulted in a tradeoff between the computational complexity and the efficiency of the algorithm. The imposed bounds resulted in runtimes 7 and 8 hours for Experiments A and B, whereas Experiments C and D ran for 13 and 26 hours respectively. \\ \newline
Inspecting the posterior surface plots for $(\sigma,h)$ in Figure \ref{SigmaHKSDensitySigma_diploid}, we can deduce that once again the method is performing well, concentrating most of its mass around the true values. As seen in the genic selection case, there is very limited statistical power to infer allele age, with the resulting posterior surface plots remaining relatively constant along the $t_0$ axis both in Figure \ref{SigmaT0KSDensitySigma_diploid} and Figure \ref{HT0KSDensitySigma_diploid}. Further plots detailing the performance of the method in terms of the mixing of inferred parameter and latent diffusion path at sampling times, as well as convergence can be found in \ref{AppendixExtraPlots}.    

\subsection{Horse coat colouration dataset}\label{HorseData}
\begin{figure}[h]
	\centering
	\includegraphics[width=\textwidth]{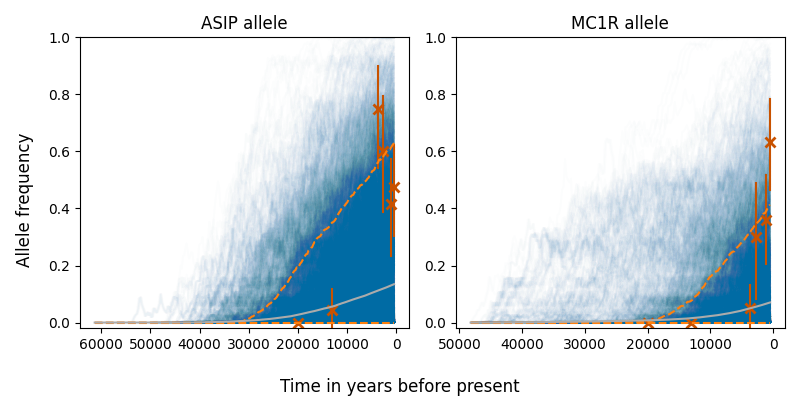}\\
	\caption[Observed and simulated sample allele frequencies for the ASIP (left) and MC1R (right) genes. The observed allele frequencies are denoted by the dark orange crosses with vertical 95\% confidence interval for the binomial sampling, whilst simulated frequencies are plotted in blue. The corresponding 95\% credible interval for the simulated paths is given by the dashed light orange line, whilst the mean is the solid grey line.]{Observed and simulated sample allele frequencies for the ASIP (left) and MC1R (right) genes. The observed allele frequencies are denoted by the dark orange crosses with vertical 95\% binomial sampling error bars, whilst simulated frequencies are plotted in blue. The corresponding 95\% credible interval for the simulated paths is given by the dashed light orange line, whilst the mean is the solid grey line.} 
	\label{Observations_horse}
\end{figure}

\begin{figure}[h]
	\centering
        \begin{subfigure}[b]{0.49\textwidth}
    		\centering
                \includegraphics[width=\textwidth]{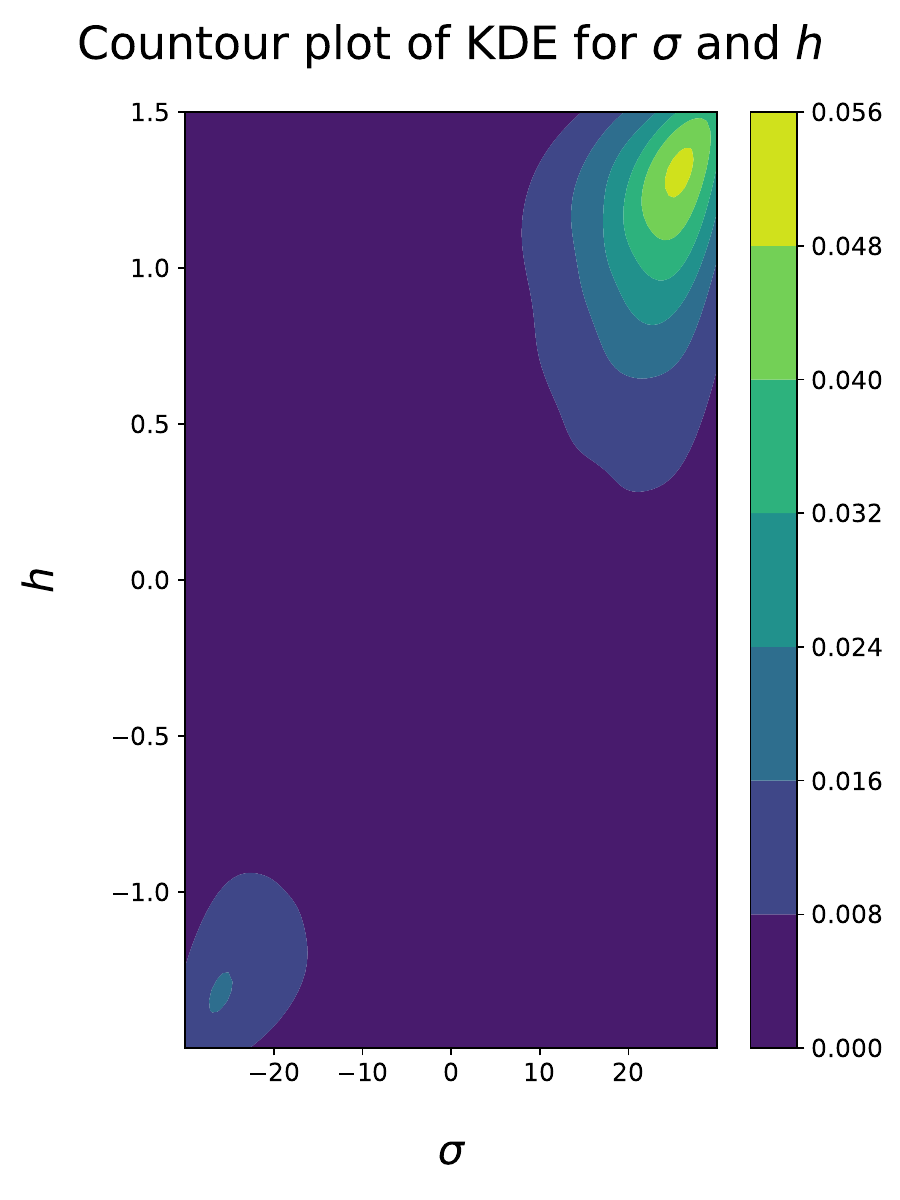}\\
    	\caption{} 
    	\label{SigmaHKSDensitySigma_ASIP}
        \end{subfigure}
	\begin{subfigure}[b]{0.49\textwidth}
            \centering
            \includegraphics[width=\textwidth]{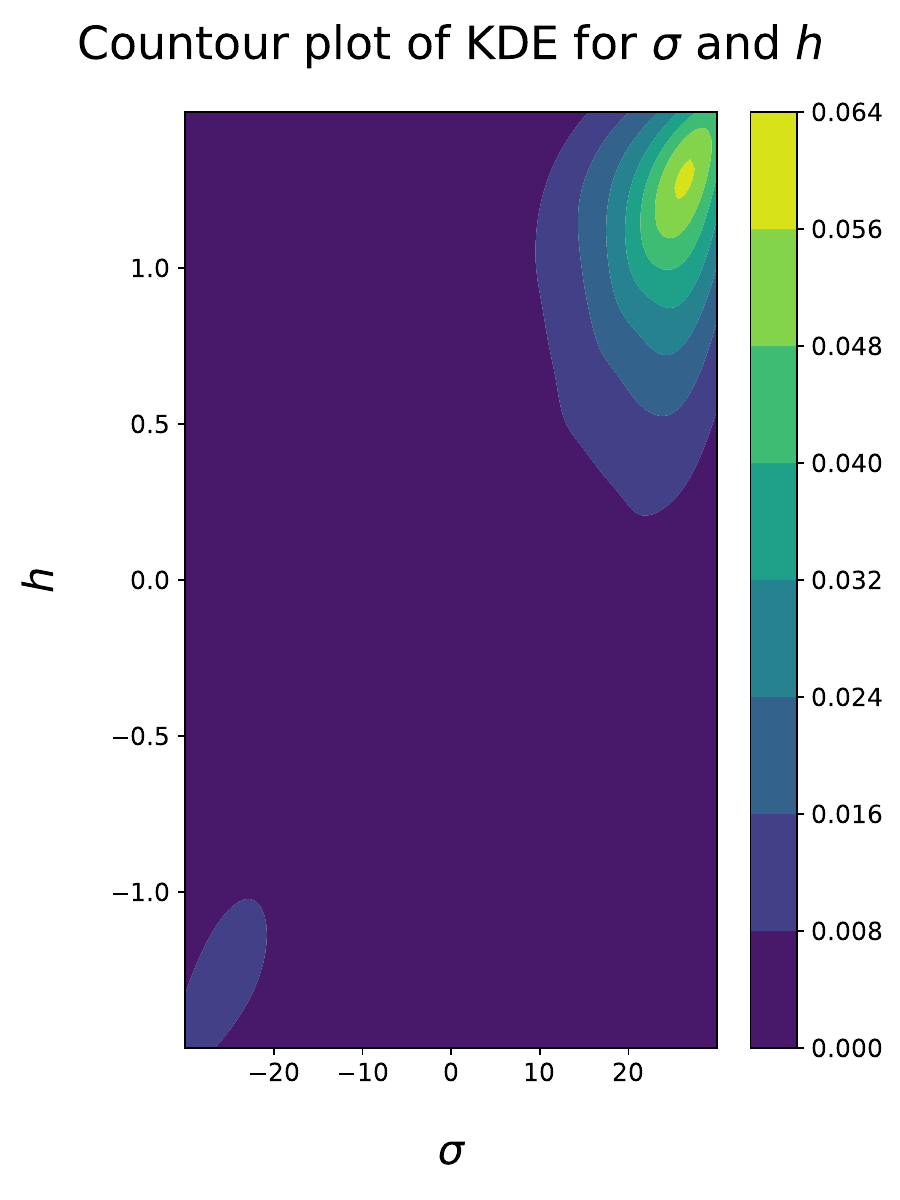}\\
            \caption{}
            \label{SigmaHKSDensitySigma_MC1R}
            \end{subfigure}
            \caption[Contour plot of the KDE for the posterior of $(\sigma, h)$ for the (a) ASIP, and (b) MC1R alleles]{Contour plot of the KDE for the posterior of $(\sigma, h)$ for the (a) ASIP, and (b) MC1R alleles.} 
            \label{SigmaHKSDensitySigma}
\end{figure}

The dataset in \citet{Ludwig} has been studied extensively in the literature, particularly in the context of inference for selection. The authors there report a time series of allele frequencies for a number of loci coding for coat colour in horses, with samples dating back to 20,000 years BC up to 500 BC. Here we report our findings when running \texttt{MCMC4WF} on the ASIP (agouti-signalling-protein) and MC1R (Melanocortin-1 receptor) alleles, present at the Agouti and Extension loci respectively, which dictate the horse coat colouration patterns in bay, chestnut, and black horses. The ASIP allele controls the distribution of black pigmentation across the horse, with the wild type distributing black pigmentation to localised regions of the horse's body, whilst the derived allele spreads black pigmentation uniformly across the entire horse. Thus horses having the wild type at the Agouti locus have a bay coat whilst those having the derived allele are black. The Extension locus on the other hand plays a key role in melanin synthesis, where the dominant type promotes the synthesis of black/brown pigmentation whilst a missense mutation results in the synthesis of yellow/red pigmentation. Thus horses with wild type at the Extension locus have a black phenotype, whilst those carrying the derived MC1R allele have chestnut coats \citep{Thiruvenkadan08}. \\ \newline
As the dataset blends multiple observations across different points in time, one would expect to see a strong signal for selection, particularly in view of the outsized exogenous effect humans have had on animal genetics through domestication. For instance, in \citet{FangEtAl09} the authors find evidence for strong artificial selection on coat coloration in pigs, whilst in \citet{Ludwig15} it is suggested that fluctuations in the frequencies of the leopard spotting complex in horses mirror shifts in coat coloration preferences. \\ \newline
In choosing the effective population size to use within our study, a sizeable discrepancy was observed across the various studies which analysed this dataset: \citet{Ludwig} assumes $N_e \in [10^{3}, 10^{6}]$, \citet{Steinrucken} assumes $N_e \in [2500,10^{4}]$, \citet{Malaspinas} opts for $N_e \in [200,5000]$, whilst \citet{Schraiber} assumes $N_e = 16,000$ in the constant demography case. We opt for a population size of $N_e = 3000$, based on the above cited literature, and to ensure that our method remains computationally feasible (as well as restricting $|\sigma| \leq 30$). \\ \newline 
Using the sample times reported in \citet{Ludwig}, together with $N_e = 3000$ and a generation time of 8 years, the time series of sample allele frequencies (together with some simulated paths (and corresponding confidence intervals) whose parameters are given by the MAP estimates inferred from the data) can be found in Figure \ref{Observations_horse}. \\ \newline
We adopted uniform priors of $\textnormal{Unif}([-30, 30])$ and $\textnormal{Unif}([-1.5, 1.5])$ for $\sigma$ and $h$ respectively, whilst for $t_0$ we used the same prior as in Section \ref{DiploidSelection}. Inspecting the posterior surface plots (obtained by suitably thinning the generated samples) Figure \ref{SigmaHKSDensitySigma}, we find strong evidence for overdominance with a large selection coefficient for both the ASIP and MC1R genes. The posterior concentrates fairly tightly around the joint MAP (with accompanying 0.8 smallest credible interval reported in brackets) of $\sigma = 28.50$ $(-9.07, 30.00)$, $h = 1.41$ $(-0.71, 1.50)$ and $t_0 = 0.002$ $(-0.21, 0.11)$ for the ASIP gene, whilst for MC1R we find a joint MAP of $\sigma = 29.69$ $(4.47, 30.00), h = 1.47$ $(-0.05, 1.50)$ and $t_0 = 0.18$ $(0.02, 0.30)$. These findings are in agreement with previous estimates: in \citet{Steinrucken} the authors report evidence of overdominance, with the same conclusion being drawn in \citet{Schraiber} for the case of constant demography in the absence of recurrent mutation (though their conclusions change to ``positive, nearly additive, selection'' when accounting for an earlier-inferred demographic model \citep{DerSarkissian15}). \\ \newline
Inspecting the posterior surface plots involving the allele age $t_0$ we observe evidence suggesting that the ASIP allele is slightly older than the MC1R allele. Translating the above reported MAP estimates for  $t_0$ from a diffusion time scale to years before present (BP), we find that the derived ASIP allele is dated to 19,904 BP (30,080 BP, 14,720 BP), whilst for MC1R we have a more recent estimate of 11,360 BP (19,040 BP, 5,600 BP).  \\ \newline
Whilst we report strong evidence for overdominance coupled with a large selection coefficient, it should be noted that the Agouti and Extension loci (at which the ASIP and MC1R genes are present respectively) are known to be in recessive epistasis \citep{Thiruvenkadan08}, which is not properly accounted for in our method. Additionally, throughout we assume a known constant population size which if incorrectly specified can be confounded with signal for selection.
\begin{figure}[h]
	\centering
        \begin{subfigure}[b]{0.49\textwidth}
    		\centering
                \includegraphics[width=\textwidth]{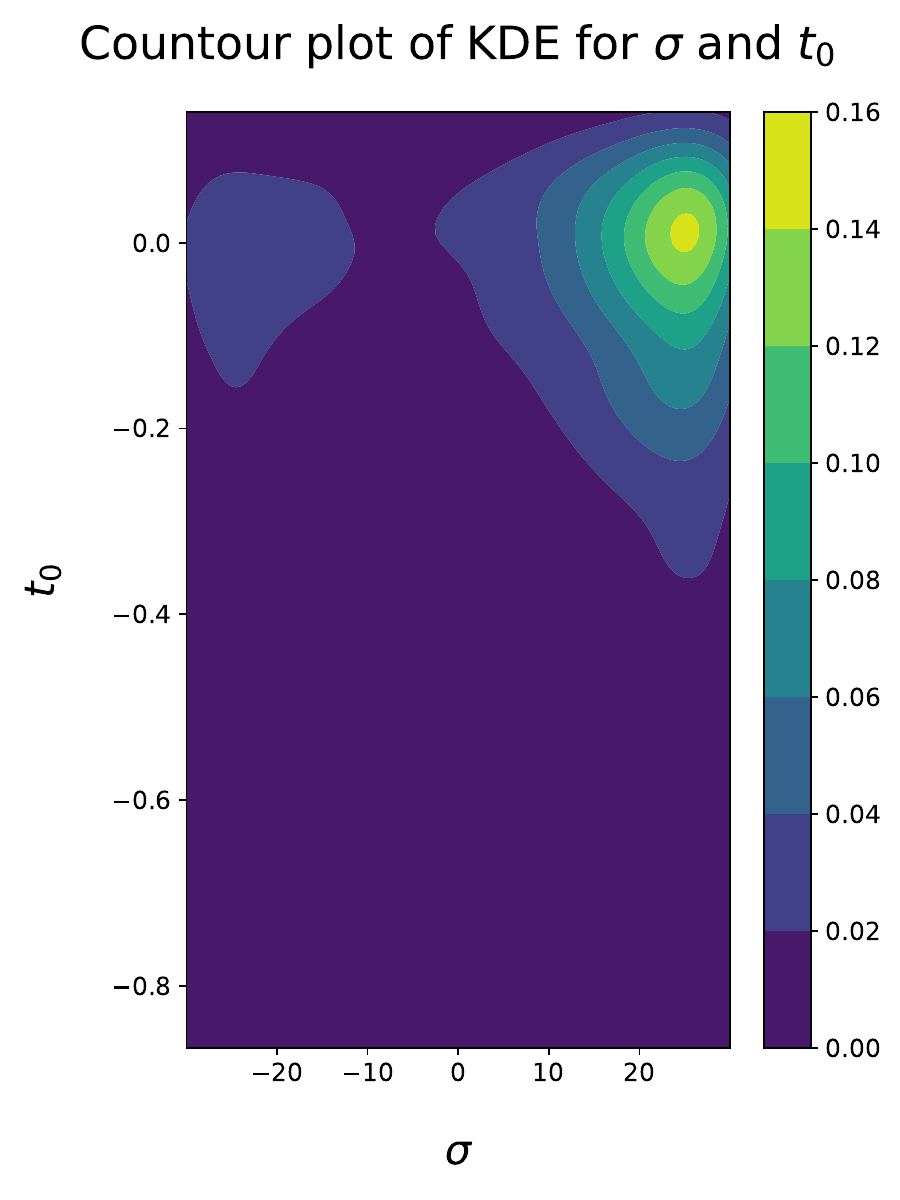}\\
                \caption{}
                \label{SigmaT0KSDensitySigma_ASIP}
        \end{subfigure}
	\begin{subfigure}[b]{0.49\textwidth}
            \centering
            \includegraphics[width=\textwidth]{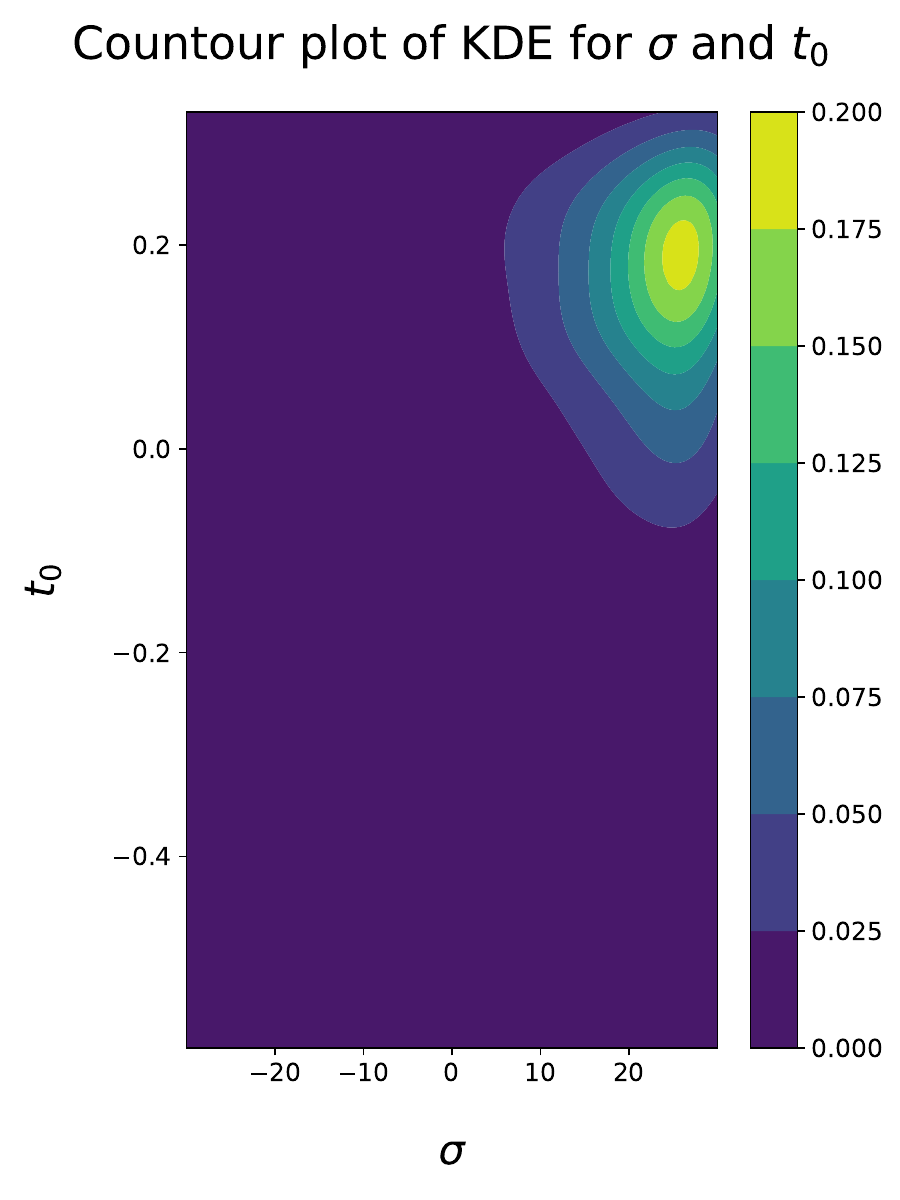}\\
            \caption{}
            \label{SigmaT0KSDensitySigma_MC1R}
        \end{subfigure}
        \caption[Contour plot of the KDE for the posterior of $(\sigma, t_0)$ for the (a) ASIP, and (b) MC1R alleles.]{Contour plot of the KDE for the posterior of $(\sigma, t_0)$ for the (a) ASIP, and (b) MC1R alleles.} 
        \label{SigmaT0KSDensitySigma}
\end{figure}
\begin{figure}[h]
	\centering
        \begin{subfigure}[b]{0.49\textwidth}
            \centering
    	\includegraphics[width=\textwidth]{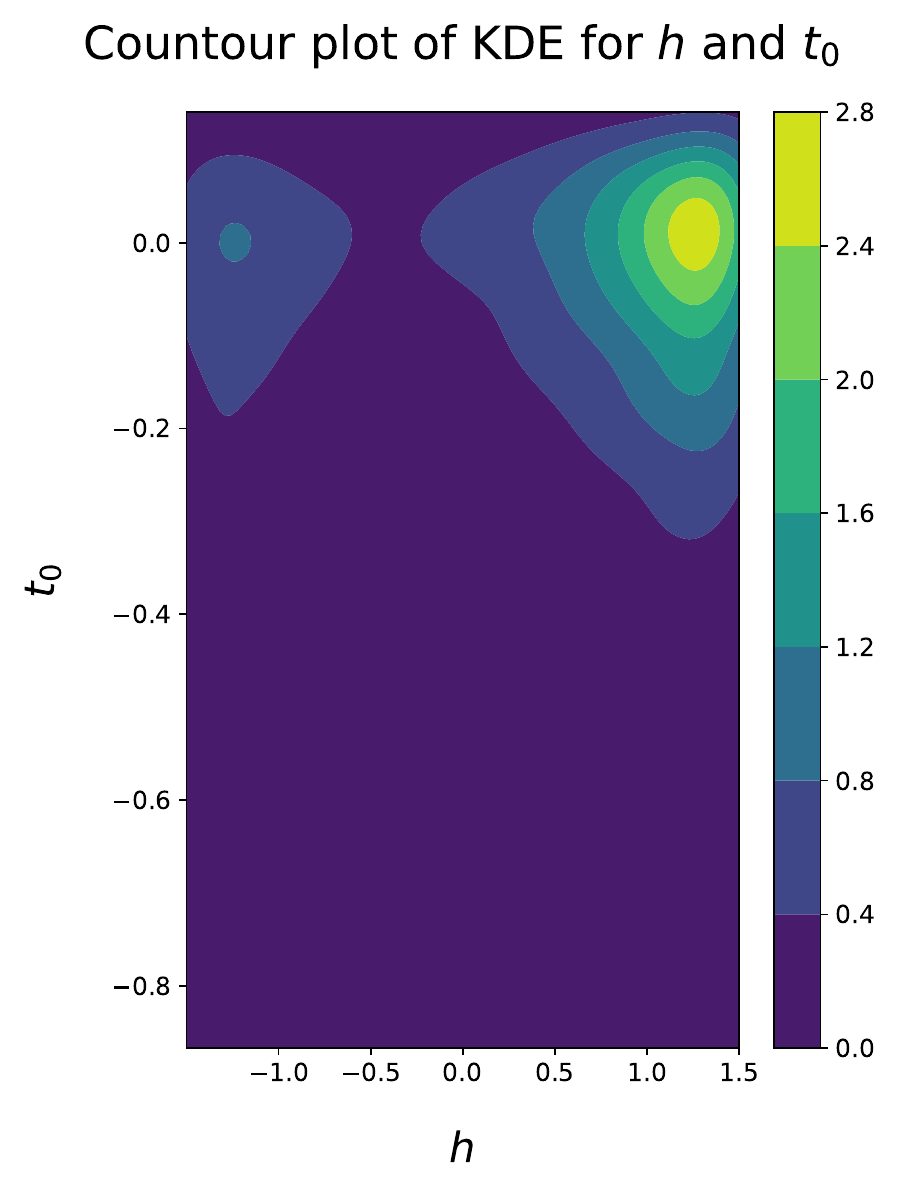}\\
    	\caption{} 
    	\label{HT0KSDensitySigma_ASIP}
        \end{subfigure}
        \begin{subfigure}[b]{0.49\textwidth}
    	\centering
    	\includegraphics[width=\textwidth]{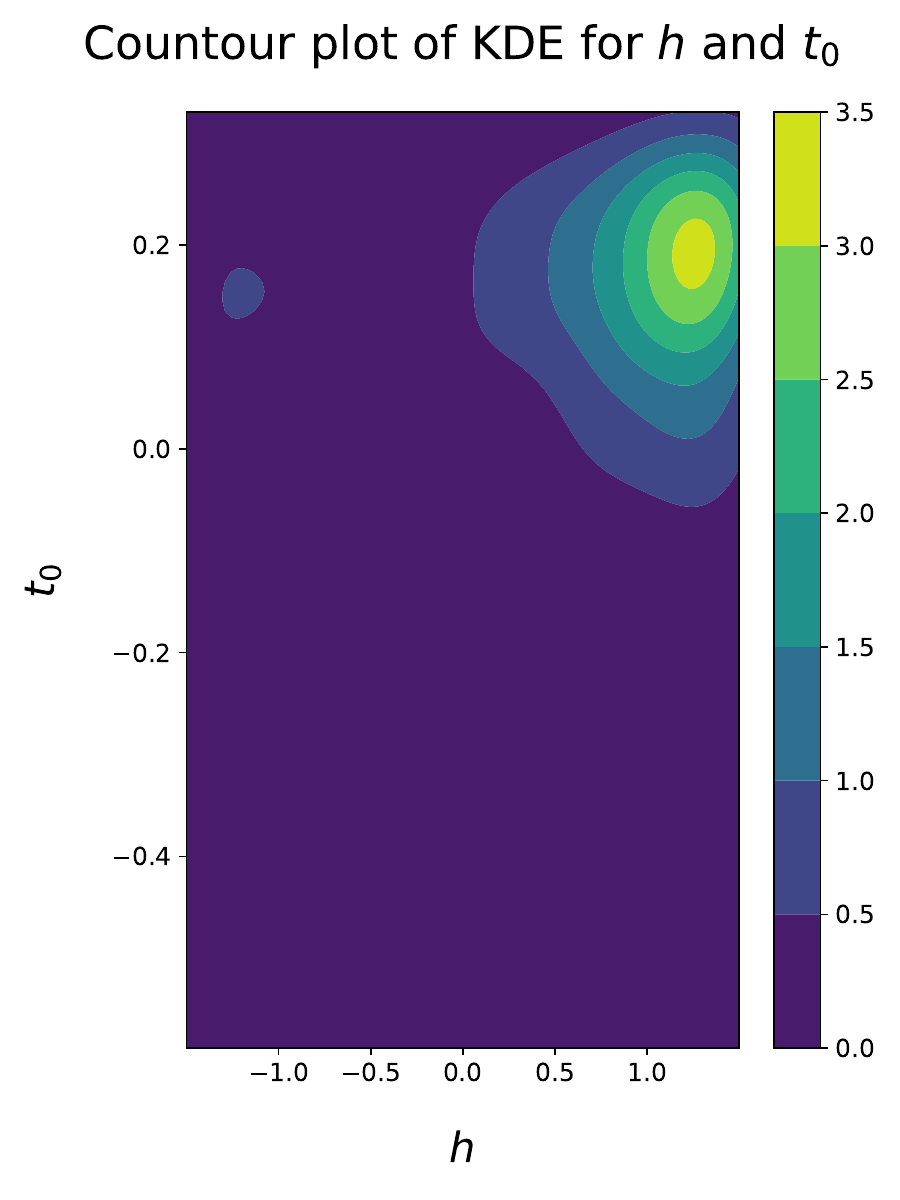}\\
    	\caption{} 
    	\label{HT0KSDensitySigma_MC1R}
        \end{subfigure}
        \caption[Contour plot of the KDE for the posterior of $(h, t_0)$ for the (a) ASIP, and (b) MC1R alleles.]{Contour plot of the KDE for the posterior of $(h, t_0)$ for the (a) ASIP, and (b) MC1R alleles.}
        \label{HT0KSDensitySigma}
\end{figure}

\section{Discussion}\label{Discussion}
\begin{figure}[H]
	\centering
	\includegraphics[width=\textwidth]{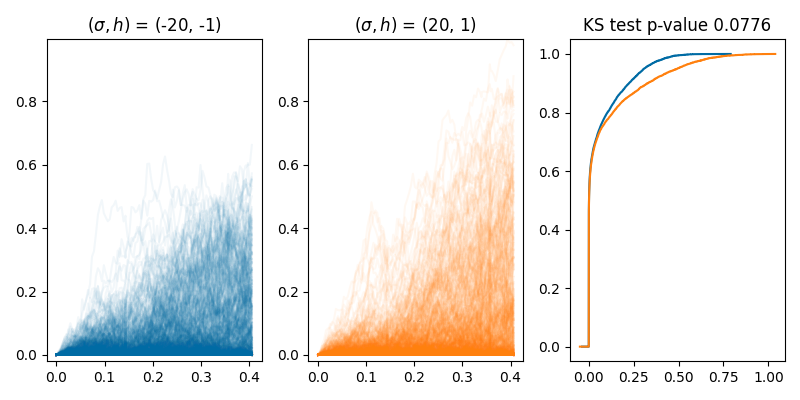}\\
	\caption[Compare paths for different $(\sigma, h)$.]{Comparison between paths generated using $(\sigma, h) = (-20, -1)$ (left panel) $(\sigma, h) = (20, 1)$ (middle panel), together with corresponding empirical CDF (right panel) for the terminal value of the generated paths. Running the Kolmogorov--Smirnov test on the terminal values of the paths returns a p-value of 0.0776, confirming the two generated sets of paths are statistically indistinguishable at level 5\%.}  
	\label{compare_paths}
\end{figure}

As the exponential growth of aDNA datasets heralds a new era in understanding the genetic histories of various populations, the development of powerful statistical methods exploiting this new temporal dimension becomes increasingly important. In this paper we achieved this for the Wright--Fisher diffusion, a fundamental evolutionary model accounting for mutation, selection, genetic drift and potentially other evolutionary phenomena, but which introduces an intractable likelihood. By leveraging recent advances in exact simulation techniques for diffusions, coupled with a suitable state space augmentation and a carefully devised updating procedure, we have developed an \emph{exact} Bayesian inferential scheme which targets the true joint posterior for the selection parameters and allele age. In doing so we avoid having to rely on any approximations or discretisation of the underlying allele frequency model, thereby ensuring that the posterior plots obtained pertain to the model we are actually using and not to some approximation of it. Our method is very general in that it allows for inference of any polynomial selection function (including genic and diploid selection as well as more general frequency-dependent effects), as well as allele age, despite the limited statistical power present in the data. \\ \newline
The inferential algorithm has been made available as the user-friendly Python package \texttt{MCMC4WF}, downloadable at \url{www.github.com/JaroSant/MCMC4WF}. The method is implemented in C++ and can be called directly from within Python. An accompanying user manual as well as example scripts illustrating how the method can be applied to a given dataset can be found in the corresponding GitHub repository. We show our method is able to efficiently recover true signal for selection from simulated data both when inferring genic as well as diploid selection under a variety of parameter regimes, but it can also be used in the more general case to detect any form of polynomial selection. We applied our method to the well-known horse coat coloration dataset in \citet{Ludwig} which has been extensively analysed in the literature. We find strong evidence for overdominance and a large selection coefficient for the ASIP gene. These results mirror those reported in \citet{Steinrucken} and \citet{Schraiber}, where the authors conduct a similar analysis but use an approximate inferential method.  \\ \newline
The bimodality of some of our posteriors deserves further comment. The surfaces plot in Figures \ref{SigmaHKSDensityD}, \ref{SigmaT0KSDensityD}, \ref{HT0KSDensityD} suggest that inference of the selection parameters suffers from a certain degree of unidentifiability. This has already been noted in the literature \citep{Maruyama74, Schraiber13}, where their likelihood (assuming the absence of mutation) is symmetric in the selection coefficient, and thus the sign of the selection coefficient is unidentifiable when conducting inference based on observations coming from a Wright--Fisher diffusion bridge. In our analysis however, we observe from \eqref{RN4WF} that signal for the selection coefficients comes from both the difference in the spatial start and end points of the diffusion $X_t-X_0$, as well as the diffusion bridge joining them. Indeed, \eqref{RN4WF} offers an explicit decomposition of the likelihood into a product of these two sources of information. Expanding out the polynomials in \eqref{RN4WF}, one observes that the only non-symmetric terms (in $\sigma, h$) are all multiplied by (at least) a linear term in $x$, and thus the larger $X_t-X_0$ the more signal in the data favouring larger values of $\sigma$ and $h$. We observe this clearly when comparing the posterior plots obtained for the simulated data in Experiments C and D (where the maximal discrepancy between the start and end points is 0.3, and the posterior indicates inability to tell apart $(\sigma, h)$ from $(-\sigma, -h)$), to those obtained when applying our method to real data (where the corresponding differences are greater than 0.45, and thus the posterior is no longer symmetric). \\ \newline
To illustrate the above, we consider two qualitatively different parameter setups in Figure \ref{compare_paths}, where we plot 10,000 paths for both $(\sigma, h) = (20, 1)$ and $(\sigma, h) = (-20, -1)$, and then compare the resulting histograms. When using an observation time interval comparable to that used for the horse dataset analysed in the next section, it was further found that the two paths are statistically indistinguishable based on the Kolmogorov--Smirnov test run on the terminal value of the diffusion paths for both sets of paths.
Understanding the full extent of identifiability or lack thereof for selection parameters requires further investigation. \\ \newline
Although our approach targets the true posterior for the selection parameters and allele age, our implementation does not consider a multi-allelic locus beyond two alleles nor does it allow for changes in demography or for interactions between multiple loci (as in the case with the ASIP and MC1R loci which are known to be in recessive epistasis \citep{Thiruvenkadan08}). Generalising the model for a multi-allelic locus should be relatively straightforward provided the mutation model leads to a reversible diffusion, since the exact simulation techniques we have exploited allow for multi-dimensional reversible diffusions \citep{JenkinsSpano}. On the other hand, accounting for demography is rather intricate as the effective population size plays a role in scaling the population level mutation and selection parameters, as well as the diffusion time units. Having a varying demography therefore means that these quantities can change from one sampling time to another, leading to substantial mathematical complications since the underlying dominating measure for the diffusion sample path distribution may no longer be independent of the parameters. Furthermore, developing an exact framework which can account for interactions between multiple linked loci would require more complex models accounting for recombination and epistasis. This presents a formidable mathematical challenge beyond the scope of this article and forms the basis of future work (see \citet{GarciaPareja2021} for some progress in this direction). Nonetheless our method is the first to conduct \emph{exact} Bayesian inference for essentially and model of selection at a single locus, completely eliminating approximation error, illustrating the power such techniques have in inferring selection parameters and allele age.

\section{Acknowledgements}
JS was supported by the ERC (Starting Grant ARGPHENO 850869). PJ is supported in part by the EPSRC via ProbAI: A Hub for the Mathematical and Computational Foundations of Probabilistic AI (EP/Y028783/1). JK was supported through the EPSRC research grant EP/V049208/1. 

\bibliographystyle{apalike}
\bibliography{bib}

\appendix

\section{Exact Simulation for the Wright--Fisher diffusion}\label{AppendixExactSimulation}
In this appendix, we provide a brief summary of how exact draws can be generated from the laws of a Wright--Fisher diffusion, as presented in \citet{JenkinsSpano}. 
\subsection{Neutral Wright--Fisher simulation}\label{EA4NWF}
The neutral Wright--Fisher diffusion admits the following classical infinite series decomposition  \citep{Griffiths79,Tavare84,EthierGriffiths93,GriffithsSpano}
\begin{align}\label{NeutralTransitionDensity}
	p_{0, \boldsymbol{0}}^{\boldsymbol{\theta}}(t,x,y) = \sum_{m=0}^{\infty}q_{m}^{\boldsymbol{\theta}}(t)\sum_{l=0}^{m}\mathcal{B}_{m,x}(l)\mathcal{D}_{\theta_1+l,\theta_2+m-l}(y),
\end{align}
where 
\begin{align}\label{q_m}
	\hspace*{-2mm}q_{m}^{\boldsymbol{\theta}}(t) = \sum_{k=m}^{\infty}(-1)^{k-m}\frac{|\boldsymbol{\theta}|+2k-1}{m!(k-m)!}\frac{\Gamma(m+|\boldsymbol{\theta}|+k-1)}{\Gamma(m+|\boldsymbol{\theta}|)}e^{\frac{-k(k+|\boldsymbol{\theta}|-1)t}{2}},
\end{align}
$|\boldsymbol{\theta}| = \theta_1 + \theta_2$ for $\boldsymbol{\theta} \in \mathbb{R}_{+}^{2}$, $x, y \in [0,1]$, $\mathcal{B}_{m,x}$ is the probability mass function for a binomial random variable with parameters $m, x$, and $\mathcal{D}_{\theta_1+l,\theta_2+m-l}$ is the probability density function for a beta random variable with parameters $\theta_1+l, \theta_2+m-l$. We observe that \eqref{q_m} defines a probability mass function on $\mathbb{N}$, and is a manifestation of the duality between the Wright--Fisher diffusion and the Kingman coalescent as \eqref{q_m} describes the number of lineages still alive at time $t$ in a typed Kingman coalescent started from infinity at time 0 \citep{Griffiths79, Tavare84}. The mixture decomposition \eqref{NeutralTransitionDensity} suggests a simple routine to return draws from the neutral Wright--Fisher diffusion:
\begin{enumerate}
	\item Draw $M$ from \eqref{q_m}
	\item Conditional on $M=m$, draw $L \sim \textnormal{Bin}(m,x)$
	\item Conditional on $M=m, L=l$, draw $Y \sim \textnormal{Beta}(\theta_1+l,\theta_2+m-l)$ 
\end{enumerate}
Only step 1 from the above is non-trivial, and can be done through the use of the `alternating series' trick \citep[Chapter 4]{Devroye}. \\ \newline
A decomposition similar to \eqref{NeutralTransitionDensity} holds for the diffusion bridge case, and thus a similar simulation strategy can be entertained---we omit the details here but direct the interested reader to \citet[Section 3]{JenkinsSpano}.
\subsection{Non-neutral Wright--Fisher simulation}\label{EA4NNWF}
Denote by $\mathbb{WF}^{(x)}_{\sigma, \boldsymbol{\eta}, \boldsymbol{\theta}}$ the law of a Wright--Fisher diffusion with mutation parameter $\boldsymbol{\theta}$ and selection parameters $\sigma, \boldsymbol{\eta}$, and started from a point $x$, on the space of continuous functions mapping $\mathbb{R}^{+}$ into $[0,1]$. Then the Radon--Nikodym derivative between the laws of a non-neutral and neutral Wright--Fisher diffusion (sharing the same mutation parameters and started from the same point $x$) \citep{dawson, JenkinsSpano, Jenkins24} is given by 
\begin{align}\label{RN4WF}
	\frac{d\mathbb{WF}_{\sigma, \boldsymbol{\eta}, \boldsymbol{\theta}}^{(x)}}{d\mathbb{WF}_{0, \boldsymbol{0}, \boldsymbol{\theta}}^{(x)}}(X^t) &= \exp\bigg\{A_{\sigma, \boldsymbol{\eta}}(X_t) - A_{\sigma, \boldsymbol{\eta}}(X_0) - \int_{0}^{t}\varphi_{\sigma, \boldsymbol{\eta}}(X_s)ds\bigg\} ,
\end{align}
where $X^t := (X_s)_{0 \leq s \leq t}$, $A_{\sigma, \boldsymbol{\eta}}(x) := \frac{\sigma}{2}\int_{0}^{x}\eta(y)dy$, 
\begin{align*}
	\varphi_{\sigma, \boldsymbol{\eta}}(x) = \frac{\sigma}{4}\left(\left(-\theta_2x + \theta_1(1-x)\right)\eta(x) + x(1-x)\left(\frac{\sigma}{2}\eta^{2}(x) + \eta'(x)\right)\right), & & x\in[0,1].
\end{align*}
In view of the fact that both $A_{\sigma, \boldsymbol{\eta}}(x)$ and $\varphi_{\sigma, \boldsymbol{\eta}}(x)$ are polynomials of finite degree in $x \in [0,1]$, we can always find $A_{\sigma, \boldsymbol{\eta}}^{+}$, $\varphi_{\sigma, \boldsymbol{\eta}}^{-}, \varphi_{\sigma, \boldsymbol{\eta}}^{+}$ such that $A_{\sigma, \boldsymbol{\eta}}(x) \leq A_{\sigma, \boldsymbol{\eta}}^{+}$, $\varphi_{\sigma, \boldsymbol{\eta}}^{-}\leq\varphi_{\sigma, \boldsymbol{\eta}}(x)\leq\varphi_{\sigma, \boldsymbol{\eta}}^{+}$, and thus we can re-write \eqref{RN4WF} as 
\begin{align}\label{RNNeutralNonNeutralForSim}
	\frac{d\mathbb{WF}_{\sigma, \boldsymbol{\eta}, \boldsymbol{\theta}}^{(x)}}{d\mathbb{WF}_{0, \boldsymbol{0}, \boldsymbol{\theta}}^{(x)}}(X^t) \propto \exp\bigg\{A_{\sigma, \boldsymbol{\eta}}(X_t) - A_{\sigma, \boldsymbol{\eta}}^{+} - \int_{0}^{t}\left(\varphi_{\sigma, \boldsymbol{\eta}}(X_s)-\varphi_{\sigma, \boldsymbol{\eta}}^{-}\right)ds\bigg\}. 
\end{align}
The term involving the integral in the exponent above can then be viewed as the probability that a unit rate Poisson point process has all points in the epigraph of $t \mapsto \varphi_{\sigma, \boldsymbol{\eta}}(X_t)-\varphi_{\sigma, \boldsymbol{\eta}}^{-}$. The remaining term on the right-hand side of \eqref{RNNeutralNonNeutralForSim} corresponds to the probability that a coin having probability $(e^{A_{\sigma, \boldsymbol{\eta}}(x) - A_{\sigma, \boldsymbol{\eta}}^{+}})$ of landing on heads does so. Thus to simulate a draw from the law of a non-neutral Wright--Fisher diffusion one simulates a neutral draw together with a corresponding Poisson point process and coin flip. One then conducts an accept/reject step, retaining only those draws for which all of the simulated Poisson points are in the epigraph of the above-mentioned map  \emph{and} the coin flip is a success. Retained draws are distributed according to the law of a non-neutral Wright--Fisher diffusion. Similar arguments allow for the simulation of non-neutral diffusion bridges through the use of neutral diffusion bridges as candidates in a rejection sampler. Further details can be found in \citet{JenkinsSpano} and \citet{GriffithsJenkinsSpano}. \\ \newline
The Radon--Nikodym derivative \eqref{RNNeutralNonNeutralForSim} also leads to a useful expression for the non-neutral transition density, which will be used in the following section to derive the tractable expression \eqref{AugmentedLikelihoodGlobal} for the likelihood. Denote by $\mathbb{WF}_{\sigma, \boldsymbol{\eta}, \boldsymbol{\theta}}^{(t, x, y)}$ the law of a Wright--Fisher diffusion bridge going from $x$ at time 0 to $y$ at time $t$, and having mutation parameter $\boldsymbol{\theta}$, and selection parameters $\sigma$ and $\boldsymbol{\eta}$. Then the Radon--Nikodym derivative between the laws of a neutral and non-neutral Wright--Fisher diffusion bridge (provided they share the same mutation parameters $\boldsymbol{\theta}=(\theta_1,\theta_2)\in\mathbb{R}_{+}^{2}$) can be obtained from \eqref{RN4WF} by conditioning on the endpoints to get
\begin{align}\label{RNForWFBridge}
	\frac{d\mathbb{WF}_{\sigma,\boldsymbol{\eta},\boldsymbol{\theta}}^{(t,x,y)}}{d\mathbb{WF}_{0,\boldsymbol{0},\boldsymbol{\theta}}^{(t,x,y)}}(X^t) &= \frac{p_{0, \boldsymbol{0}}^{\boldsymbol{\theta}}(t,x,y)}{p_{\sigma, \boldsymbol{\eta}}^{\boldsymbol{\theta}}(t,x,y)}\frac{d\mathbb{WF}_{\sigma,\boldsymbol{\eta},\boldsymbol{\theta}}^{(x)}}{d\mathbb{WF}_{0,\boldsymbol{0},\boldsymbol{\theta}}^{(x)}}(X^t) \nonumber \\
	&\propto \exp\left\{ - \int_{0}^{t}\left(\varphi_{\sigma, \boldsymbol{\eta}}(X_s)-\varphi_{\sigma, \boldsymbol{\eta}}^{-}\right)ds\right\}.
\end{align}
Re-arranging and integrating both sides with respect to the bridge measure $\mathbb{WF}_{0,\boldsymbol{0},\boldsymbol{\theta}}^{(t,x,y)}$, we get that
\begin{align}\label{WFNonNeutralTransitionDecomposition}
	p_{\sigma, \boldsymbol{\eta}}^{\boldsymbol{\theta}}(t,x,y) &= p_{0,\boldsymbol{0}}^{\boldsymbol{\theta}}(t,x,y)e^{A_{\sigma, \boldsymbol{\eta}}(y)-A_{\sigma, \boldsymbol{\eta}}(x)-t\varphi_{\sigma, \boldsymbol{\eta}}^{-}}\mathbb{E}_{\mathbb{WF}_{0,\boldsymbol{0},\boldsymbol{\theta}}^{(t,x,y)}}\left[e^{-\int_{0}^{t}\left(\varphi_{\sigma, \boldsymbol{\eta}}(X_s)-\varphi_{\sigma, \boldsymbol{\eta}}^{-}\right)ds}\right] \nonumber \\
	&=: p_{0,\boldsymbol{0}}^{\boldsymbol{\theta}}(t,x,y)e^{A_{\sigma, \boldsymbol{\eta}}(y)-A_{\sigma, \boldsymbol{\eta}}(x)-t\varphi_{\sigma, \boldsymbol{\eta}}^{-}}a(t,x,y,\sigma,\boldsymbol{\eta}).
\end{align}
Note that whilst the first two quantities on the right-hand side admit explicit representations in terms of known functions, the term $a(t,x,y,\sigma,\boldsymbol{\eta})$ is intractable in view of the Lebesgue integral involved and the fact that it is an average over the space of continuous functions with respect to the measure $\mathbb{WF}_{0,\boldsymbol{0},\boldsymbol{\theta}}^{(t,x,y)}$.

\section{Deriving of a tractable likelihood}\label{AppendixTractableLikelihood}
In this appendix we derive the tractable expression \eqref{AugmentedLikelihoodGlobal}. We start by considering a simple case when we only have two observations, and then generalise to the case with $K$ observations. \\ \newline
Assume that there is only one observation interval $[0,t]$, and assume that the latent diffusion assumes the values $X_{0} = x$ and $X_t = y$. Let $(\Psi,\Gamma,\Xi)$ define a unit rate Poisson point process on $(0,t)\times[0,1]\times(0,\infty)$, such that for any bounded $A \subset (0,\infty)$, the restriction of this process to $(0,t)\times[0,1]\times A$ (which we denote $(\Psi,\Gamma,\Xi)|_{A}$) gives that $\Psi|_{A} := \{ \psi_{j} : \xi_{j} \in A \}\overset{\textnormal{iid}}{\sim}\textnormal{Unif}((0,t))$, $\Gamma|_{A} = \{\gamma_{j} : \xi_j \in A\}\sim\textnormal{Unif}([0,1])$, $\Xi|_{A} = \{ \xi_{j} : \xi_j \in A \}\overset{\textnormal{iid}}{\sim}\textnormal{Unif}(A)$. The role of the extra variate $\Xi$ here is to allow for a non-centred re-parametrisation of the problem, where the Poisson point process is decoupled from the parameter of interest (here $\sigma$ and $\boldsymbol{\eta}$), thereby leading to a more efficient sampler. See \citet{Papaspiliopoulos2003} as well as \citet[Section 4]{Sermaidis} for further details. \\ \newline
Let $\omega \sim \mathbb{WF}_{0,\boldsymbol{0},\boldsymbol{\theta}}^{(t,x,y)}$ denote a neutral path from a Wright--Fisher bridge going from $x$ to $y$ in time $t$. Then, as detailed in \ref{EA4NNWF}, given a realisation of $(\Psi,\Gamma,\Xi)$ and $\omega$, one checks that the generated points lie in the epigraph of the function $s\mapsto \frac{\varphi_{\sigma, \boldsymbol{\eta}}(X_s)-\varphi_{\sigma, \boldsymbol{\eta}}^{-}}{\varphi_{\sigma, \boldsymbol{\eta}}^{+}-\varphi_{\sigma, \boldsymbol{\eta}}^{-}}$, and if this is the case, one defines $\omega^{\Psi} := \{ \omega_{\psi_{j}} \}$ (i.e.\ $\omega^{\Psi}$ corresponds to the values of the path $\omega$ at the timestamps given by $\Psi$), and stores the generated points by setting $\Phi:=(\Psi,\Xi,\omega^{\Psi})$ which will be henceforth termed the skeleton points. Such an accepted configuration $(\Psi,\Xi,\omega^{\Psi})$ has density given by
\begin{align}\label{SkeletonDensity}
	p(\Psi,\Xi,\omega^{\Psi}) = \frac{\displaystyle\prod_{\{j : \xi_j \leq \lambda_{\sigma}\}}\frac{\varphi_{\sigma, \boldsymbol{\eta}}^{+}-\varphi_{\sigma, \boldsymbol{\eta}}(\omega_{\psi_{j}})}{\varphi_{\sigma, \boldsymbol{\eta}}^{+}-\varphi_{\sigma, \boldsymbol{\eta}}^{-}}}{a(t,x,y,\sigma,\boldsymbol{\eta})}
\end{align}
with respect to $\mathbb{PP}^{(t)}\otimes\mathbb{WF}_{0,\boldsymbol{0},\boldsymbol{\theta}}^{(t,x,y)}$, where $\mathbb{PP}^{(t)}$ denotes the law of a unit rate Poisson point process on $(0,t)\times(0,\infty)$. Note that we have marginalised out the uniform marks $\{\gamma_{j}\}$. Observe now that the denominator in \eqref{SkeletonDensity} matches the same term appearing in the numerator of \eqref{WFNonNeutralTransitionDecomposition} if one considers the joint distribution of the latent diffusion and the skeleton points. \\ \newline
We further point out that the dominating measure $\mathbb{PP}^{(t)}\otimes\mathbb{WF}_{0,\boldsymbol{0},\boldsymbol{\theta}}^{(t,x,y)}$ depends on both the endpoints $x$ and $y$, as well as the time interval $t$. Since we want to target the posterior of both the allele age and selection parameters through the use of a Metropolis-within-Gibbs scheme, we need to ensure that the dominating measure does not depend on any of the parameters of interest nor on any of the auxiliary variables \citep{RobertsStramer01, Sermaidis}. The dependence on the time increment $t$ will only be problematic when $t_0$ is involved as all other timestamps $t_i$ for $i\in\{1,\dots,n\}$ will be fixed. For the time interval $[t_0, t_1]$ we make use of a unit rate Poisson point process on $(0, \infty)^{2}$ as dominating measure (whose law we denote by $\mathbb{PP}$), and thin accordingly to retain the corresponding Poisson points. More precisely, for any given $t_0$, we retain only those Poisson points that fall within the interval $[t_0, t_1]$, thereby ensuring that the resulting points have the appropriate law $\mathbb{PP}^{(t_0, t_1)}$ and that one only ever generates a finite number of points. \\ \newline
To deal with the dependence of the reference diffusion bridge measure on the endpoints $x$ and $y$, we start by changing the dominating measure from that of a diffusion bridge to that of a diffusion. By conditioning on the endpoint $X_t = y$, we have that 
\begin{align}\label{ChangingMeasure}
	\frac{d\mathbb{WF}_{0,\boldsymbol{0},\boldsymbol{\theta}}^{(t,x,y)}}{d\mathbb{WF}_{0,\boldsymbol{0}, \boldsymbol{\theta}}^{(x)}}\left(X^{t}\right) = \frac{1}{p_{0, \boldsymbol{0}}^{\boldsymbol{\theta}}(t,x,y)}
\end{align}
and thus by multiplying \eqref{SkeletonDensity} by the resulting Radon--Nikodym derivative we get that the dominating Wright--Fisher measure for \eqref{SkeletonDensity} is now $\mathbb{WF}_{0,\boldsymbol{0},\boldsymbol{\theta}}^{(x)}$. This still depends on the left-hand bridge endpoint $x$, however when updating the latent path, these measures are chained to one another, such that the global dominating measure is given by $\mathbb{WF}_{0,\boldsymbol{0},\boldsymbol{\theta}}^{(0)}$ (see Subsection \ref{Updating} for details on how the updating procedure is executed). In particular, we observe that
\begin{align*}
    \frac{d\left(\mathbb{WF}_{0,\boldsymbol{0},\boldsymbol{\theta}}^{(s,0,x)}\otimes\mathbb{WF}_{0,\boldsymbol{0},\boldsymbol{\theta}}^{(t,x,y)}\right)}{d\mathbb{WF}_{0,\boldsymbol{0},\boldsymbol{\theta}}^{(s+t,0,y)}}\left(X^{s+t}\right) = \frac{p_{0, \boldsymbol{0}}^{\boldsymbol{\theta}}(s+t, 0, y)}{p_{0, \boldsymbol{0}}^{\boldsymbol{\theta}}(s, 0, x)p_{0, \boldsymbol{0}}^{\boldsymbol{\theta}}(t, x, y)}
\end{align*}
and making use of \eqref{ChangingMeasure} together with the above allows us to deduce that
\begin{align}\label{BridgeToDiffusion}
\frac{d\left(\mathbb{WF}_{0,\boldsymbol{0},\boldsymbol{\theta}}^{(s,0,x)}\otimes\mathbb{WF}_{0,\boldsymbol{0},\boldsymbol{\theta}}^{(t,x,y)}\right)}{d\mathbb{WF}_{0,\boldsymbol{0},\boldsymbol{\theta}}^{(0)}}\left(X^{s+t}\right) &= \frac{d\left(\mathbb{WF}_{0,\boldsymbol{0},\boldsymbol{\theta}}^{(s,0,x)}\otimes\mathbb{WF}_{0,\boldsymbol{0},\boldsymbol{\theta}}^{(t,x,y)}\right)}{d\mathbb{WF}_{0,\boldsymbol{0},\boldsymbol{\theta}}^{(s+t,0,y)}}\left(X^{s+t}\right)\frac{d\mathbb{WF}_{0,\boldsymbol{0},\boldsymbol{\theta}}^{(s+t,0,y)}}{d\mathbb{WF}_{0,\boldsymbol{0},\boldsymbol{\theta}}^{(0)}}\left(X^{s+t}\right) \nonumber \\
    &= \frac{1}{p_{0,\boldsymbol{0}}^{\boldsymbol{\theta}}(s, 0, x)p_{0, \boldsymbol{0}}^{\boldsymbol{\theta}}(t, x, y)}.
\end{align}
Putting \eqref{WFNonNeutralTransitionDecomposition}, \eqref{SkeletonDensity} and the above two changes of measure together, we note that the joint density of  $(X_t, \Phi)$ is given by
\begin{align*}
	p(X_t, \Phi) = e^{A_{\sigma, \boldsymbol{\eta}}(X_t)-A_{\sigma, \boldsymbol{\eta}}(x)-\varphi_{\sigma, \boldsymbol{\eta}}^{-}t}\displaystyle\prod_{\{j : \xi_j \leq \lambda_{\sigma}\}}\frac{\varphi_{\sigma, \boldsymbol{\eta}}^{+}-\varphi_{\sigma, \boldsymbol{\eta}}(\omega_{\psi_{j}})}{\varphi_{\sigma, \boldsymbol{\eta}}^{+}-\varphi_{\sigma, \boldsymbol{\eta}}^{-}}
\end{align*}
with respect to $\textnormal{Leb}([0,1])\otimes\mathbb{PP}\otimes\mathbb{WF}_{0,\boldsymbol{0},\boldsymbol{\theta}}^{(x)}$, where the intractable $a(t,x,y,\sigma,\boldsymbol{\eta})$ as well as the infinite sum $p_{0, \boldsymbol{0}}^{\boldsymbol{\theta}}(t,x,y)$ cancel out when combining \eqref{WFNonNeutralTransitionDecomposition}, \eqref{SkeletonDensity} and \eqref{ChangingMeasure}. \\ \newline
Thus the joint density of the datum $y_{1}$, the selection coefficient $\sigma$, the selection parameters $\boldsymbol{\eta}$, the latent diffusion at the observation time $X_{1}$, and the skeleton point $\Phi$ (we ignore the allele age here), is given by
\begin{align*}
    p(y_1, \sigma, \boldsymbol{\eta}, X_1, \Phi) = \mathcal{B}_{n_{1}, X_1}(y_{1})e^{A_{\sigma, \boldsymbol{\eta}}(X_t) - A_{\sigma, \boldsymbol{\eta}}(x) - \varphi_{\sigma, \boldsymbol{\eta}}^{-}t}\displaystyle\prod_{\{j : \xi_j \leq \lambda_{\sigma}\}}\frac{\varphi_{\sigma, \boldsymbol{\eta}}^{+}-\varphi_{\sigma, \boldsymbol{\eta}}(\omega_{\psi_{j}})}{\varphi_{\sigma, \boldsymbol{\eta}}^{+}-\varphi_{\sigma, \boldsymbol{\eta}}^{-}}
\end{align*}
with respect to $\Sigma(n_{1})\otimes\textnormal{Leb}(\mathbb{R}\times\mathbb{R}^{d+1}\times[0,1])\otimes\mathbb{PP}\otimes\mathbb{WF}_{0,\boldsymbol{0},\boldsymbol{\theta}}^{(x)}$, where $\Sigma(n_{1})$ denotes the counting measures on $\{0, 1, \dots, n_{1}\}$, and we recall $d$ is the degree of the polynomial $\eta(\cdot)$. \\ \newline
Extending to the case when we have $K$ observations $\boldsymbol{y} = \{y_{i}\}_{i=1}^{K}$ is straightforward. Denoting by $\Phi_i = (\Psi_i , \Xi_i, \omega_i^{\Psi_i})$ the skeleton points over the time interval $[t_{i-1},t_i]$ for $i=1,\dots,K$ (and assuming that $y_{1} > 0$), the joint density of the data $\boldsymbol{y}$, the selection coefficient $\sigma$, selection parameters $\boldsymbol{\eta}$, the allele age $t_0$, the value of the diffusion at the sampling times $\boldsymbol{X} = \{ X_{i} \}_{i=1}^{K}$, and the collection of skeleton points $\{ \Phi_i \}_{i=1}^{K}$ is
\begin{align}\label{AugmentedLikelihoodGlobalExplicit}
    p(\boldsymbol{y}, \sigma, \boldsymbol{\eta}, t_0, \boldsymbol{X}, \{\Phi_i\}_{i=1}^{K}) &= q_{1}(\sigma)q_{2}(\boldsymbol{\eta})q_{3}(t_0)e^{A_{\sigma, \boldsymbol{\eta}}(X_K) - \varphi_{\sigma, \boldsymbol{\eta}}^{-}\left(t_{K}-t_{0}\right)} \nonumber \\
    &{}\quad{}\times\prod_{i=1}^{n}\mathcal{B}_{n_{i},X_i}\left(y_{i}\right)   \displaystyle\prod_{\left\{\substack{j : \xi_{i,j} \leq \lambda_{\sigma}, \\ \psi_{1,j} < t_1-t_0 }\right\}}\frac{\varphi_{\sigma, \boldsymbol{\eta}}^{+}-\varphi_{\sigma, \boldsymbol{\eta}}(\omega_{i,\psi_{i,j}})}{\varphi_{\sigma, \boldsymbol{\eta}}^{+}-\varphi_{\sigma, \boldsymbol{\eta}}^{-}}
\end{align}
with respect to the dominating measure 
\begin{align*}
\Sigma(\otimes_{i=1}^{K}n_{i})\otimes\textnormal{Leb}\left(E_{t_1}^{n}\right)\otimes\mathbb{PP}\otimes\mathbb{PP}^{(t_1, t_{K})}\otimes\mathbb{WF}_{0,\boldsymbol{0},\boldsymbol{\theta}}^{(0)}\otimes,
\end{align*}
where we set $E_{t_1}^{K} := \mathbb{R}\times\mathbb{R}^{d+1}\times(-\infty, t_1)\times[0,1]^{K}$, $q_1(\cdot), q_2(\cdot), q_3(\cdot)$ represent the prior densities on $\sigma$, $\boldsymbol{\eta}$ and $t_0$ respectively with respect to Lebesgue measure, $\mathbb{WF}_{0,\boldsymbol{0},\boldsymbol{\theta}}^{(0)}$ denotes the law of a neutral Wright--Fisher diffusion started from frequency $0$, $\mathbb{PP}^{(t_1, t_{K})}$ is the law of a unit rate Poisson point process on $(t_{1},t_{K})\times(0,\infty)$, $\mathbb{PP}$ is the law of a unit rate Poisson point process on $(0,\infty)^{2}$, and $\Sigma(\otimes_{i=1}^{K}n_{i})$ is the counting measure over $\otimes_{i=1}^{K}\{0,\dots,n_{i}\}$. Note that conditioning on $X_i = x_i$ for each $i \in \{1, \dots, K\}$, we applied the following change of measure (which follows by iterating \eqref{BridgeToDiffusion}) to ensure that the dominating measure for the skeleton points does not depend on the values $\boldsymbol{X}$: 
\begin{align*}
	&\frac{d\left(\underset{i=1}{\overset{K}{\otimes}}\mathbb{WF}_{0,\boldsymbol{0},\boldsymbol{\theta}}^{(t_{i}-t_{i-1},x_{{i-1}},x_{i})}\right)}{d\mathbb{WF}_{0,\boldsymbol{0},\boldsymbol{\theta}}^{(0)}}\left((X_t)_{t=t_0}^{t_K}\right) 
    = \prod_{i=1}^{K}\frac{1}{p_{0}^{\boldsymbol{\theta}}(t_i-t_{i-1},x_{{i-1}},x_{i})}.
\end{align*}
which follows by iterating \eqref{ChangingMeasure} over subsequent observation time intervals. 

\section{Path updates and likelihood contributions}\label{AppendixUpdatingProcedure}
In Subsection \ref{Updating} we illustrated how different path segments are updated in our algorithm and provided the relevant acceptance probabilities. Here we compute the separate likelihood contributions and proposal densities for each path segment, thereby deriving said acceptance probabilities. 
\subsection{Initial path segments}\label{InitialPathsDerivation}
Fixing the right endpoint $X_2 = x_{2}$, we can extract from \eqref{AugmentedLikelihoodGlobalExplicit} the joint density of the data $y_{1}$, the selection coefficient $\sigma$, selection parameters $\boldsymbol{\eta}$, the allele age $t_0$, the diffusion value $X_1$ at the first observation time, and corresponding skeleton points $\Phi_1, \Phi_2$, which conditional on $x_2$ is given by
\begin{align}\label{AugmentedLikelihoodInitial}
    p(y_1, \sigma, \boldsymbol{\eta}, t_0, X_1, \{\Phi_i\}_{i=1}^{2}| X_2 = x_2) &= q_{1}(\sigma)q_{2}(\boldsymbol{\eta})q_{3}(t_0)\mathcal{B}_{n_{1},X_1}\left(y_{1}\right) e^{\frac{\sigma}{2}x_{{2}} - \varphi_{\sigma, \boldsymbol{\eta}}^{-}\left(t_{2}-t_{0}\right)} \nonumber \\
    &\quad{}\times\prod_{i=1}^{2}\displaystyle\prod_{\substack{\{j : \xi_{i,j} \leq \lambda_{\sigma}, \\ \psi_{1,j} < t_1-t_0\} }}\frac{\varphi_{\sigma, \boldsymbol{\eta}}^{+}-\varphi_{\sigma, \boldsymbol{\eta}}(\omega_{i,\psi_{i,j}})}{\varphi_{\sigma, \boldsymbol{\eta}}^{+}-\varphi_{\sigma, \boldsymbol{\eta}}^{-}}
\end{align}
with respect to the dominating measure 
\begin{align*}
	\Sigma(n_{1})\otimes\textnormal{Leb}\left(E_{t_1}^{1}\right)\otimes\mathbb{PP}\otimes\mathbb{PP}^{(t_1, t_2)}\otimes\mathbb{WF}_{0,\boldsymbol{0},\boldsymbol{\theta}}^{(0)}.
\end{align*}
A proposal $(\widetilde{t}_0, \widetilde{X}_{1}, \widetilde{\Phi}_1, \widetilde{\Phi}_{2})$ generated according to the routine described in Subsection \ref{InitialPathSegmentUpdate} has density
\begin{align}\label{ProposalXtct1}
    p(\widetilde{t}_0, \widetilde{X}_1, \{\widetilde{\Phi}_i\}_{i=1}^{2}) &= g_{\textnormal{age}}(\widetilde{t}_0|\cdot)\frac{e^{\frac{\sigma}{2}x_2 - \varphi_{\sigma, \boldsymbol{\eta}}^{-}\left(t_2-\widetilde{t}_0\right)}}{p_{\sigma, \boldsymbol{\eta}}^{\boldsymbol{\theta}}(t_2-\widetilde{t}_0,0,x_2)}\prod_{i=1}^{2}\displaystyle\prod_{\substack{\{j : \widetilde{\xi}_{i,j} \leq \lambda_{\sigma}, \\ \widetilde{\psi}_{1,j} < t_1-\widetilde{t}_0\}}}\frac{\varphi_{\sigma, \boldsymbol{\eta}}^{+}-\varphi_{\sigma, \boldsymbol{\eta}}(\widetilde{\omega}_{i,\widetilde{\psi}_{i,j}})}{\varphi_{\sigma, \boldsymbol{\eta}}^{+}-\varphi_{\sigma, \boldsymbol{\eta}}^{-}}
\end{align}
with respect to 
\begin{align*}
	\textnormal{Leb}((-\infty,t_1)\times[0,1])\otimes\mathbb{PP}\otimes\mathbb{PP}^{(t_1, t_2)}\otimes\mathbb{WF}_{0,\boldsymbol{0},\boldsymbol{\theta}}^{(0)}.
\end{align*}
Combining \eqref{AugmentedLikelihoodInitial} with the above \eqref{ProposalXtct1}, we get \eqref{AcceptanceT0}. The only problematic term here is the ratio of intractable quantities of the form $a(t,x,y,\sigma,\boldsymbol{\eta})$ which cannot be evaluated exactly but can be estimated unbiasedly through the use of a Poisson estimator (see \ref{AppendixPoissonEstimator} for further details). We point out that whilst the ratio of neutral transition densities present in \eqref{AcceptanceT0} cannot be evaluated exactly, it can be easily targetted via a refinement scheme. In particular, one can show that for a sufficiently large number of terms, there exist monotonically tightening upper and lower bounds for the neutral transition density which converge to the true value \citep[Proposition 4]{JenkinsSpano}, and thus one can keep on refining either bound until a decision can be made on whether to accept or reject the proposed move. This is an instance of the `alternating series' method \citep[Chapter 4]{Devroye}.
\subsection{Updating internal path segments}
Take $i\in\{2,\dots,K-1\}$, fix $X_{i-1} = x_{{i-1}}$ and $X_{i+1} = x_{{i+1}}$, and denote by $\Phi_i, \Phi_{i+1}$ the collection of skeleton points over the time intervals $[t_{i-1},t_i], [t_i,t_{i+1}]$ respectively. Then from \eqref{AugmentedLikelihoodGlobalExplicit} the joint density of the data $y_{i}$, the selection coefficients $\sigma$ and $\boldsymbol{\eta}$, the value of the diffusion $X_i$ and corresponding skeleton points $\Phi_i,\Phi_{i+1}$ is given by
\begin{align}\label{AugmentedLikelihoodInterior}
    p(y_i, \sigma, \boldsymbol{\eta}, X_i, \{\Phi_i\}_{i=i}^{i+1} &| X_{i-1} = x_{i-1}, X_{i+1} = x_{i+1}) \nonumber \\
    &= q_{1}(\sigma)\mathcal{B}_{n_{i},X_i}\left(y_{i}\right) e^{A_{\sigma, \boldsymbol{\eta}}(x_{{i+1}}) - A_{\sigma, \boldsymbol{\eta}}(x_{{i-1}}) - \varphi_{\sigma, \boldsymbol{\eta}}^{-}\left(t_{i+1}-t_{i-1}\right)} \nonumber \\
    &{}\quad{}\times\prod_{k=i}^{i+1}\displaystyle\prod_{\{j : \xi_{k,j} \leq \lambda_{\sigma}\}}\frac{\varphi_{\sigma, \boldsymbol{\eta}}^{+}-\varphi_{\sigma, \boldsymbol{\eta}}(\omega_{k,\psi_{k,j}})}{\varphi_{\sigma, \boldsymbol{\eta}}^{+}-\varphi_{\sigma, \boldsymbol{\eta}}^{-}}
\end{align}
with respect to the dominating measure 
\begin{align*}
	\Sigma(n_{i})\otimes\textnormal{Leb}\left(\mathbb{R}\times\mathbb{R}^{d+1}\times[0,1]\right)\otimes\mathbb{PP}^{(t_{i-1},  t_{i+1})}\otimes\mathbb{WF}_{0,\boldsymbol{0},\boldsymbol{\theta}}^{(x_{{i-1}})}.
\end{align*} 
A proposal $(\widetilde{X}_i,\widetilde{\Phi}_i,\widetilde{\Phi}_{i+1})$ generated by the mechanism in Section \ref{InteriorEndPathSegmentUpdate} has density given by
\begin{align}\label{ProposalXti}
    p(\widetilde{X}_i,\widetilde{\Phi}_i,\widetilde{\Phi}_{i+1}) &= \frac{e^{\frac{\sigma}{2}\left(x_{i+1}-x_{i-1}\right)-\varphi_{\sigma, \boldsymbol{\eta}}^{-}\left(t_{i+1}-t_{i-1}\right)}}{p_{\sigma, \boldsymbol{\eta}}^{\boldsymbol{\theta}}(t_{i+1}-t_{i-1},x_{i-1},x_{i+1})}\prod_{k=i}^{i+1}\displaystyle\prod_{\{j : \widetilde{\xi}_{k,j} \leq \lambda_{\sigma}\}}\frac{\varphi_{\sigma, \boldsymbol{\eta}}^{+}-\varphi_{\sigma, \boldsymbol{\eta}}(\widetilde{\omega}_{k,\widetilde{\psi}_{k,j}})}{\varphi_{\sigma, \boldsymbol{\eta}}^{+}-\varphi_{\sigma, \boldsymbol{\eta}}^{-}}
\end{align}
with respect to $\mathbb{PP}^{(t_{i}, t_{i+1})}\otimes\mathbb{WF}_{0,\boldsymbol{0}, \boldsymbol{\theta}}^{(x_{i-1})}\otimes\textnormal{Leb}([0,1])$, matching the corresponding part of the dominating measure of \eqref{AugmentedLikelihoodInterior}. Since the non-neutral transition density appearing in the denominator of \eqref{ProposalXti} does not depend on any of the simulated quantities, we deduce that it cancels out when computing the Metropolis--Hastings acceptance probability for this proposal, leading to the expression \eqref{AcceptanceXti}. 
\subsection{Updating an end path segment}\label{UpdateXtn}
Fixing $X_{K-1} = x_{{K-1}}$, we get from \eqref{AugmentedLikelihoodGlobalExplicit} that the joint density of the data $y_{K}$, the selection coefficient $\sigma$ and selection parameter $\boldsymbol{\eta}$, the diffusion endpoint $X_K$ and corresponding skeleton points $\Phi_K$ is given by
\begin{align}\label{AugmentedLikelihoodEnd}
	p(y_{K}, \sigma, \boldsymbol{\eta}, X_K, \Phi_K | X_{K-1} = x_{K-1}) &= q_{1}(\sigma)q_2(\boldsymbol{\eta})\mathcal{B}_{n_{K},X_K}\left(y_{K}\right) e^{\frac{\sigma}{2}\left(X_K-x_{{K-1}}\right) - \varphi_{\sigma, \boldsymbol{\eta}}^{-}\left(t_{K}-t_{K-1}\right)}\nonumber \\
    &{}\quad{}\times\displaystyle\prod_{\{j : \xi_{K,j} \leq \lambda_{\sigma}\}}\frac{\varphi_{\sigma, \boldsymbol{\eta}}^{+}-\varphi_{\sigma, \boldsymbol{\eta}}(\omega_{K,\psi_{K,j}})}{\varphi_{\sigma, \boldsymbol{\eta}}^{+}-\varphi_{\sigma, \boldsymbol{\eta}}^{-}}
\end{align}
with respect to the dominating measure 
\begin{align*}
	\Sigma(n_{K})\otimes\textnormal{Leb}\left(\mathbb{R}\times\mathbb{R}^{d+1}\times[0,1]\right)\otimes\mathbb{PP}^{(t_{K-1},t_{K})}\otimes\mathbb{WF}_{0,\boldsymbol{\theta}}^{(x_{{K-1}})}.
\end{align*}
A proposal $(\widetilde{X}_K,\widetilde{\Phi}_K)$ generated by the procedure described in Subsection \ref{EndPathSegmentUpdates} has density
\begin{align}\label{ProposalXtn}
    p(\widetilde{X}_K,\widetilde{\Phi}_K) &= e^{A_{\sigma, \boldsymbol{\eta}}(\widetilde{X}_K)-A_{\sigma, \boldsymbol{\eta}}(x_{K-1})-\varphi_{\sigma, \boldsymbol{\eta}}^{-}\left(t_K-t_{K-1}\right)}\displaystyle\prod_{\{j : \widetilde{\xi}_{K,j} \leq \lambda_{\sigma}\}}\frac{\varphi_{\sigma, \boldsymbol{\eta}}^{+}-\varphi_{\sigma, \boldsymbol{\eta}}(\widetilde{\omega}_{K,\widetilde{\psi}_{K,j}})}{\varphi_{\sigma, \boldsymbol{\eta}}^{+}-\varphi_{\sigma, \boldsymbol{\eta}}^{-}}
\end{align}
with respect to the dominating measure $\textnormal{Leb}([0,1])\otimes\mathbb{PP}^{(t_{K-1}, t_{K})}\otimes\mathbb{WF}_{0,\boldsymbol{\theta}}^{(x_{K-1})}$, leading to \eqref{AcceptanceXtn}.

\section{Poisson Estimator and Pseudo-Marginal Algorithm}\label{AppendixPoissonEstimator}
We now provide some details on how the Poisson estimator \citep{Wagner88, Wagner89} can be used to the deal with the ratio of intractable quantities $a(t,x,y,\sigma,\boldsymbol{\eta})$, and how this estimator is implemented within our MCMC scheme. Further details on the Poisson estimator can be found in \citet{Beskos2006}. \\ \newline
We start by introducing the quantities $\kappa \sim \textnormal{Pois}(\lambda t)$, $\tau = \{ \tau_j \}_{j=1}^{\kappa} \overset{\textnormal{iid}}{\sim} \textnormal{Unif}((0,t))$ and $\zeta \sim \mathbb{WF}_{0,\boldsymbol{0},\boldsymbol{\theta}}^{(t,x,y)}$, which we combine together to obtain
\begin{align*}
	e^{\left(\lambda-c\right)t}\prod_{j=1}^{\kappa}\left(\frac{c-\left(\varphi_{\sigma, \boldsymbol{\eta}}(\zeta_{\tau_{j}})-\varphi_{\sigma, \boldsymbol{\eta}}^{-}\right)}{\lambda}\right),
\end{align*}
where $\lambda \in \mathbb{R}_{+}, c>0$ are two arbitrary constants. This is an unbiased estimator for $a(t, x, y, \sigma, \boldsymbol{\eta})$ \citep{Wagner88, Beskos2006}, and has second moment given by 
\begin{align}\label{SecondMomentPoisson}
	e^{\left(\lambda-2c\right)t}\mathbb{E}_{\mathbb{WF}_{0,\boldsymbol{0},\boldsymbol{\theta}}^{(t,x,y)}}\left[ \exp\left\{ \int_{0}^{t}\frac{\left(c-\varphi_{\sigma, \boldsymbol{\eta}}(X_s)-\varphi_{\sigma, \boldsymbol{\eta}}^{-}\right)^{2}}{\lambda}ds \right\} \right],
\end{align}
which makes choosing a pair $(\lambda,c)$ which minimises the variance non-trivial due to the associated Lebesgue integrals being incomputable. Instead, as discussed in \citet[Section 7]{Beskos2006}, we upper-bound the terms in the exponent in \eqref{SecondMomentPoisson} by $\exp\left\{\lambda -2c + \frac{c^{2}}{\lambda}  \right\}$ and minimise this bound in $c$ to deduce that $c = \lambda$, and subsequently set $\lambda = \varphi_{\sigma, \boldsymbol{\eta}}^{+}-\varphi_{\sigma, \boldsymbol{\eta}}^{-}$. Setting $\zeta_{\tau}:=\{ \zeta_{\tau_{j}} \}_{j=1}^{\kappa}$, this choice leads to the unbiased estimator
\begin{align}\label{PoissonEstimator}
	\widetilde{a}(t,x,y,\sigma,\boldsymbol{\eta},\tau,\zeta_{\tau}) = \prod_{j=1}^{\kappa}\left(\frac{\varphi_{\sigma, \boldsymbol{\eta}}^{+}-\varphi_{\sigma, \boldsymbol{\eta}}(\zeta_{\tau_{j}})}{\varphi_{\sigma, \boldsymbol{\eta}}^{+}-\varphi_{\sigma, \boldsymbol{\eta}}^{-}}\right).
\end{align}
It now remains to detail how and when we employ the above unbiased estimate. If one re-estimates $a(t,x,y,\sigma,\boldsymbol{\eta})$ using \eqref{PoissonEstimator} each time the quantity needs to be evaluated, the resulting MCMC scheme does not target the correct posterior (see \citet{Beaumont2003} and \citet[Table 1]{Andrieu}). Instead, if the estimator is recalculated solely when a new proposal is introduced with the resulting estimate being stored for an accepted proposal to be used in the subsequent iterations (and thus we augment the state space further with the variables $\tau$ and $\zeta_{\tau}$, and recycle these in the next iteration), then the resulting pseudo-marginal scheme does indeed target the correct distribution. By making use of this \emph{pseudo-marginal} formulation and expression for $\widetilde{a}(t,x,y,\sigma,\boldsymbol{\eta},\tau,\zeta_{\tau})$, the pseudo-marginal Metropolis--Hastings acceptance probability at iteration $m$ becomes computable:
\begin{align}\label{PseudoAP}
	\widetilde{\alpha}_{\textnormal{init}} &= \min\Bigg\{1,\frac{q_2(\widetilde{t}_0)}{q_2(t_{0}^{(m-1)})}\frac{g_{\textnormal{age}}(t_0^{(m-1)}|\widetilde{t}_0)}{g_{\textnormal{age}}(\widetilde{t}_0|t_{0}^{(m-1)})}e^{ -\varphi_{\sigma, \boldsymbol{\eta}}^{-}\left(t_{0}^{(m-1)}-\widetilde{t}_0\right) } \frac{p_{0}^{\boldsymbol{\theta}}(t_2-\widetilde{t}_0,0,x_2)}{p_{0}^{\boldsymbol{\theta}}(t_2-t_0^{(m-1)},0,x_2)} \nonumber\\
	&\qquad{}\qquad{}\times\frac{\widetilde{a}(t_2-\widetilde{t}_0,0,x_2,\sigma,\boldsymbol{\eta},\tau,\zeta_{\tau})}{\widetilde{a}(t_2-t_0^{(m-1)},0,x_2,\sigma,\boldsymbol{\eta},\tau^{(m-1)},\zeta_{\tau^{(m-1)}}^{(m-1)})} \Bigg\}
\end{align}
where $\tau,\zeta_{\tau}$ are the variables generated jointly with the proposals $t_0$, $X_1$, $\Phi_1$, and $\Phi_2$ to unbiasedly estimate $a(t_2-t_0,0,x_2,\sigma, \boldsymbol{\eta})$, whilst $\tau^{(k)},\zeta_{\tau^{(k)}}^{(k)}$ are the stored values which were used in the previous iteration. So replacing the acceptance probability \eqref{AcceptanceT0} by \eqref{PseudoAP} leads to an implementable pseudo-marginal Metropolis--Hastings update over the initial path segment. 

\section{Extra Plots}\label{AppendixExtraPlots}
We include here some further plots to illustrate the algorithm's performance.

\subsection{Output when inferring genic selection}

\begin{figure}[H]
\centering
\begin{subfigure}[b]{0.495\textwidth}
	\centering
	\includegraphics[width=\textwidth]{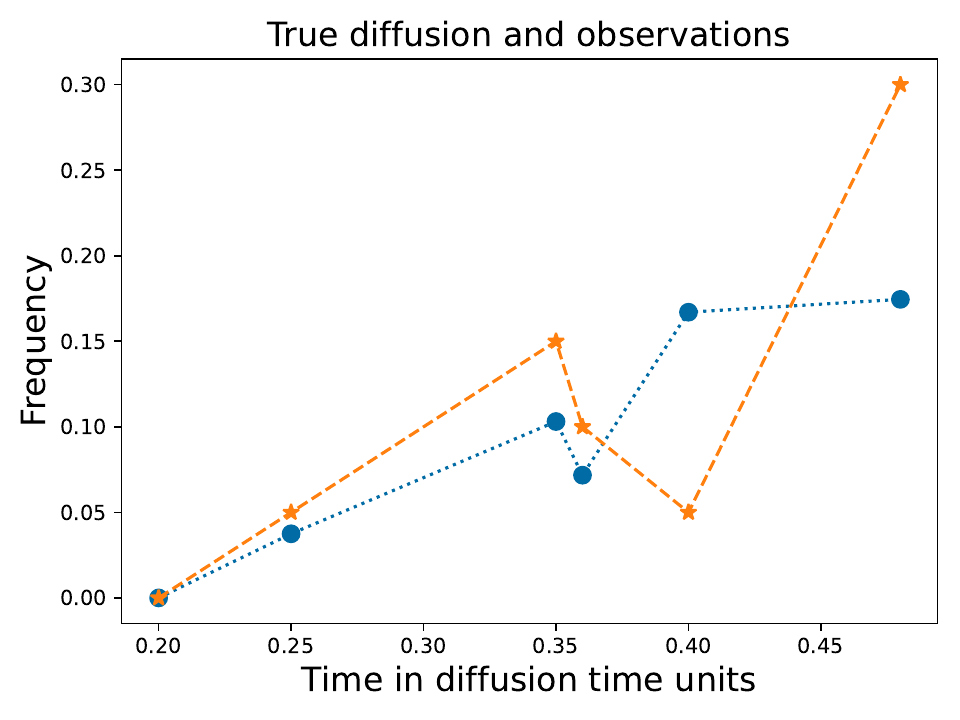}\\
	\caption{A}  
	\label{ObsA}
\end{subfigure}
\hfill
\begin{subfigure}[b]{0.495\textwidth}
	\centering
	\includegraphics[width=\textwidth]{simulations/plots/2024-06-12-14-37Observations.pdf}\\
	\caption{B}  
	\label{ObsB}
\end{subfigure}
\caption[The simulated dataset generated via the parameter configurations in the first two rows of Table \ref{SetupTable}]{Simulated data generated for the case when inferring solely genic selection and allele age, corresponding to experiments A and B in Table \ref{SetupTable}. Blue circles denote exact draws from the Wright--Fisher diffusion generated via \texttt{EWF}, whilst the orange stars are the binomial draws obtained. Dotted lines between observations are a linear interpolation.}
\label{Observations_genic}
\end{figure}

\begin{figure}[H]
	\centering
	\begin{subfigure}[b]{0.495\textwidth}
		\centering
		\includegraphics[width=\textwidth]{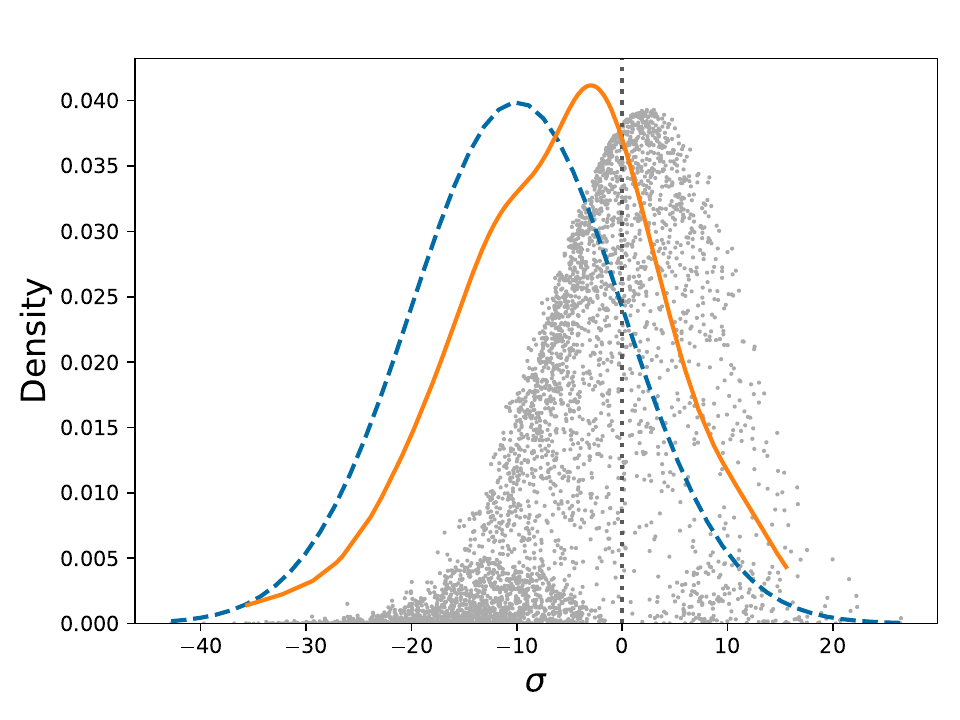}\\
		\caption{}  
		\label{SigmaKSDensityGenicSigmaT0ExpB}
	\end{subfigure}
	\hfill
	\begin{subfigure}[b]{0.495\textwidth}  
		\centering 
		\includegraphics[width=\textwidth]{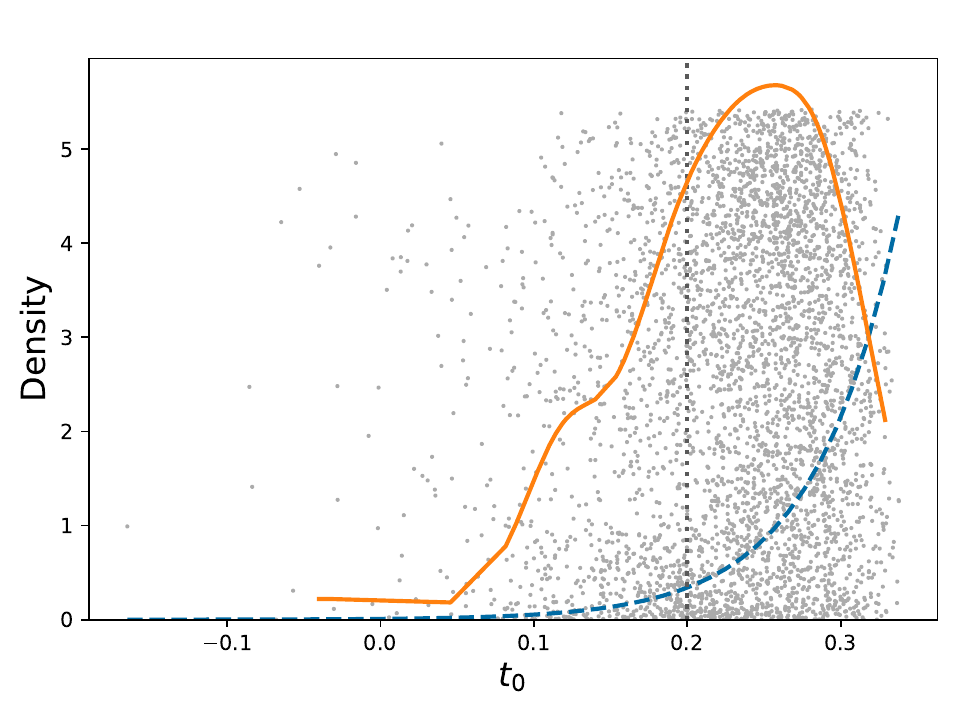}\\
		\caption{}  
		\label{T0KSDensityGenicSigmaT0ExpB}
	\end{subfigure}\\
	\caption[Plots of the prior, likelihood, posterior and true value of the selection coefficient and allele age]{Plots of the prior (dashed blue line), likelihood (grey dots), kernel smoothed posterior (solid orange line) and the truth (dotted vertical black line) for: (a) the selection coefficient $\sigma$, and (b) the allele age $t_0$ for Experiment B in Table \ref{SetupTable}. The prior and smoothed posterior are plotted against the left axis whilst the likelihood is plotted against the right axis.} 
	\label{KSDensityGenicSigmaT0ExpB}
\end{figure}

\begin{figure}[H]
	\centering
	\begin{subfigure}[b]{0.495\textwidth}
		\centering
		\includegraphics[width=\textwidth]{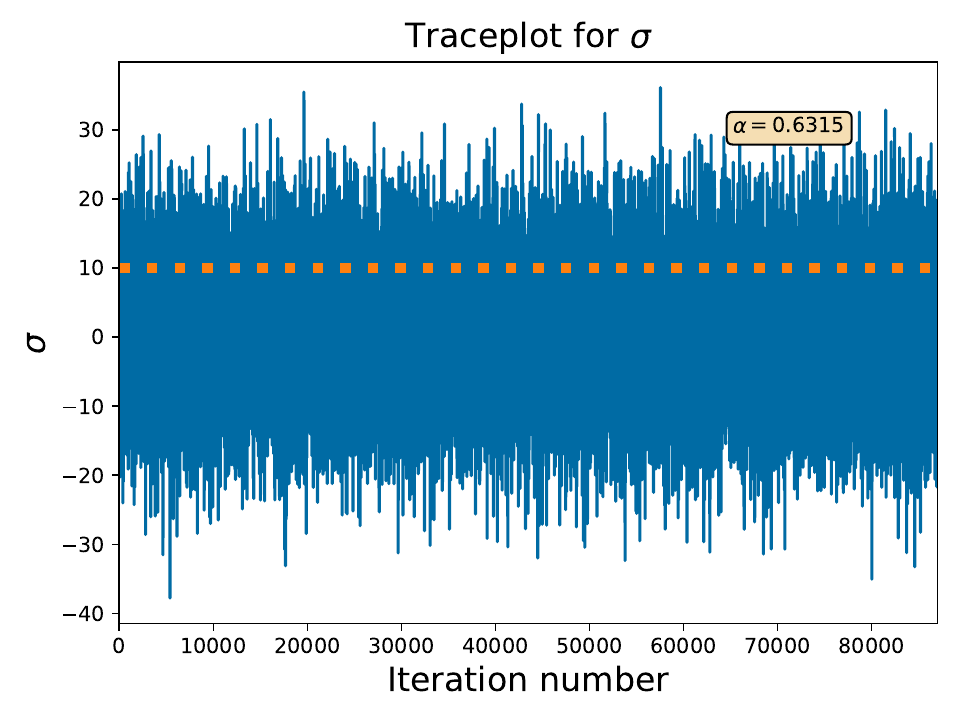}\\
		\caption{}  
		\label{TraceplotSigmaA}
	\end{subfigure}
	\hfill
	\begin{subfigure}[b]{0.495\textwidth}  
		\centering 
		\includegraphics[width=\textwidth]{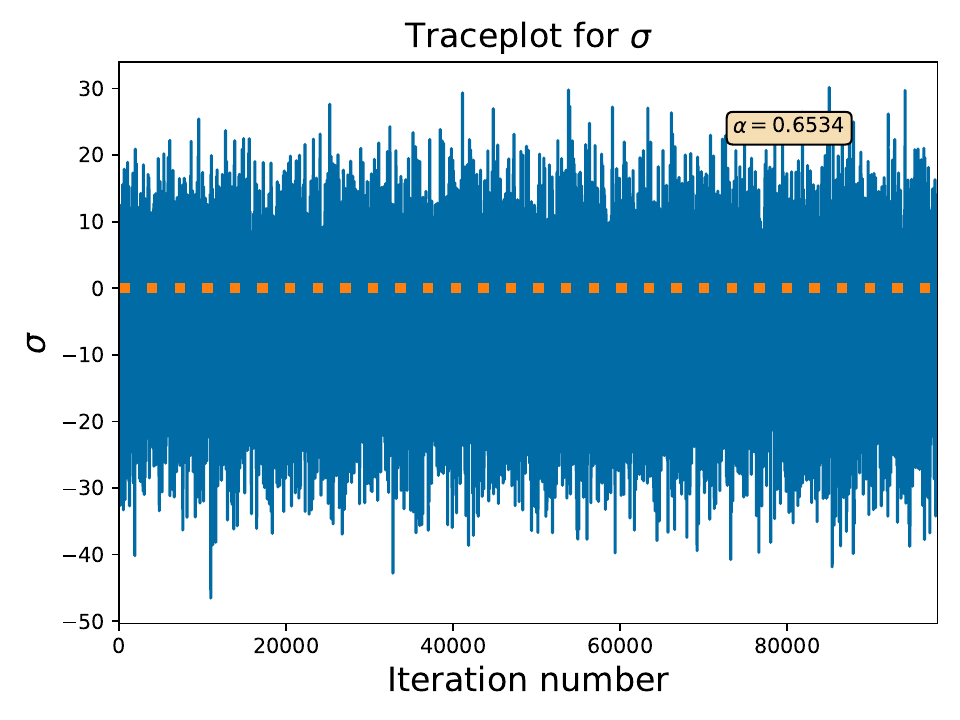}\\
		\caption{}  
		\label{TraceplotSigmaB}
	\end{subfigure}\\
	\caption[Traceplots of the selection coefficient]{Traceplots for $\sigma$ for: (a) Experiment A, and (b) Experiment B. The true value is denoted by the horizontal dashed orange line, whilst $\alpha$ denotes the mean acceptance probability.}
	\label{TraceplotsGenicSigma}
\end{figure}

\begin{figure}[H]
	\centering
	\begin{subfigure}[b]{0.495\textwidth}
		\centering
		\includegraphics[width=\textwidth]{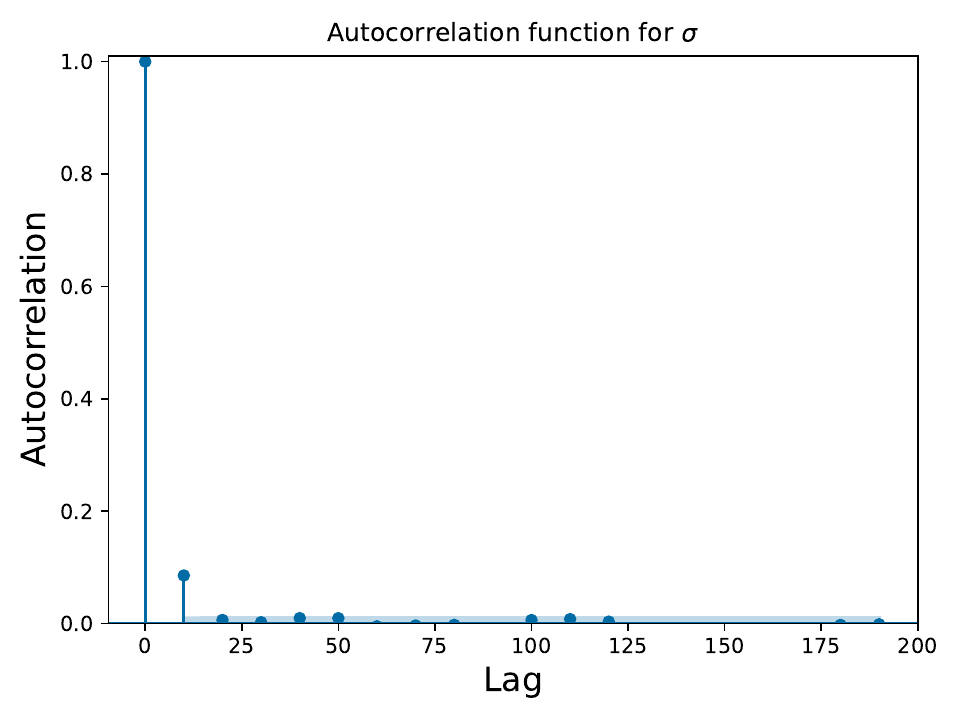}\\
		\caption{}  
		\label{AutoCorrSigmaA}
	\end{subfigure}
	\hfill
	\begin{subfigure}[b]{0.495\textwidth}  
		\centering 
		\includegraphics[width=\textwidth]{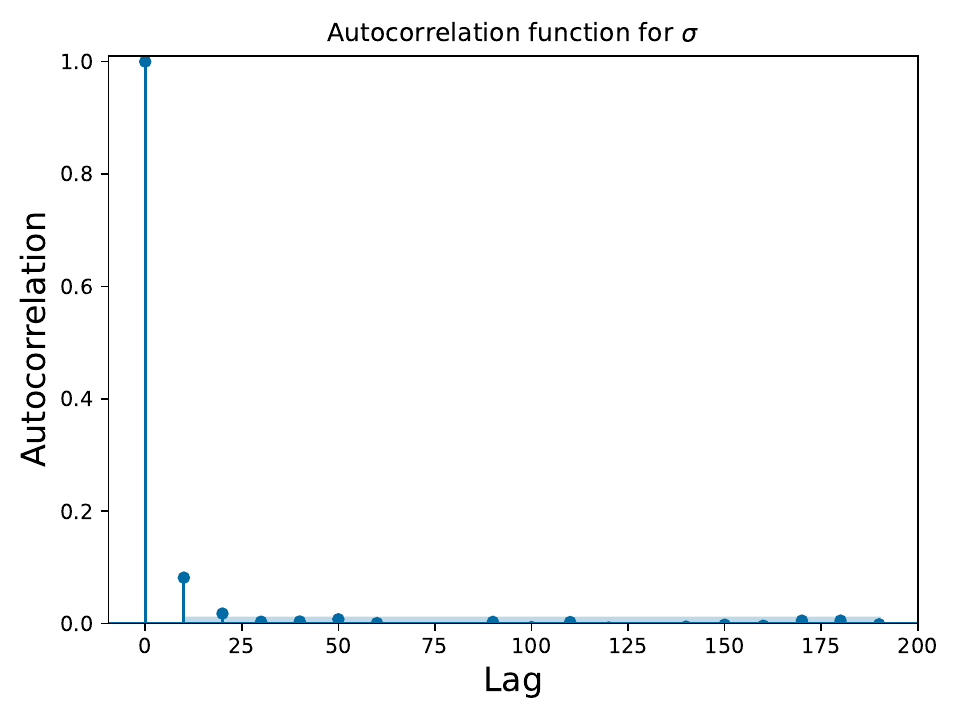}\\
		\caption{}  
		\label{AutoCorrSigmaB}
	\end{subfigure}\\
	\caption[Autocorrelation function for the selection coefficient]{Autocorrelation function plot for $\sigma$ for: (a) Experiment A, and (b) Experiment B} 
	\label{AutoCorrGenicSigma}
\end{figure}

\begin{figure}[H]
	\centering
	\begin{subfigure}[b]{0.495\textwidth}
		\centering
		\includegraphics[width=\textwidth]{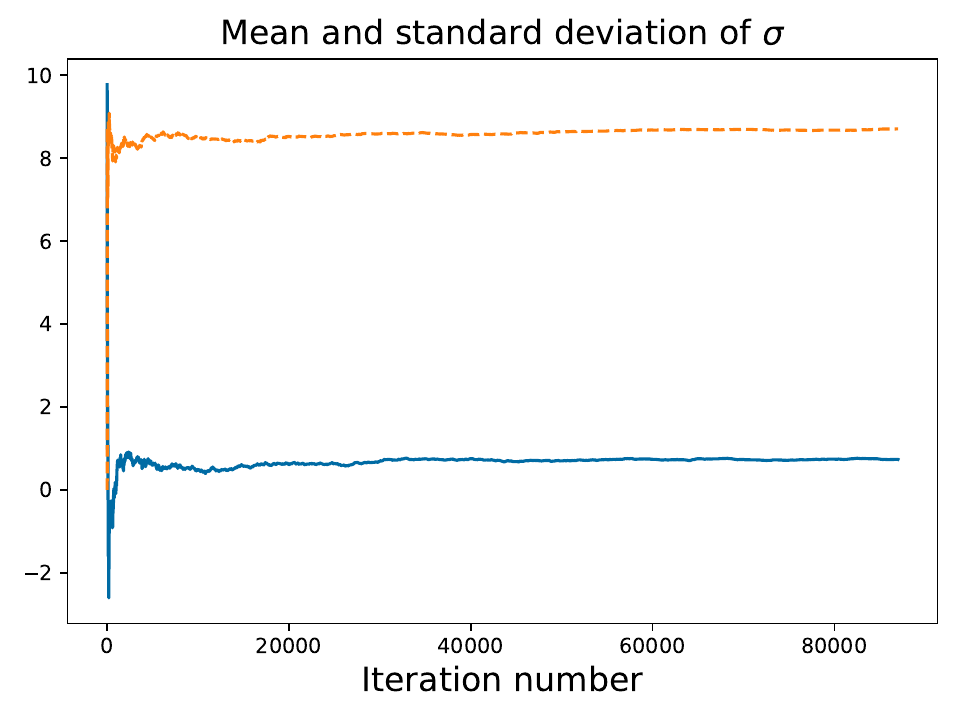}\\
		\caption{}  
		\label{SigmaMeanA}
	\end{subfigure}
	\hfill
	\begin{subfigure}[b]{0.495\textwidth}  
		\centering 
		\includegraphics[width=\textwidth]{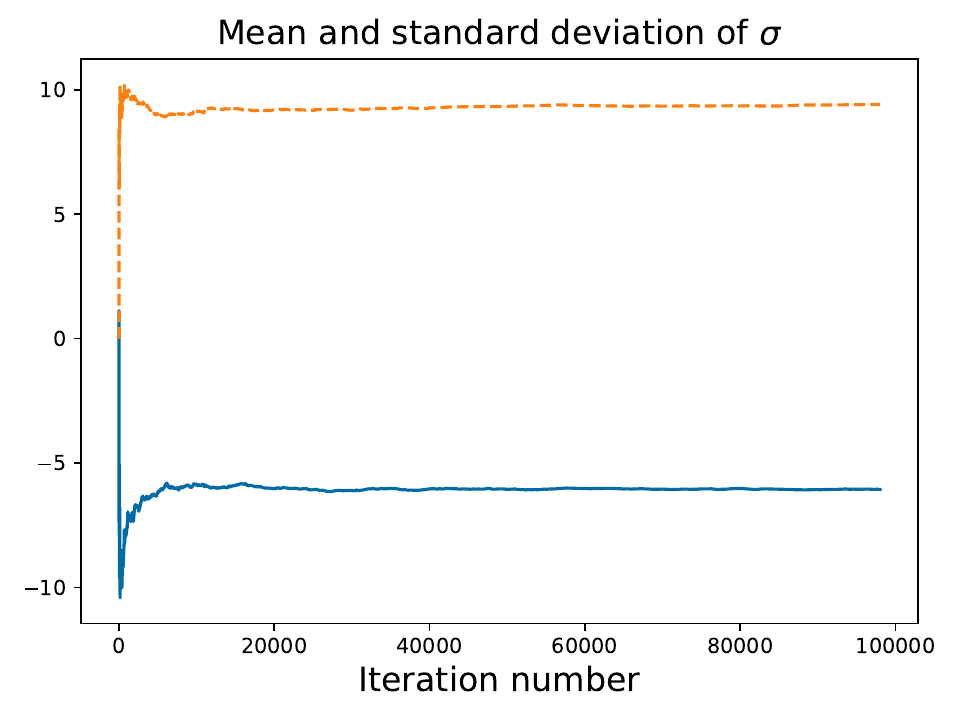}\\
		\caption{}  
		\label{SigmaMeanB}
	\end{subfigure}\\
	\caption[Plots of the mean and standard deviation of the selection coefficient]{Plot of the mean (solid blue line) and standard deviation (dashed orange line) of $\sigma$ for: (a) Experiment A, and (b) Experiment B.} 
	\label{SelMeanSigma}
\end{figure}

\begin{figure}
	\centering
	\begin{subfigure}[b]{0.495\textwidth}
		\centering
		\includegraphics[width=\textwidth]{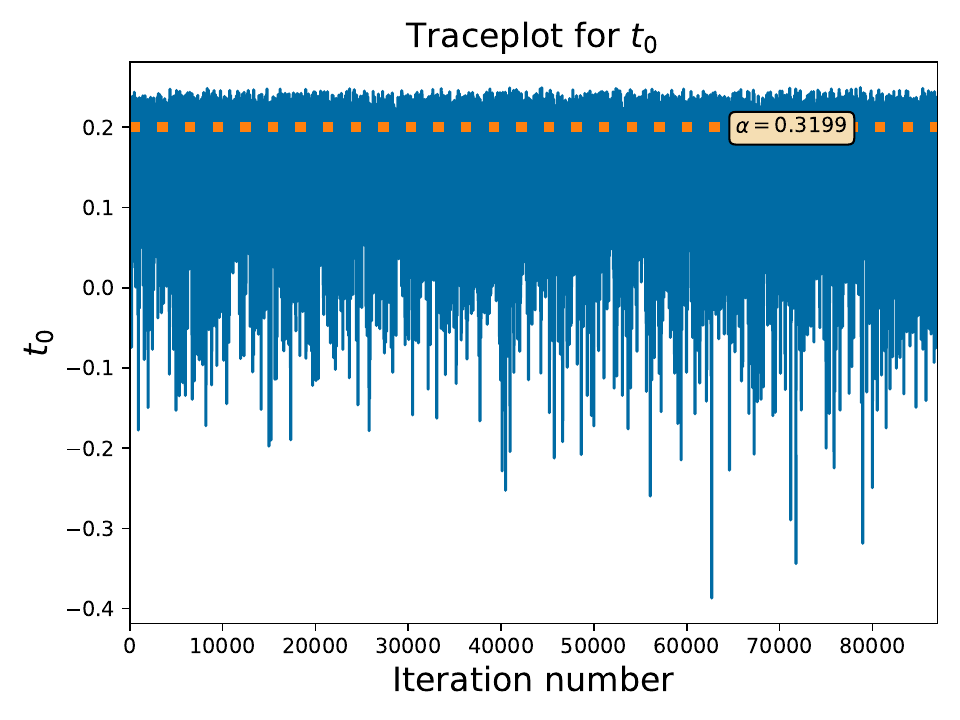}\\
		\caption{}  
		\label{TraceplotT0A}
	\end{subfigure}
	\hfill
	\begin{subfigure}[b]{0.495\textwidth}  
		\centering 
		\includegraphics[width=\textwidth]{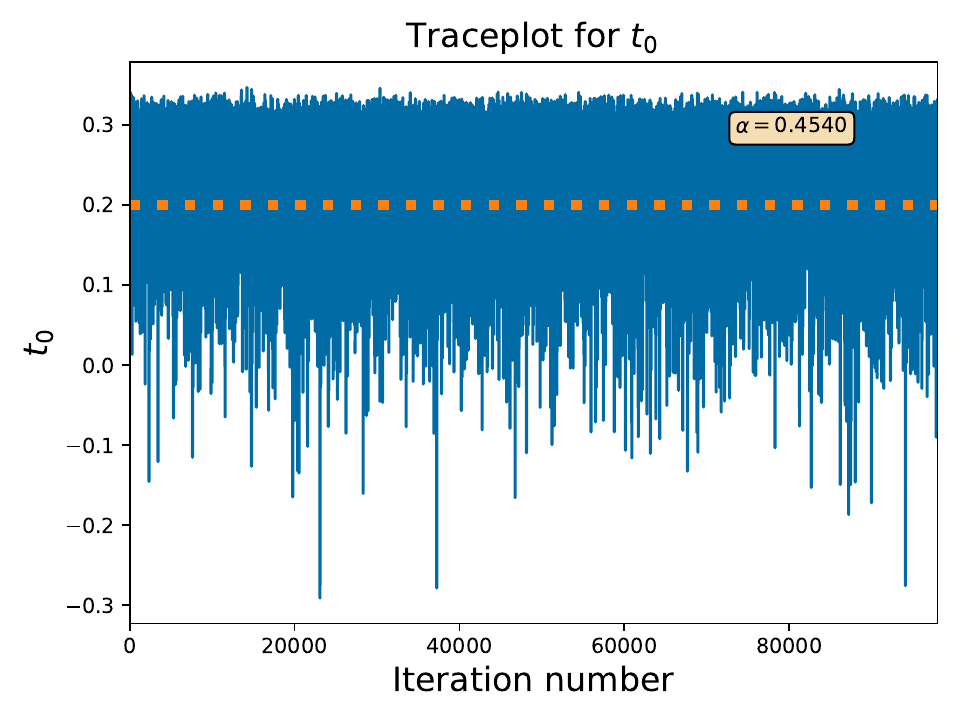}\\
		\caption{}  
		\label{TraceplotT0B}
	\end{subfigure}\\
	\caption[Traceplots of the allele age]{Traceplots for $t_0$ for: (a) Experiment A, and (b) Experiment B. The true value is denoted by the horizontal dashed orange line, whilst $\alpha$ denotes the mean acceptance probability.}
	\label{TraceplotsT0}
\end{figure}

\begin{figure}[H]
	\centering
	\begin{subfigure}[b]{0.495\textwidth}
		\centering
		\includegraphics[width=\textwidth]{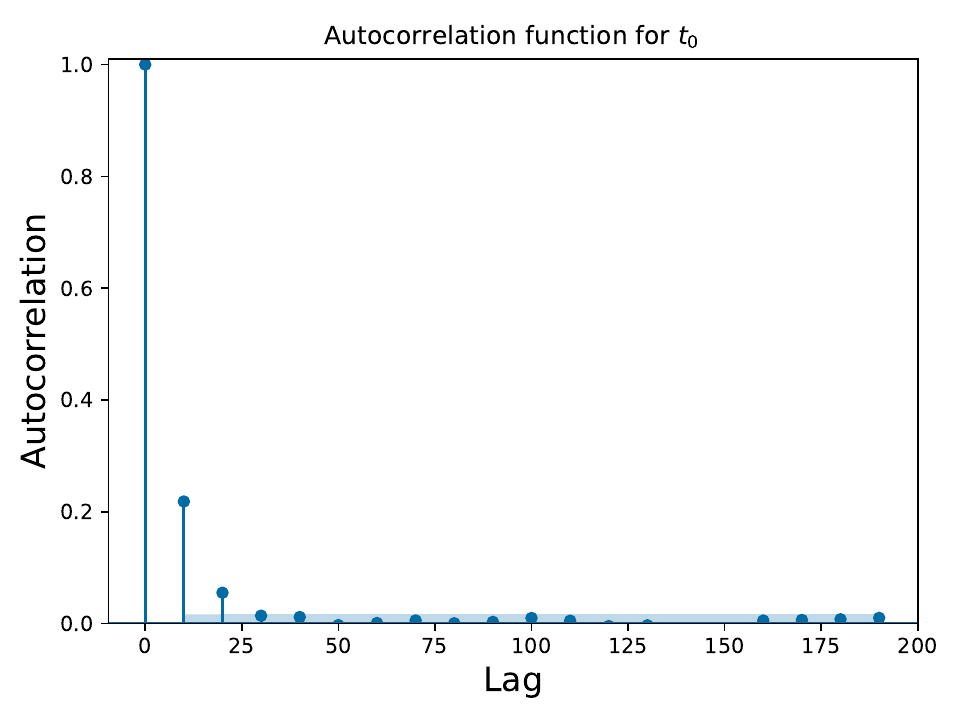}\\
		\caption{}  
		\label{AutoCorrT0A}
	\end{subfigure}
	\hfill
	\begin{subfigure}[b]{0.495\textwidth}  
		\centering 
		\includegraphics[width=\textwidth]{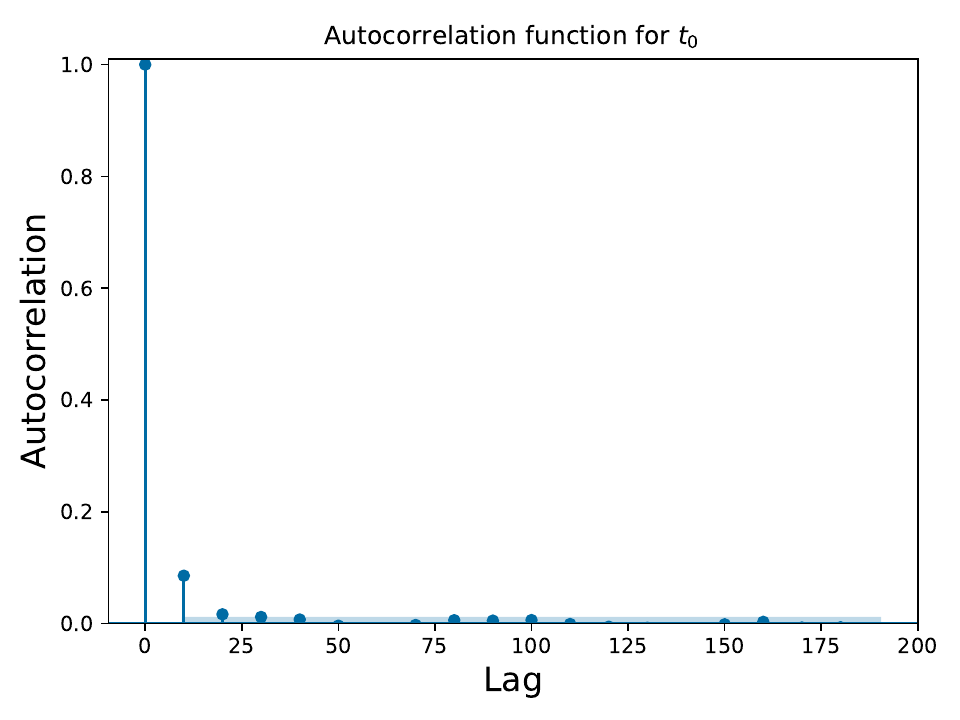}\\
		\caption{}  
		\label{AutoCorrT0B}
	\end{subfigure}\\
	\caption[Autocorrelation function for the allele age]{Autocorrelation function plot for $t_0$ for: (a) Experiment A, and (b) Experiment B} 
	\label{AutoCorrT0}
\end{figure}

\begin{figure}[H]
	\centering
	\begin{subfigure}[b]{0.495\textwidth}
		\centering
		\includegraphics[width=\textwidth]{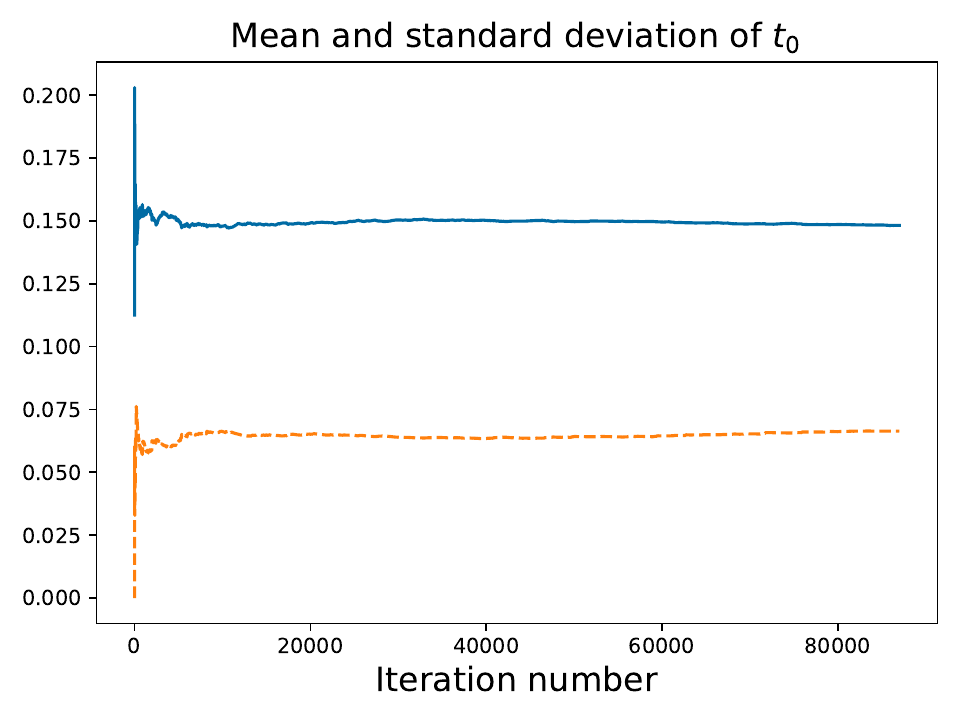}\\
		\caption{}  
		\label{T0MeanA}
	\end{subfigure}
	\hfill
	\begin{subfigure}[b]{0.495\textwidth}  
		\centering 
		\includegraphics[width=\textwidth]{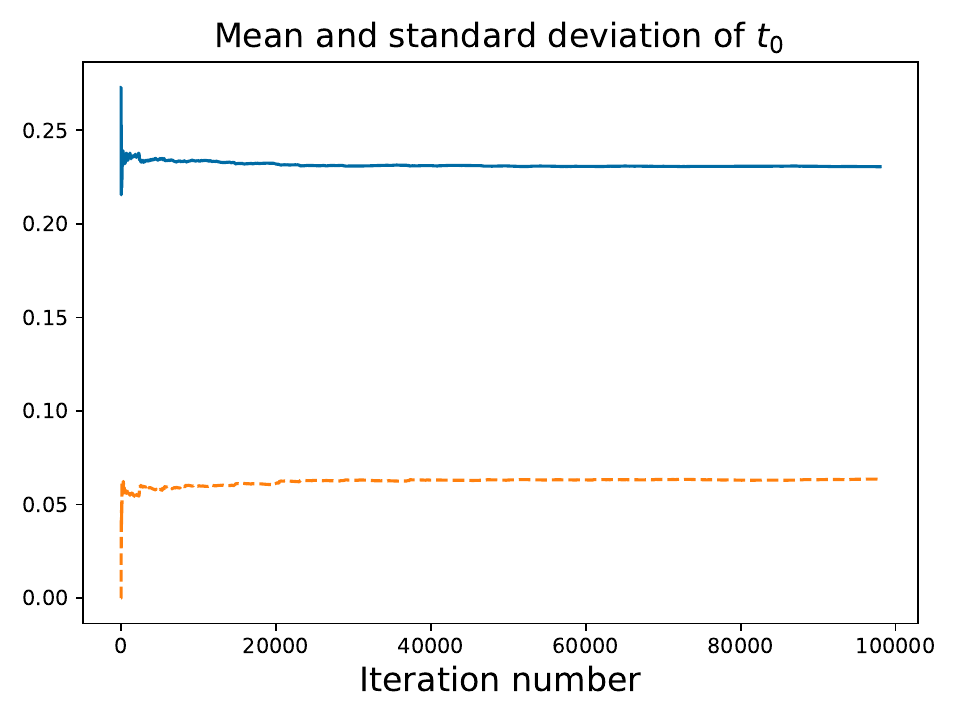}\\
		\caption{}  
		\label{T0MeanB}
	\end{subfigure}\\
	\caption[Plots of the mean and standard deviation of the allele age]{Plot of the mean (solid blue line) and standard deviation (dashed orange line) of $\sigma$ for: (a) Experiment A, and (b) Experiment B.} 
	\label{T0MeanSigma}
\end{figure}

\begin{figure}[H]
	\centering
	\begin{subfigure}[b]{\textwidth}
		\centering
		\includegraphics[width=\textwidth]{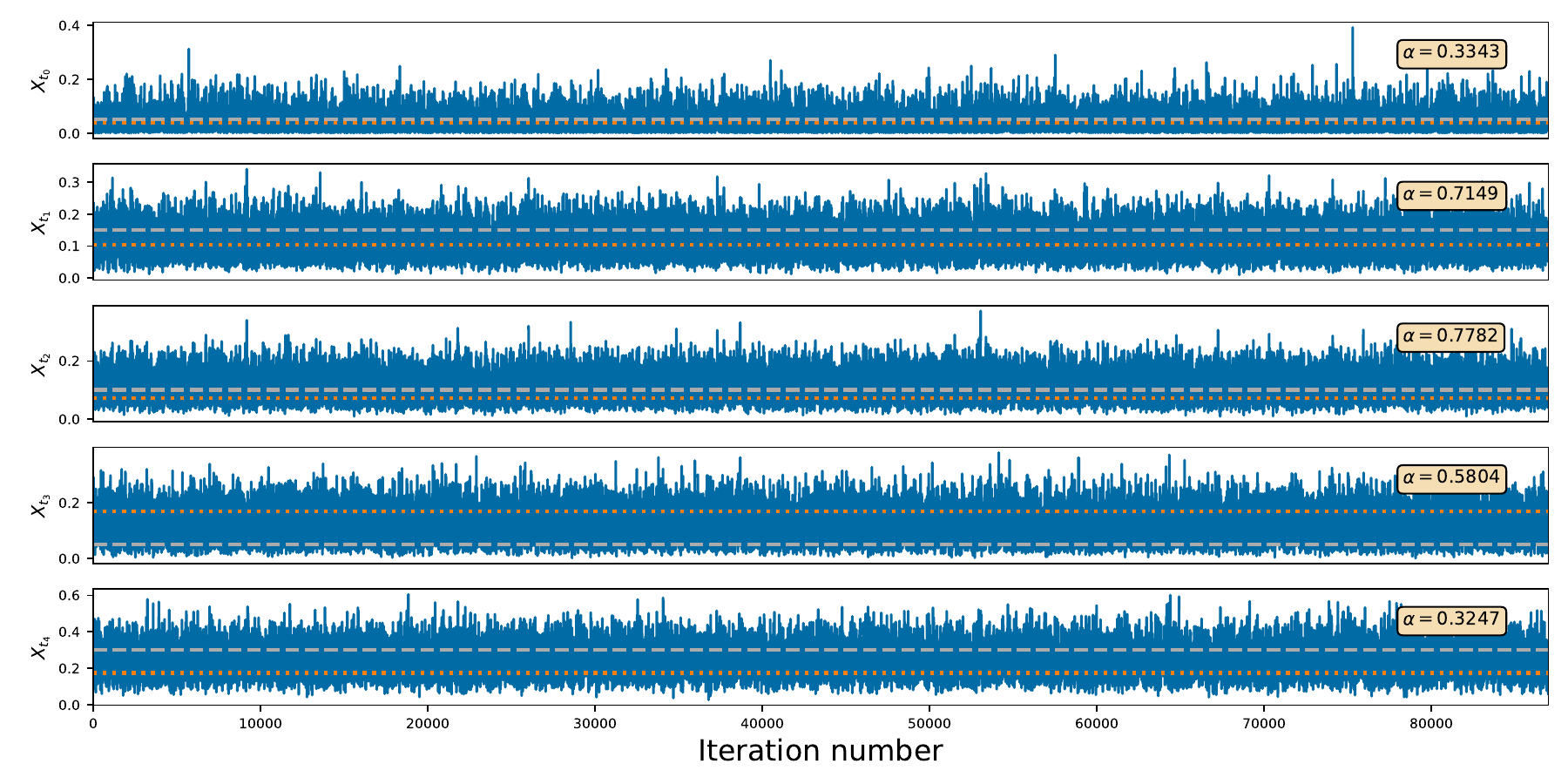}\\
		\caption{}  
		\label{MixingXA}
	\end{subfigure}
	\begin{subfigure}[b]{\textwidth}  
		\centering 
		\includegraphics[width=\textwidth]{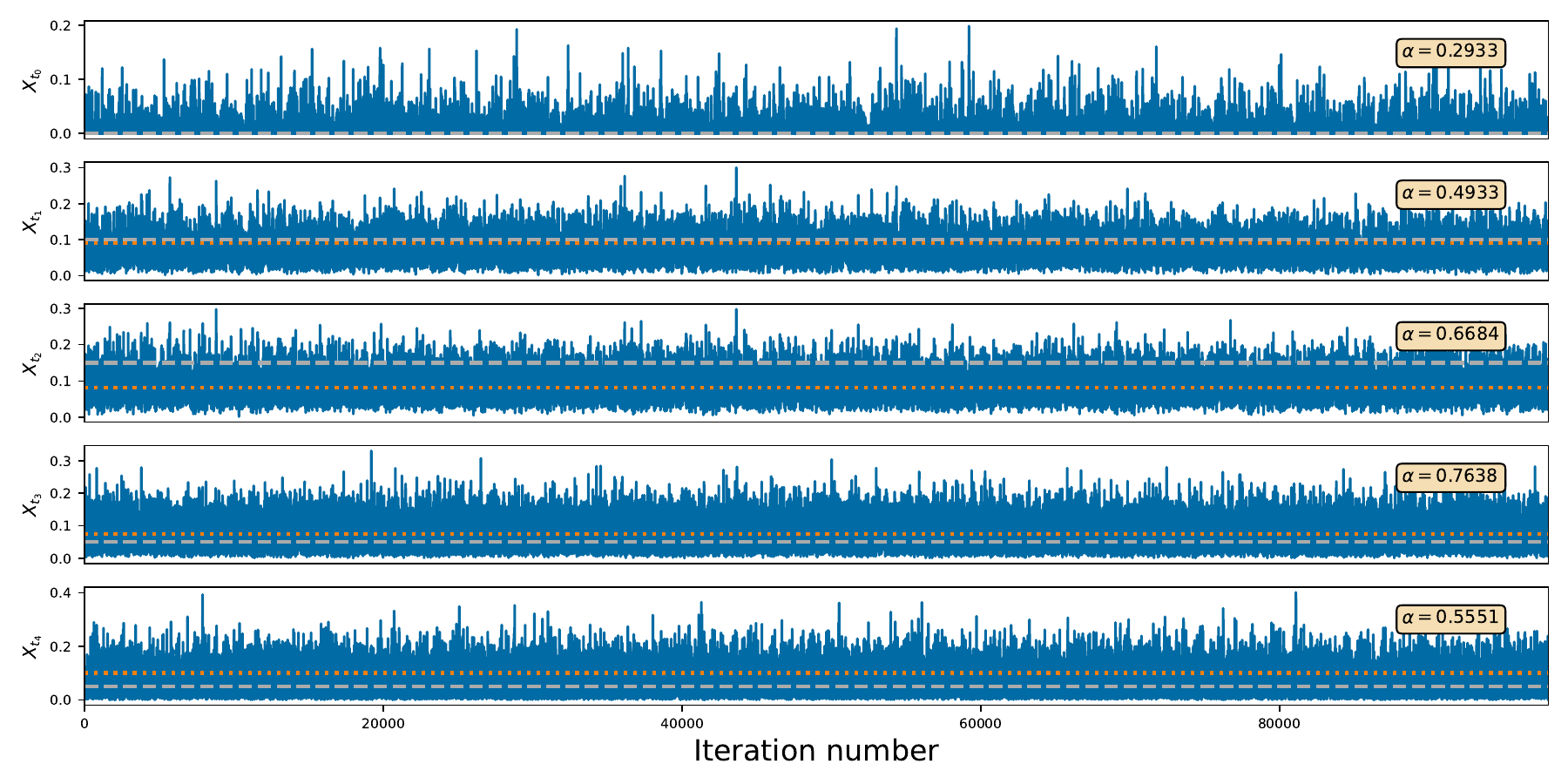}\\
		\caption{}  
		\label{MixingXB}
	\end{subfigure}\\
	\caption[Traceplots for the latent paths]{Traceplots of the latent diffusion for: (a) Experiment A, and (b) Experiment B. The true allele frequency is given by the dotted orange horizontal line, the observed frequency is given by the dashed grey line, whilst $\alpha$ denotes the mean acceptance probability.} 
\end{figure}

\subsection{Output when inferring diploid selection}

\begin{figure}[H]
	\centering
	\begin{subfigure}[b]{0.495\textwidth}
		\centering
		\includegraphics[width=\textwidth]{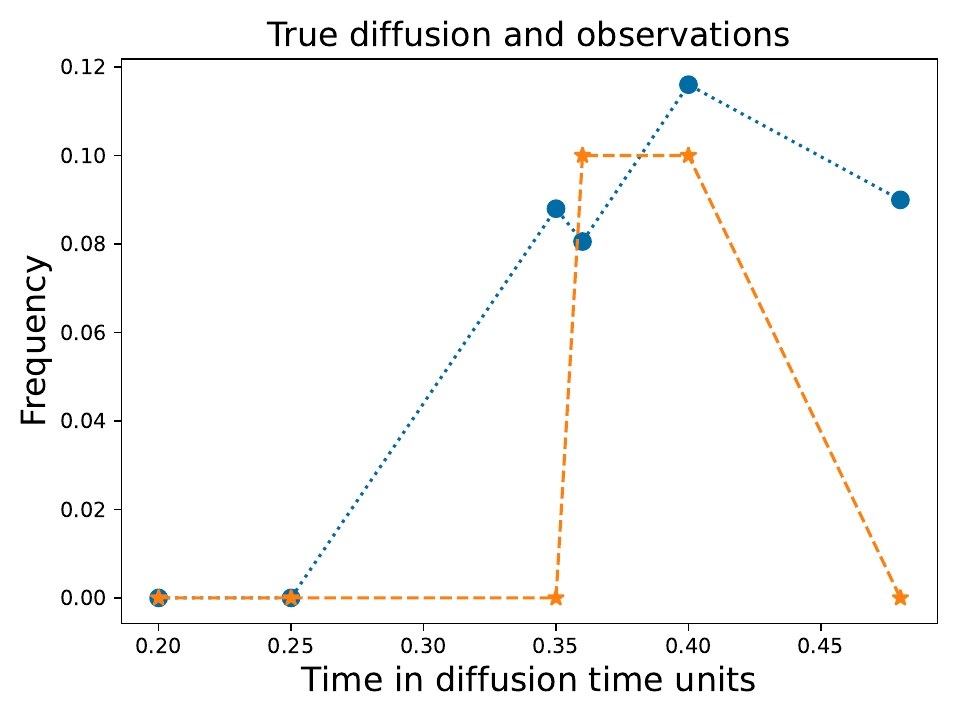}\\
		\caption{}  
		\label{ObsC}
	\end{subfigure}
	\hfill
	\begin{subfigure}[b]{0.495\textwidth}
		\centering
		\includegraphics[width=\textwidth]{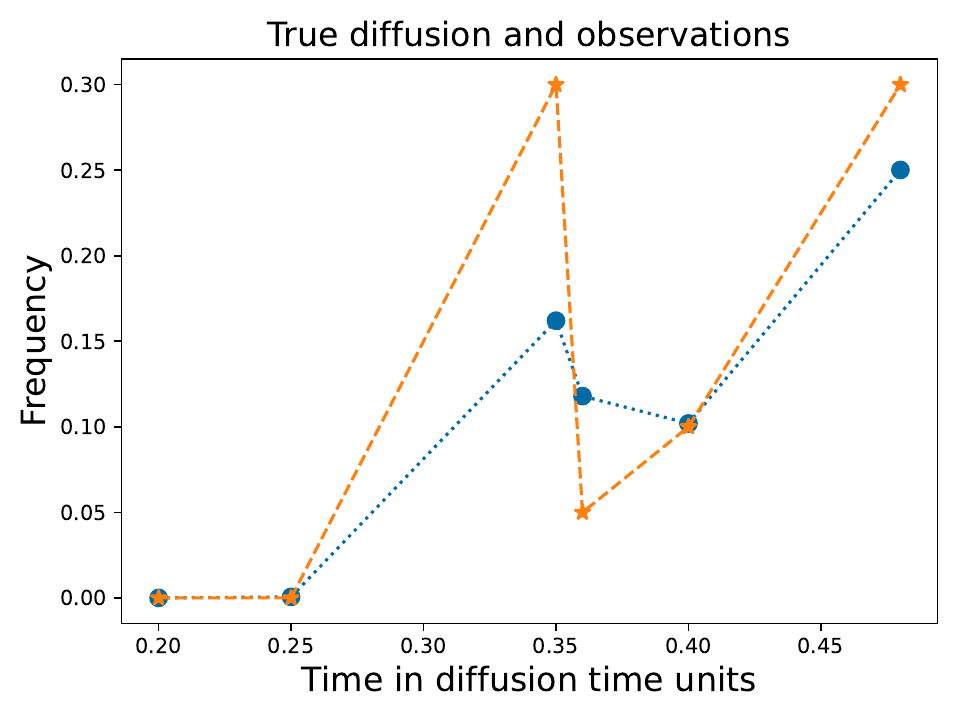}\\
		\caption{}  
		\label{ObsD}
	\end{subfigure}
	\caption[The simulated dataset generated via the parameter configurations in the last two rows of Table \ref{SetupTable}]{Simulated data generated for the case when inferring diploid selection and allele age, corresponding to Experiments C and D in Table \ref{SetupTable}. Blue circles denote exact draws from the Wright--Fisher diffusion generated via \texttt{EWF}, whilst the orange stars are the binomial draws obtained. Dotted lines between observations are a linear interpolation.}
	\label{Observations_diploid}
\end{figure}

\begin{figure}[H]
	\centering
	\begin{subfigure}[b]{0.49\textwidth}
		\centering
		\includegraphics[width=\textwidth]{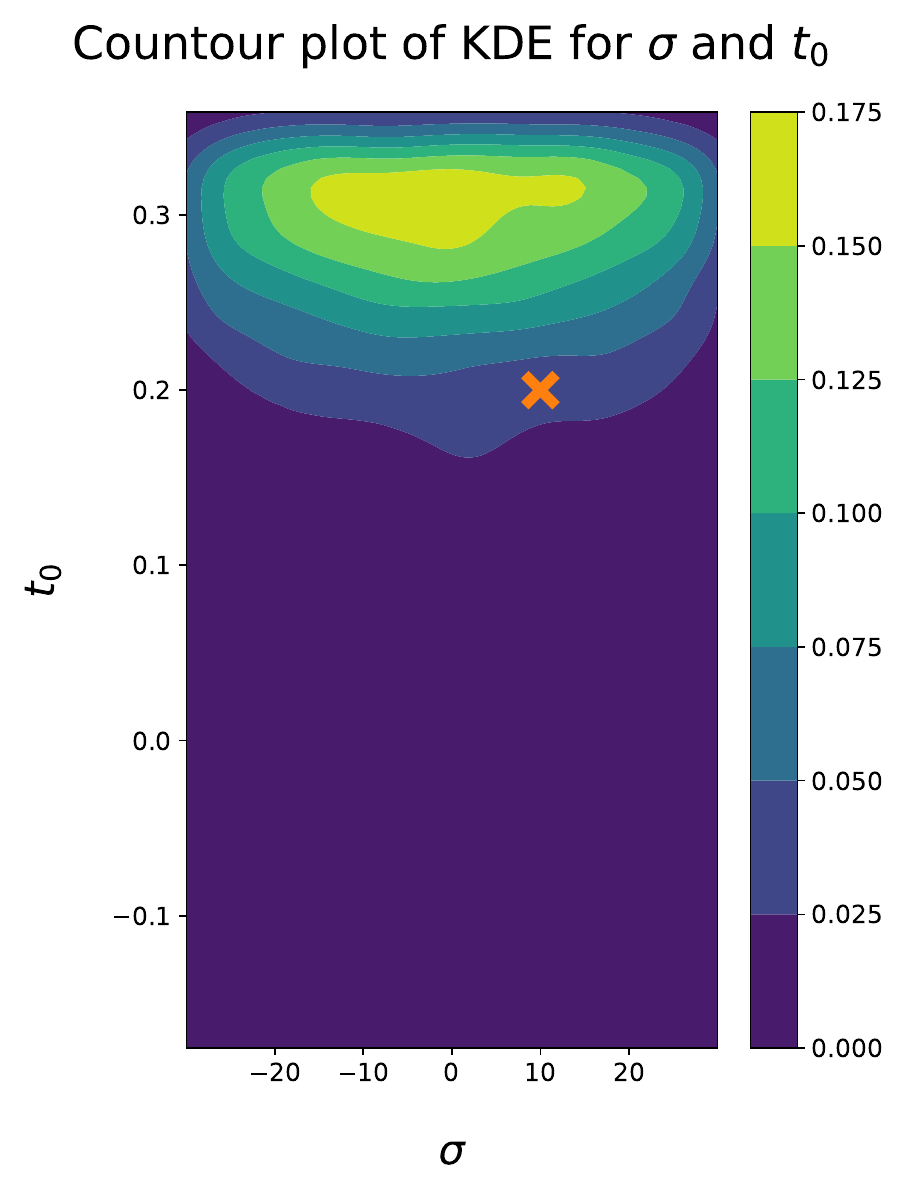}\\
		\caption{}  
		\label{SigmaT0KSDensityC}
	\end{subfigure}
	\begin{subfigure}[b]{0.49\textwidth}  
		\centering 
		\includegraphics[width=\textwidth]{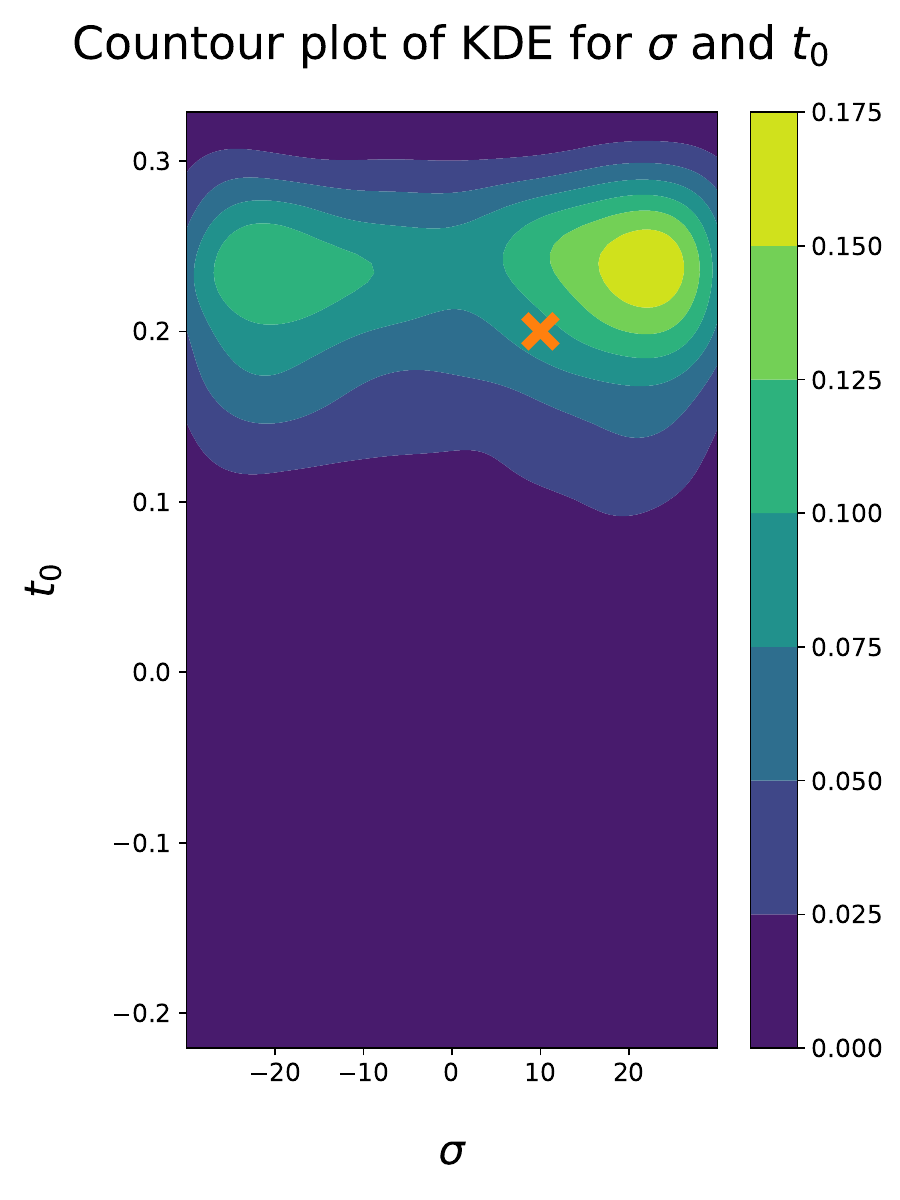}\\
		\caption{}  
		\label{SigmaT0KSDensityD}
	\end{subfigure}\\
	\caption[Contour plot for the KDE of the posterior of $(\sigma, t_0)$ for (a) Experiment C, and (b) Experiment D.]{Contour plot of the KDE for the posterior of $(\sigma, t_0)$ for (a) Experiment C, and (b) Experiment D. The true value is denoted by an orange cross.} 
	\label{SigmaT0KSDensitySigma_diploid}
\end{figure}

\begin{figure}[H]
	\centering
	\begin{subfigure}[b]{0.49\textwidth}
		\centering
		\includegraphics[width=\textwidth]{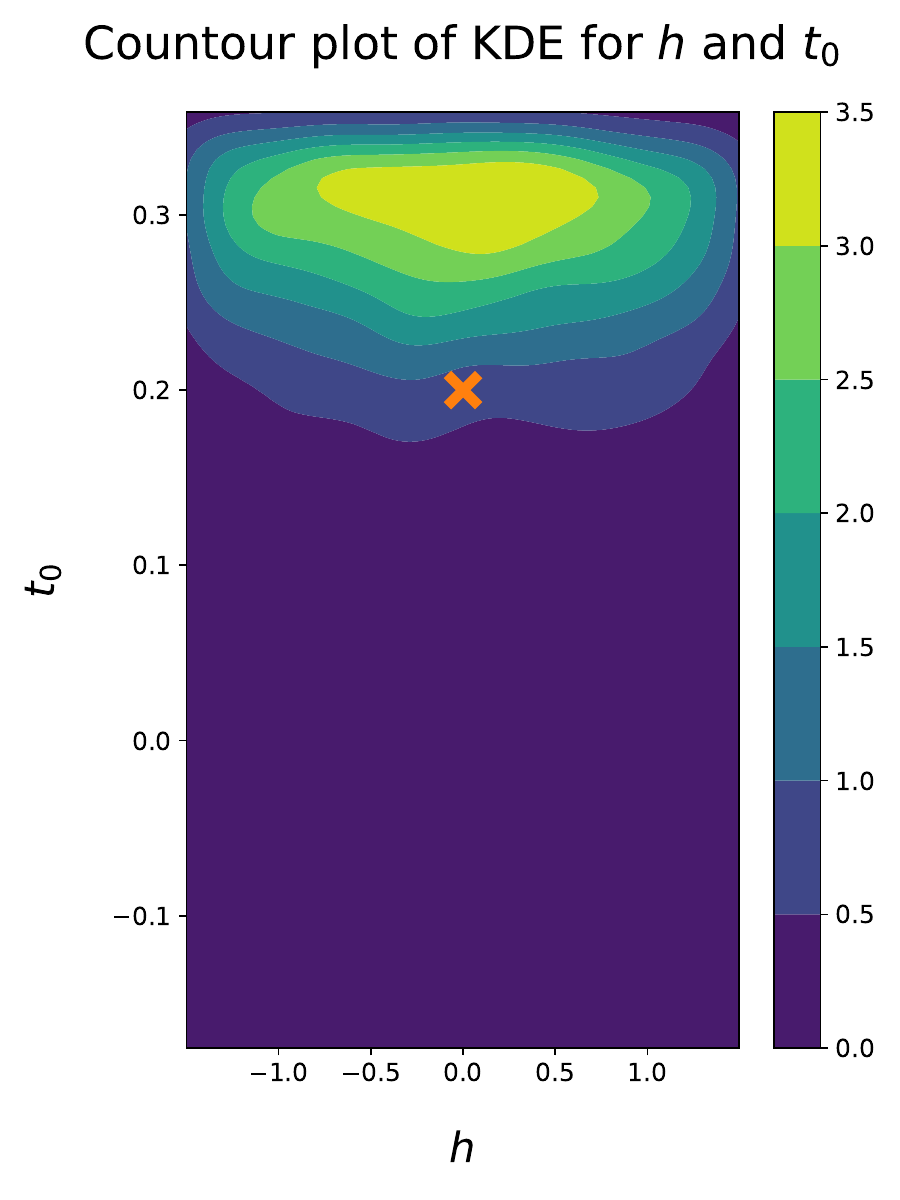}\\
		\caption{}  
		\label{HT0KSDensityC}
	\end{subfigure}
	\begin{subfigure}[b]{0.49\textwidth}  
		\centering 
		\includegraphics[width=\textwidth]{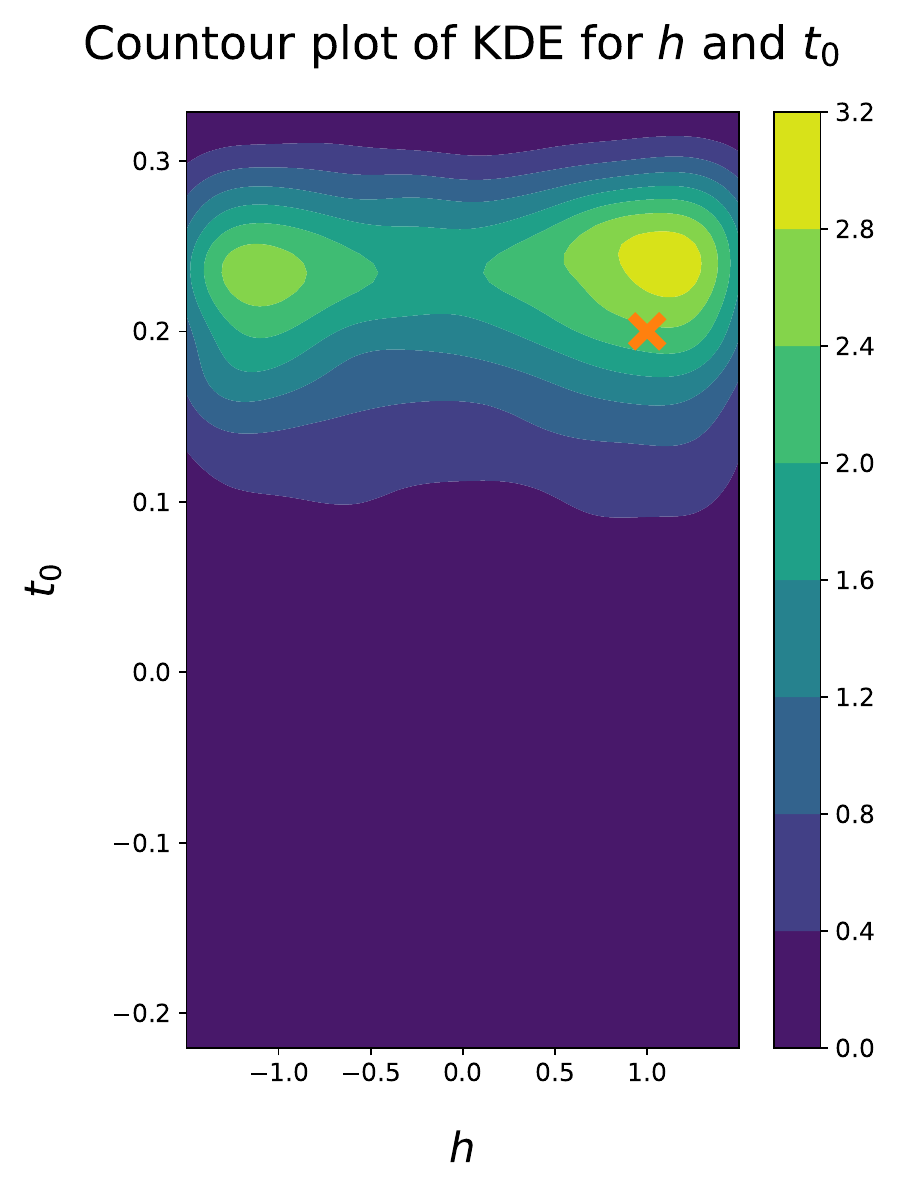}\\
		\caption{}  
		\label{HT0KSDensityD}
	\end{subfigure}\\
	\caption[Contour plot for the KDE of the posterior of $(h, t_0)$ for (a) Experiment C, and (b) Experiment D.]{Contour plot of the KDE for the posterior of $(h, t_0)$ for: (a) Experiment C, and (b) Experiment D. The true value is denoted by an orange cross.} 
	\label{HT0KSDensitySigma_diploid}
\end{figure}

\begin{figure}[H]
	\centering
	\begin{subfigure}[b]{0.495\textwidth}
		\centering
		\includegraphics[width=\textwidth]{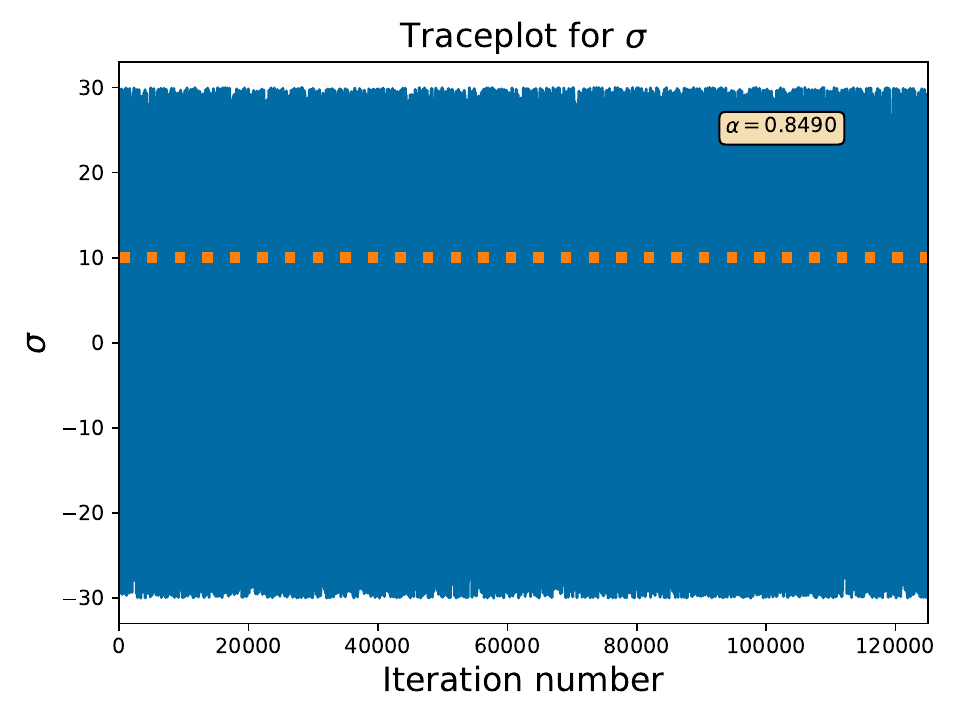}\\
		\caption{}  
		\label{TraceplotSigmaC}
	\end{subfigure}
	\hfill
	\begin{subfigure}[b]{0.495\textwidth}  
		\centering 
		\includegraphics[width=\textwidth]{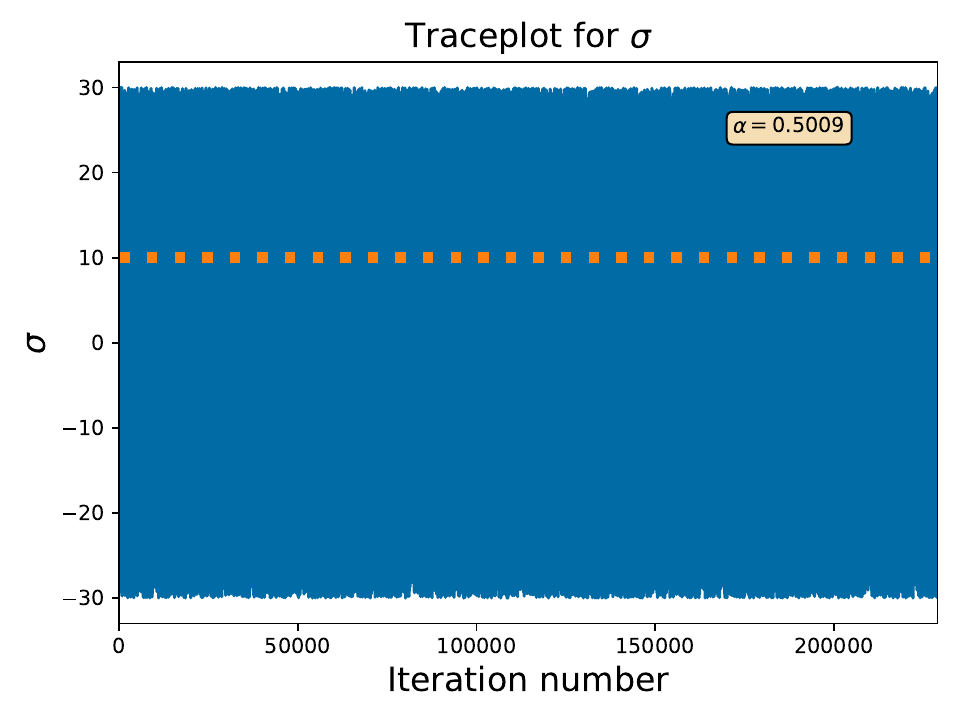}\\
		\caption{}  
		\label{TraceplotSigmaD}
	\end{subfigure}\\
	\caption[Traceplots of the selection coefficient]{Traceplots for $\sigma$ for: (a) Experiment C, and (b) Experiment D. The true value is denoted by the horizontal dashed orange line, whilst $\alpha$ denotes the mean acceptance probability.}
        \label{TraceplotsSigma_diploid}
\end{figure}

\begin{figure}[H]
	\centering
	\begin{subfigure}[b]{0.495\textwidth}
		\centering
		\includegraphics[width=\textwidth]{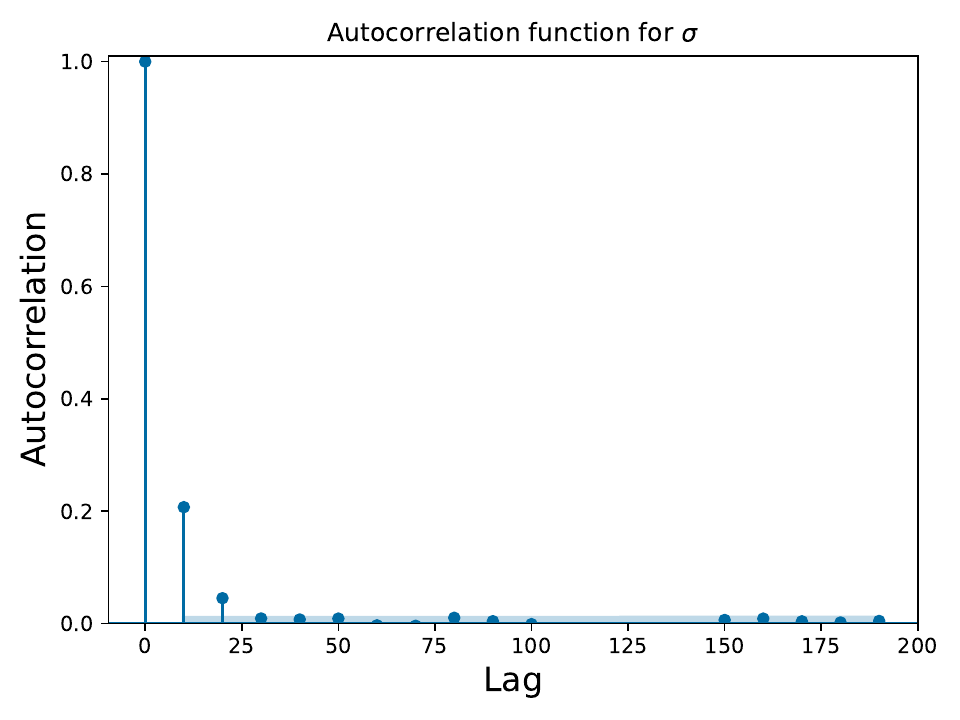}\\
		\caption*{}  
		\label{AutoCorrSigmaC}
	\end{subfigure}
	\hfill
	\begin{subfigure}[b]{0.495\textwidth}  
		\centering 
		\includegraphics[width=\textwidth]{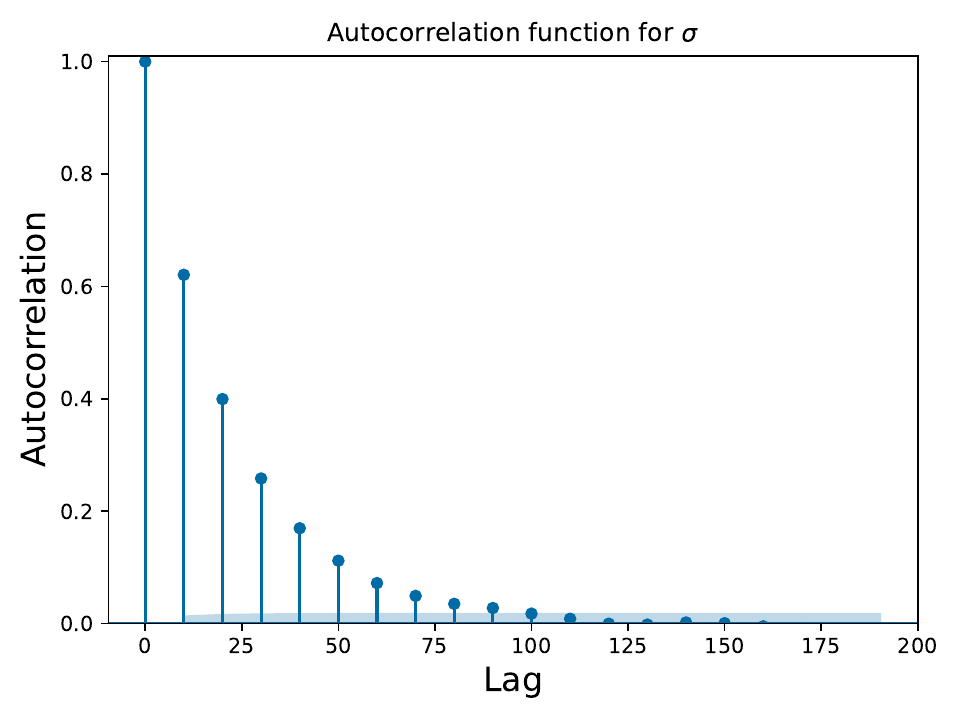}\\
		\caption*{}  
		\label{AutoCorrSigmaD}
	\end{subfigure}\\
	\caption[Autocorrelation function for the selection coefficient]{Autocorrelation function plot for $\sigma$ for: (a) Experiment C, and (b) Experiment D}  
	\label{AutoCorrSigma_diploid}
\end{figure}

\begin{figure}[H]
	\centering
	\begin{subfigure}[b]{0.495\textwidth}
		\centering
		\includegraphics[width=\textwidth]{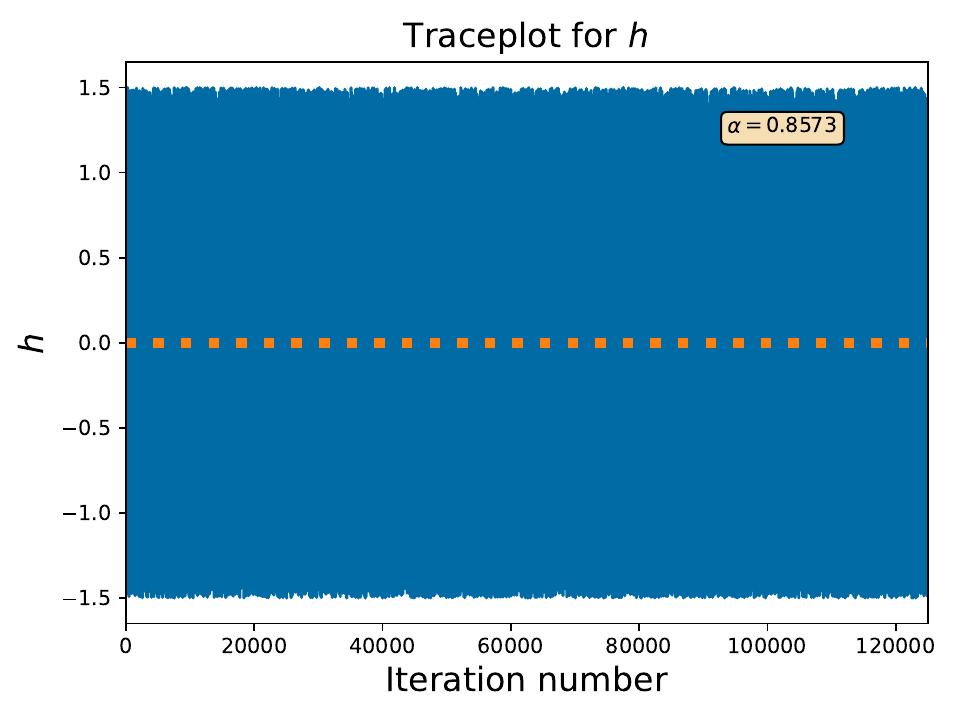}\\
		\caption{}  
		\label{TraceplotHC}
	\end{subfigure}
	\hfill
	\begin{subfigure}[b]{0.495\textwidth}  
		\centering 
		\includegraphics[width=\textwidth]{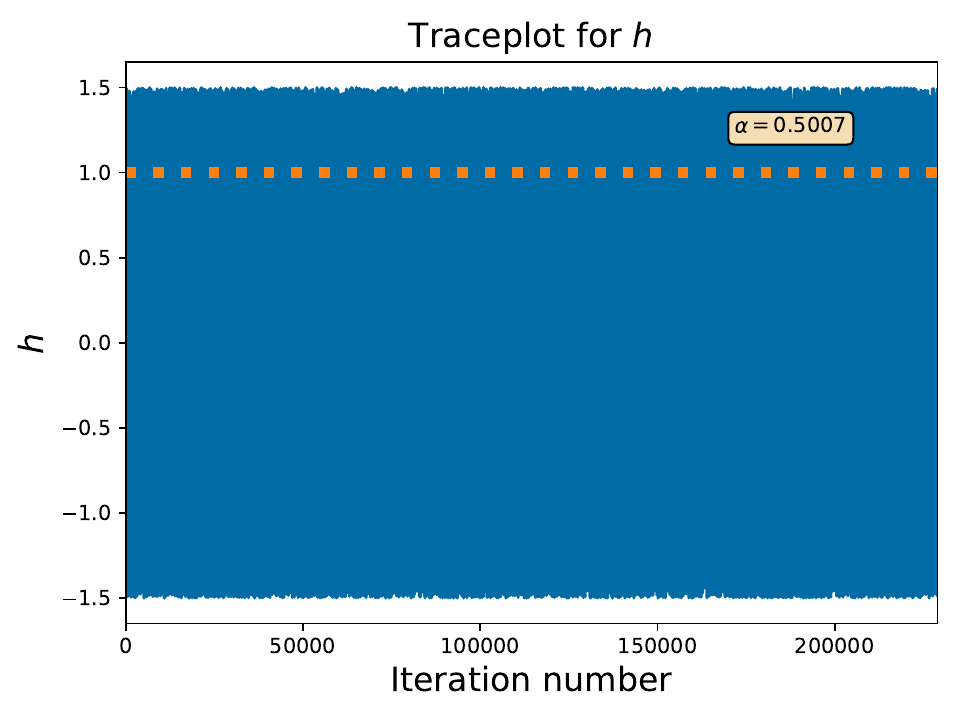}\\
		\caption{}  
		\label{TraceplotHD}
	\end{subfigure}\\
	\caption[Traceplots of the dominance parameter]{Traceplots for $h$ for: (a) Experiment C, and (b) Experiment D. The true value is denoted by the horizontal dashed orange line, whilst $\alpha$ denotes the mean acceptance probability.}
	\label{TraceplotsH_diploid}
\end{figure}

\begin{figure}[H]
	\centering
	\begin{subfigure}[b]{0.495\textwidth}
		\centering
		\includegraphics[width=\textwidth]{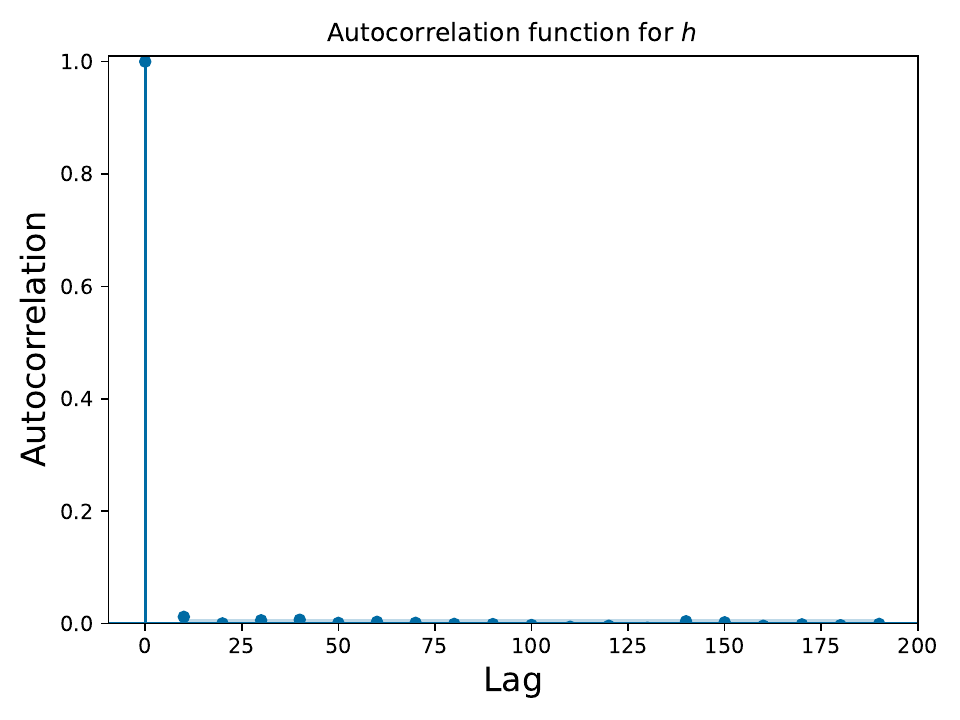}\\
		\caption{}  
		\label{AutoCorrHC}
	\end{subfigure}
	\hfill
	\begin{subfigure}[b]{0.495\textwidth}  
		\centering 
		\includegraphics[width=\textwidth]{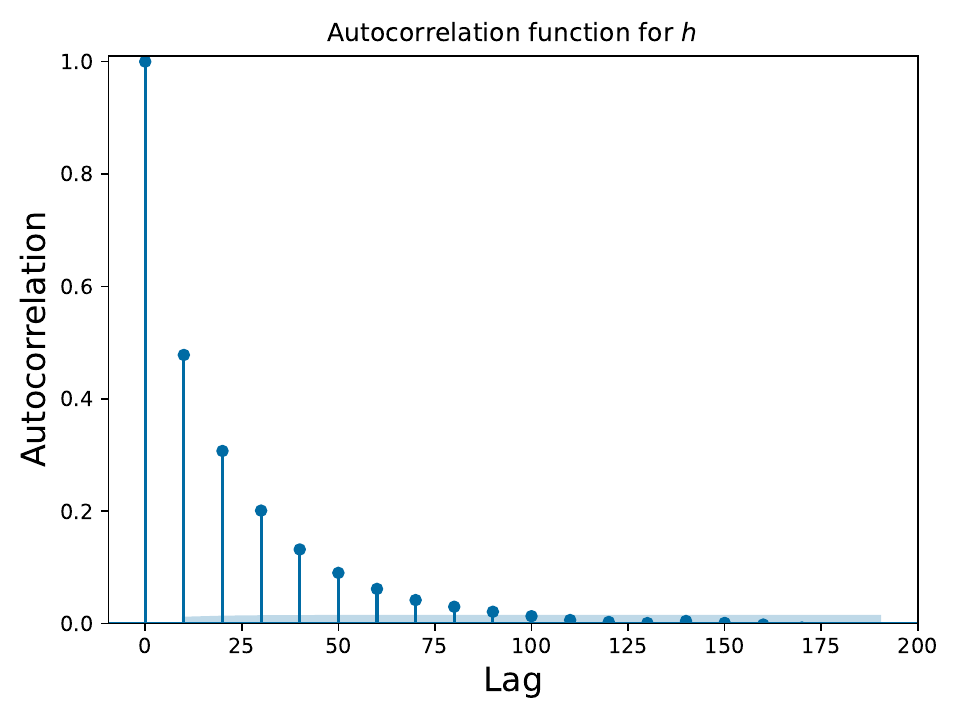}\\
		\caption{}  
		\label{AutoCorrHD}
	\end{subfigure}\\
	\caption[Autocorrelation function for the dominance parameter]{Autocorrelation function plot for $h$ for: (a) Experiment C, and (b) Experiment D} 
	\label{AutoCorrH_diploid}
\end{figure}

\begin{figure}[H]
	\centering
	\begin{subfigure}[b]{0.495\textwidth}
		\centering
		\includegraphics[width=\textwidth]{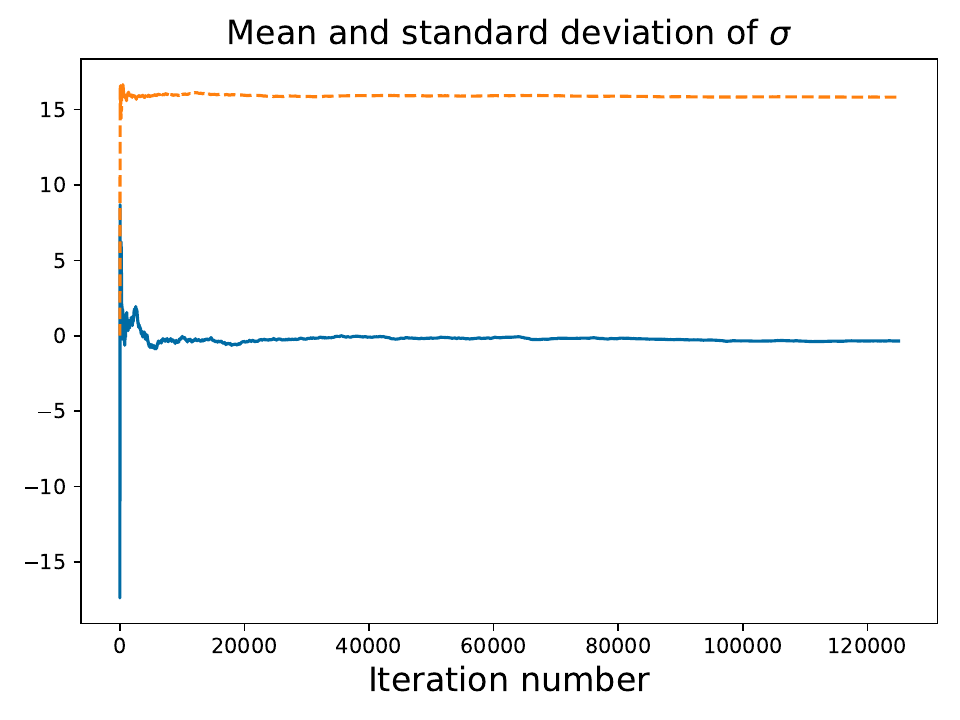}\\
		\caption{}  
		\label{SigmaMeanC}
	\end{subfigure}
	\hfill
	\begin{subfigure}[b]{0.495\textwidth}  
		\centering 
		\includegraphics[width=\textwidth]{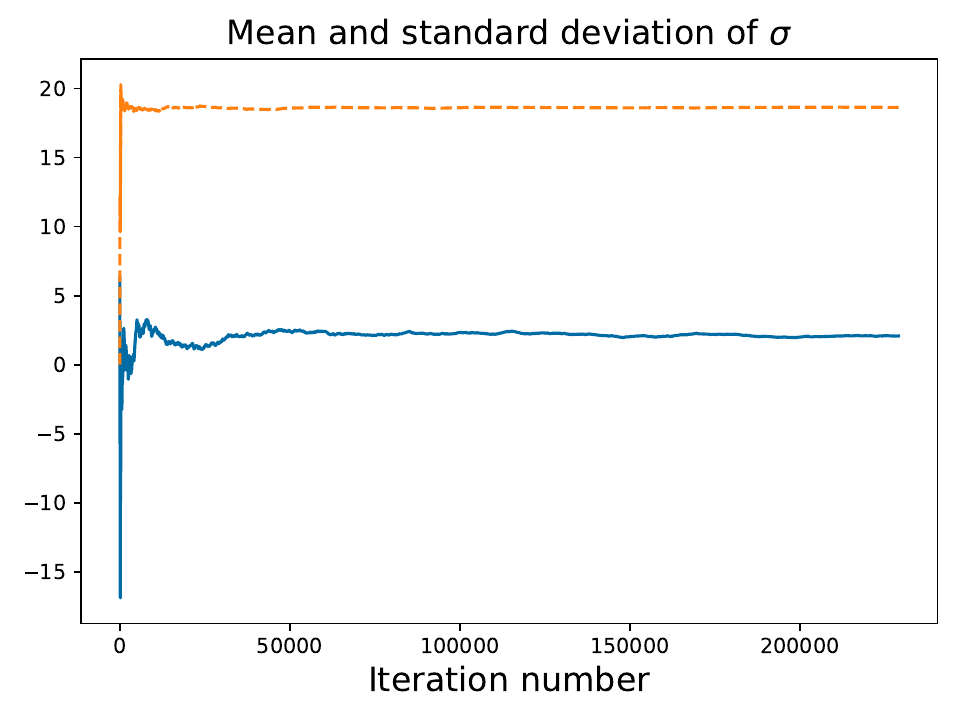}\\
		\caption{}  
		\label{SigmaMeanD}
	\end{subfigure}\\
	\caption[Plots of the mean and standard deviation of the selection coefficient]{Plot of the mean (solid blue line) and standard deviation (dashed orange line) of $\sigma$ for: (a) Experiment C, and (b) Experiment D.} 
	\label{SelMeanSigma_diploid}
\end{figure}

\begin{figure}[H]
	\centering
	\begin{subfigure}[b]{0.495\textwidth}
		\centering
		\includegraphics[width=\textwidth]{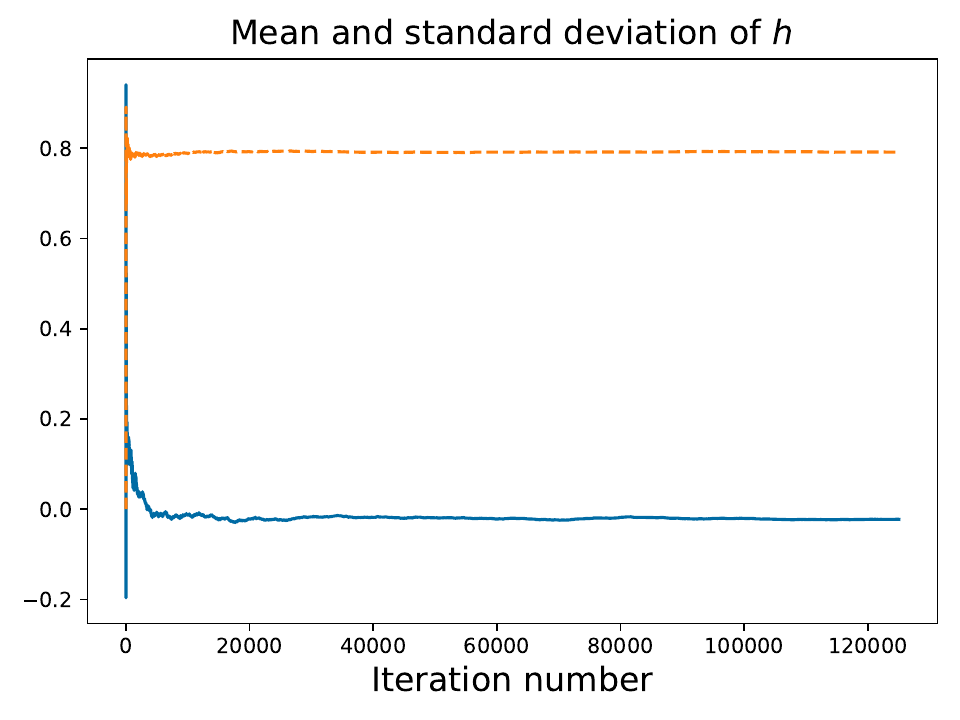}\\
		\caption{}  
		\label{HMeanC}
	\end{subfigure}
	\hfill
	\begin{subfigure}[b]{0.495\textwidth}  
		\centering 
		\includegraphics[width=\textwidth]{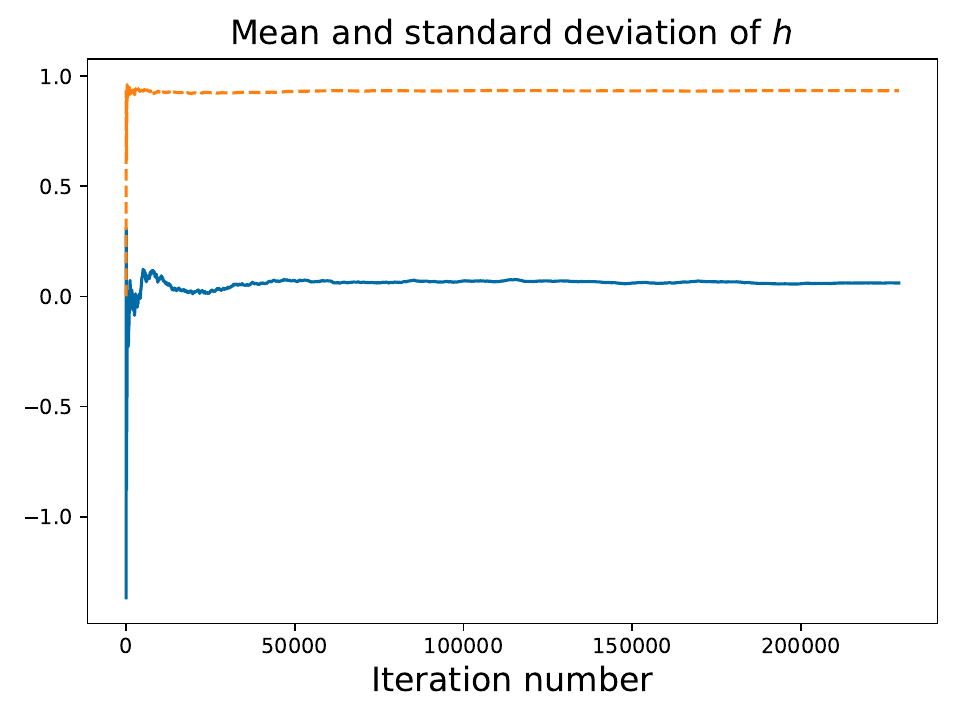}\\
		\caption{}  
		\label{HMeanD}
	\end{subfigure}\\
	\caption[Plots of the mean and standard deviation of the dominance parameter]{Plot of the mean (solid blue line) and standard deviation (dashed orange line) of $h$ for: (a) Experiment C, and (b) Experiment D.} 
	\label{HMeanSigma_diploid}
\end{figure}

\begin{figure}[H]
	\centering
	\begin{subfigure}[b]{0.495\textwidth}
		\centering
		\includegraphics[width=\textwidth]{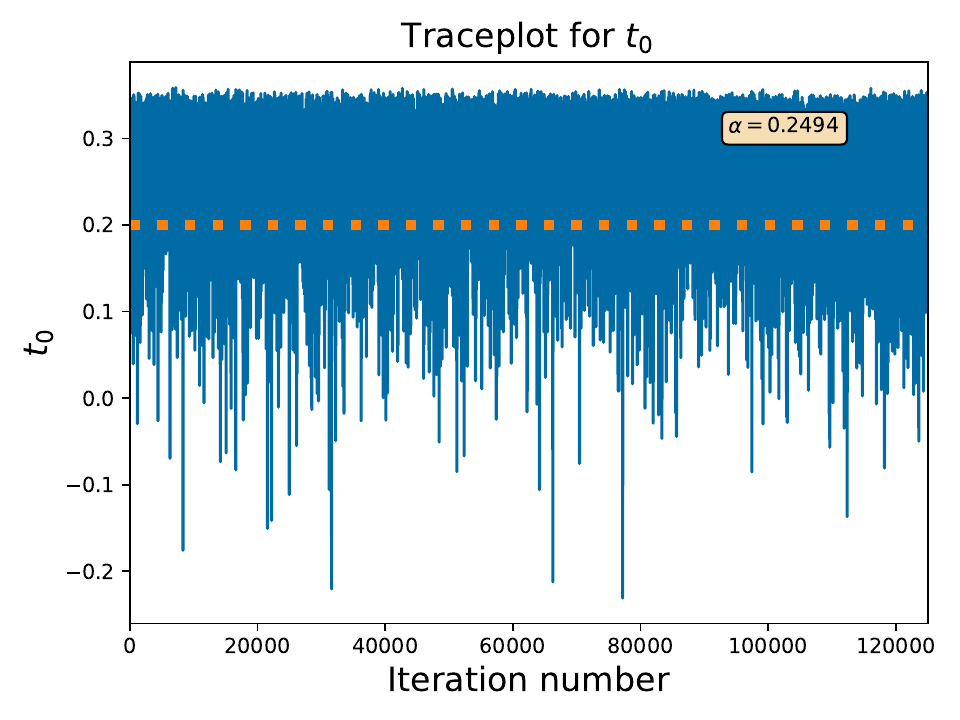}\\
		\caption{}  
		\label{TraceplotT0C}
	\end{subfigure}
	\hfill
	\begin{subfigure}[b]{0.495\textwidth}  
		\centering 
		\includegraphics[width=\textwidth]{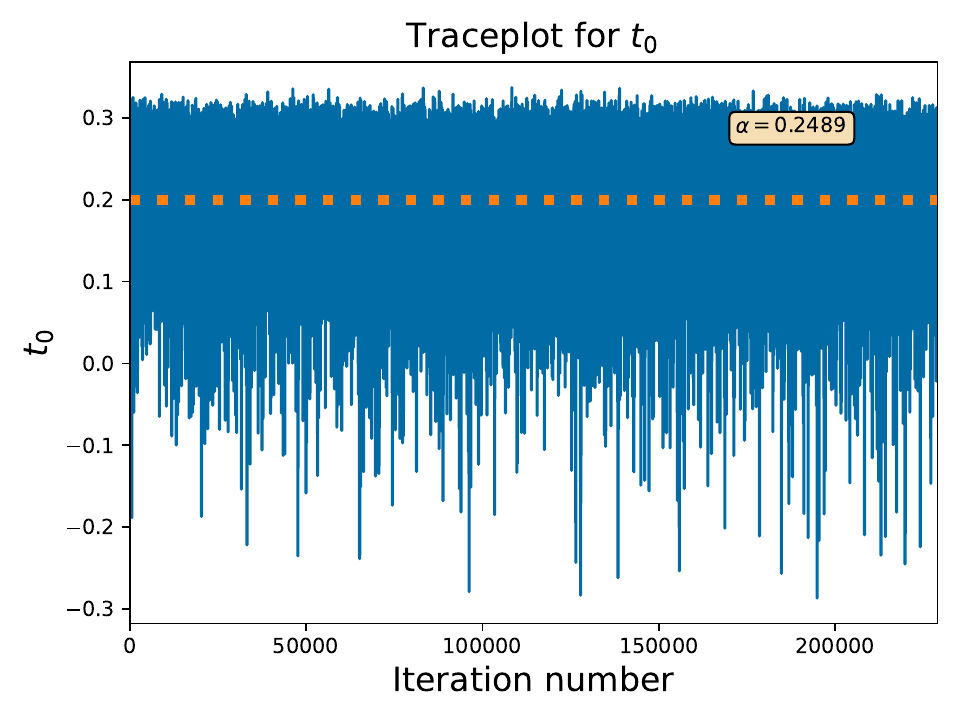}\\
		\caption{}  
		\label{TraceplotT0D}
	\end{subfigure}\\
	\caption[Traceplots of the allele age]{Traceplots for $t_0$ for: (a) Experiment C, and (b) Experiment D. The true value is denoted by the horizontal dashed orange line, whilst $\alpha$ denotes the mean acceptance probability.}
	\label{TraceplotsT0_diploid}
\end{figure}

\begin{figure}[H]
	\centering
	\begin{subfigure}[b]{0.495\textwidth}
		\centering
		\includegraphics[width=\textwidth]{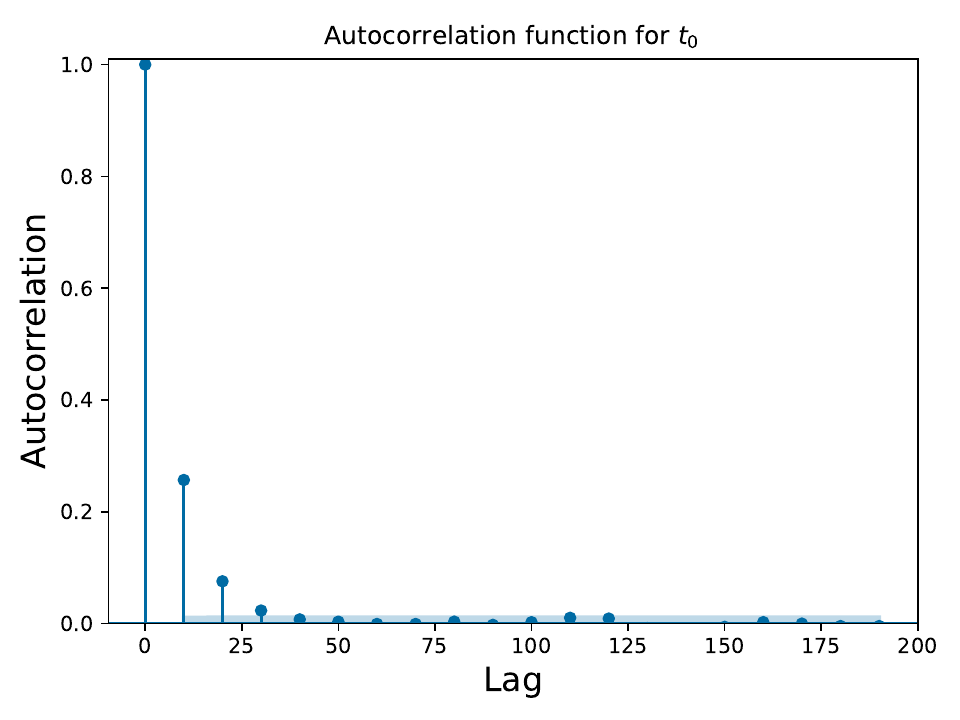}\\
		\caption{}  
		\label{AutoCorrT0C}
	\end{subfigure}
	\hfill
	\begin{subfigure}[b]{0.495\textwidth}  
		\centering 
		\includegraphics[width=\textwidth]{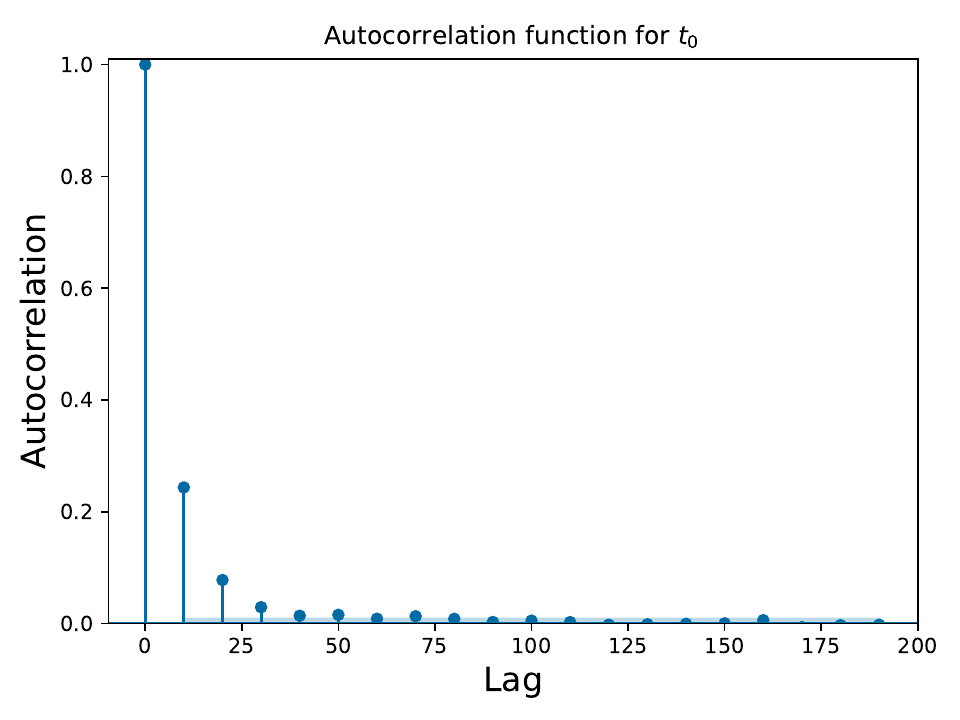}\\
		\caption{}  
		\label{AutoCorrT0D}
	\end{subfigure}\\
	\caption[Autocorrelation function for the allele age]{Autocorrelation function plot for $t_0$ for: (a) Experiment C, and (b) Experiment D} 
	\label{AutoCorrT0_diploid}
\end{figure}

\begin{figure}[H]
	\centering
	\begin{subfigure}[b]{0.495\textwidth}
		\centering
		\includegraphics[width=\textwidth]{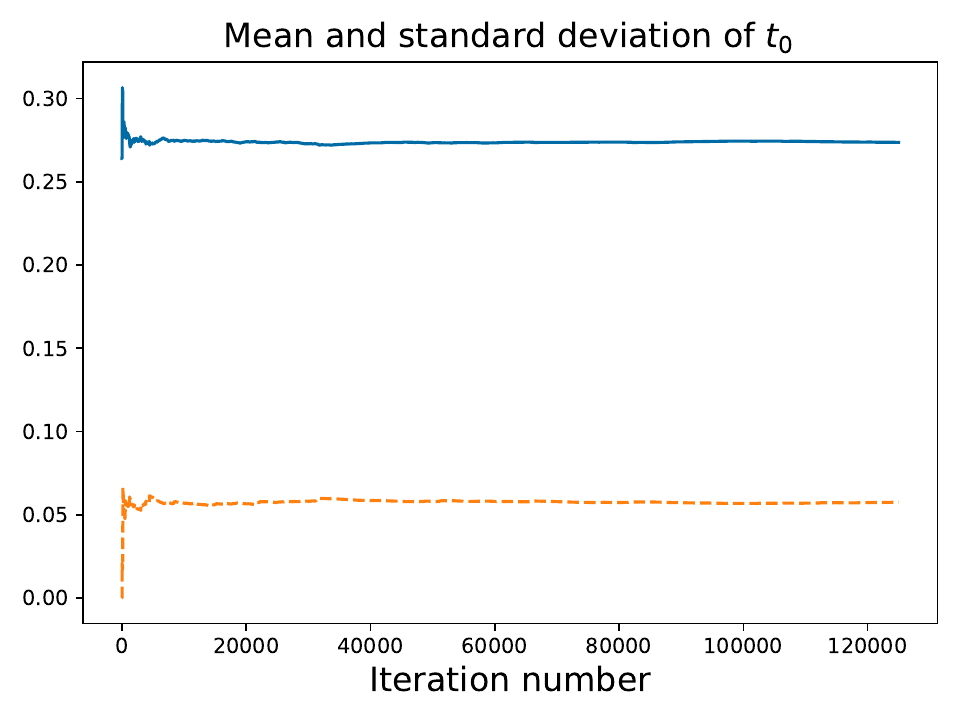}\\
		\caption{}  
		\label{T0MeanC}
	\end{subfigure}
	\hfill
	\begin{subfigure}[b]{0.495\textwidth}  
		\centering 
		\includegraphics[width=\textwidth]{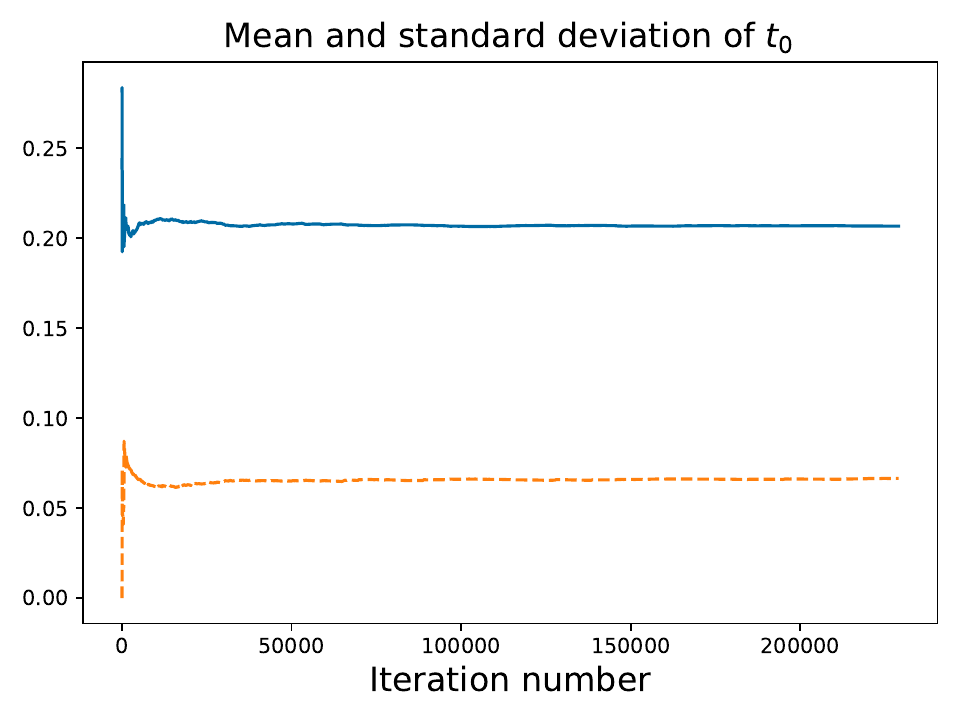}\\
		\caption{}  
		\label{T0MeanD}
	\end{subfigure}\\
	\caption[Plots of the mean and standard deviation of the allele age]{Plot of the mean (solid blue line) and standard deviation (dashed orange line) of $t_0$ for: (a) Experiment C, and (b) Experiment D.}
	\label{T0MeanSigma_diploid}
\end{figure}

\begin{figure}[H]
	\centering
	\begin{subfigure}[b]{\textwidth}
		\centering
		\includegraphics[width=\textwidth]{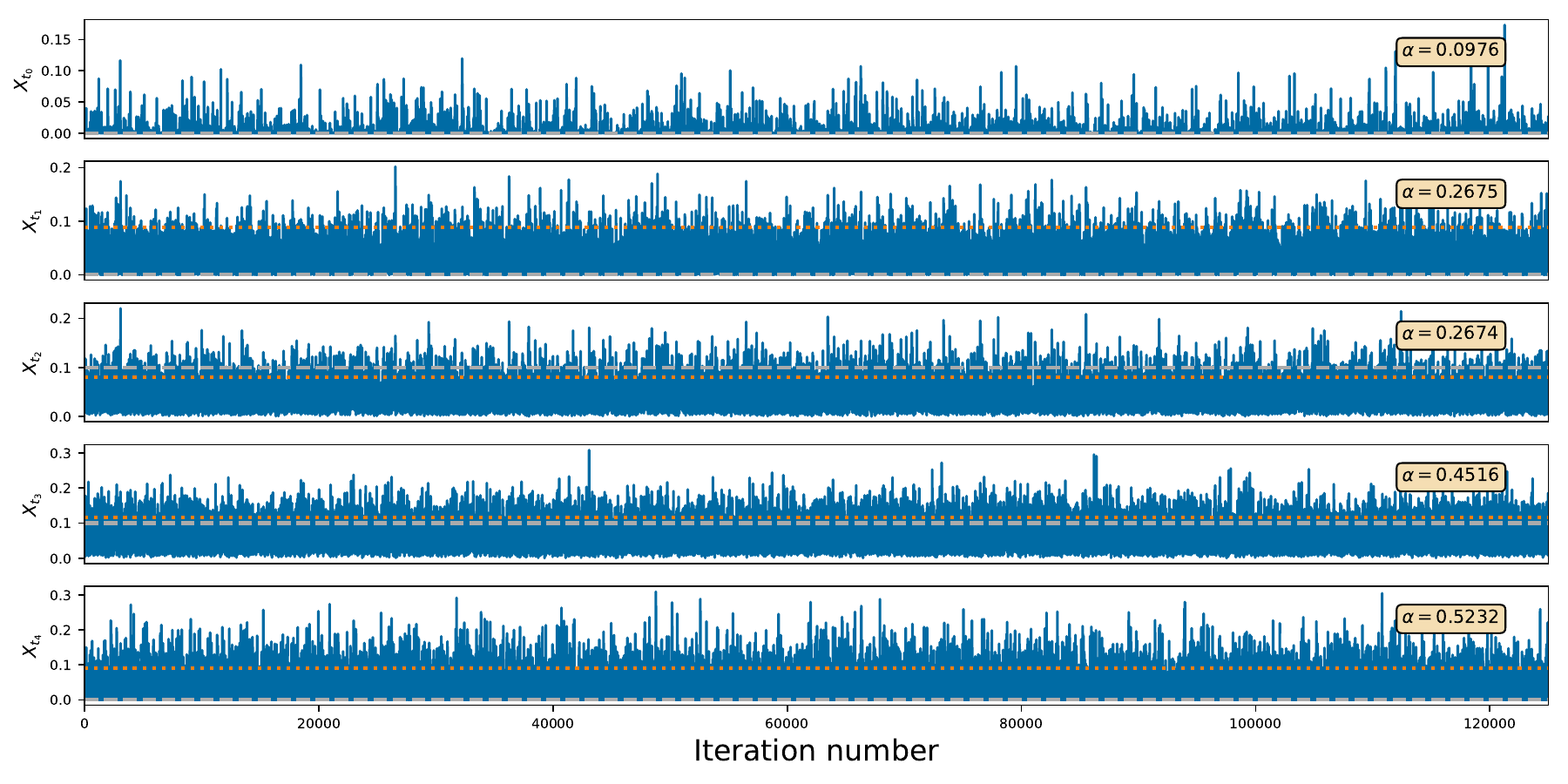}\\
		\caption{}  
		\label{MixingXC}
	\end{subfigure}
	\begin{subfigure}[b]{\textwidth}  
		\centering 
		\includegraphics[width=\textwidth]{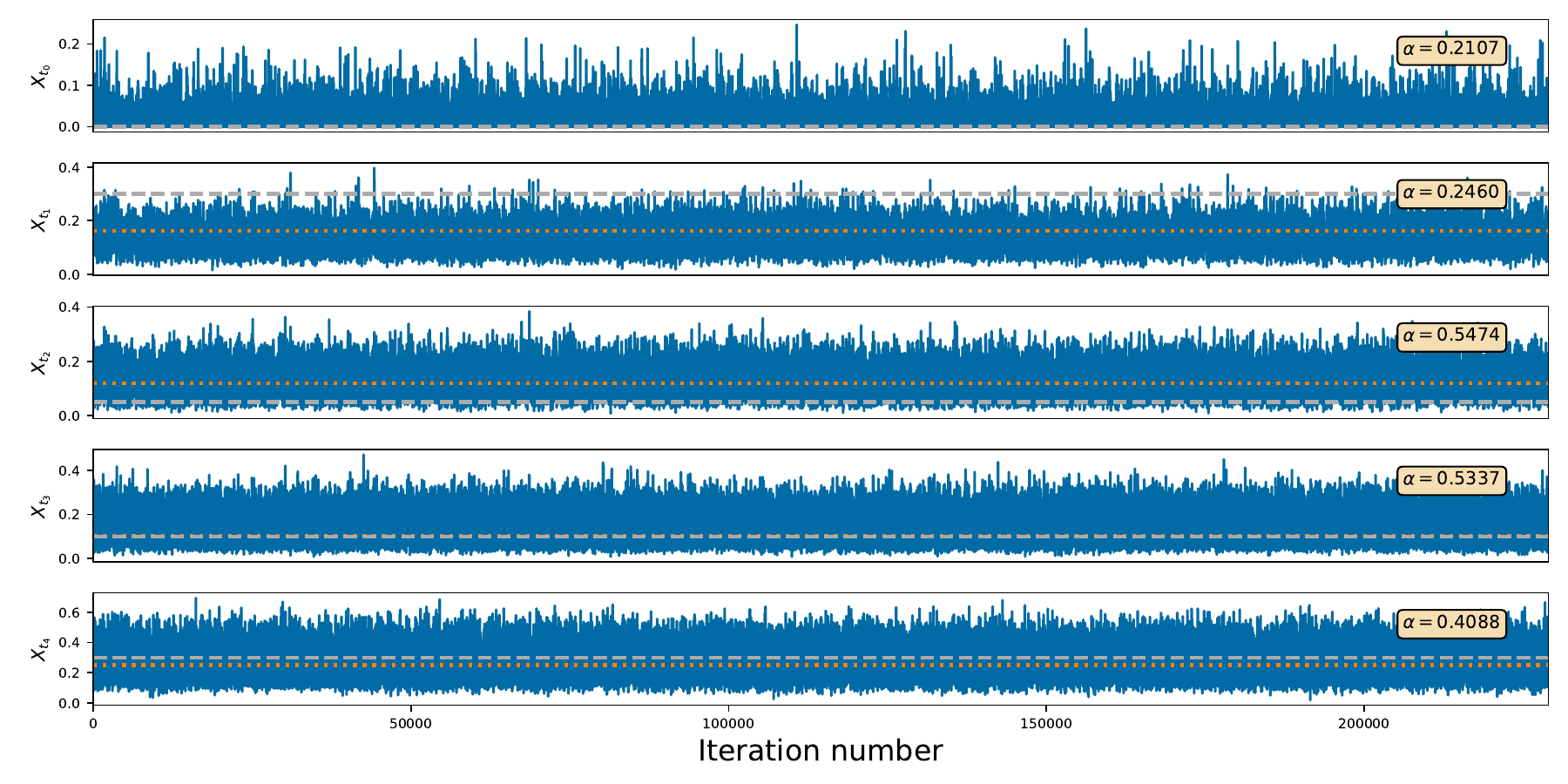}\\
		\caption{}  
		\label{MixingXD}
	\end{subfigure}\\
	\caption[Traceplots for the latent paths]{Traceplots of the latent diffusion for: (a) Experiment C, and (b) Experiment D. The true allele frequency is given by the dotted orange horizontal line, the observed frequency is given by the dashed grey line, whilst $\alpha$ denotes the mean acceptance probability.} 
\end{figure}

\subsection{Output for the ASIP gene}

\begin{figure}[H]
	\centering
	\begin{subfigure}[b]{0.495\textwidth}
		\centering
		\includegraphics[width=\textwidth]{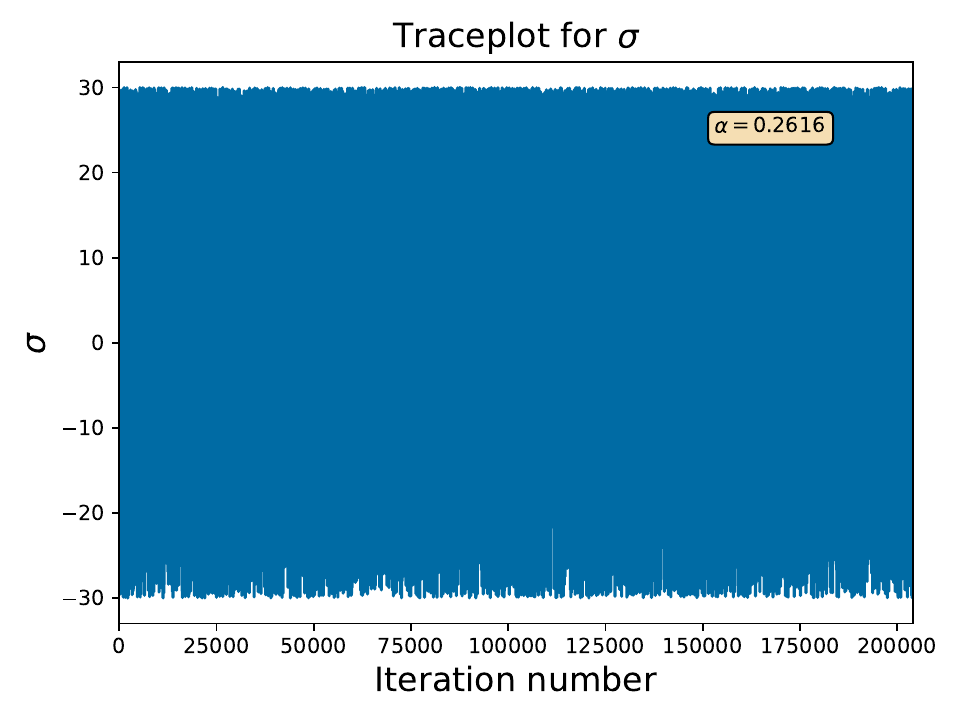}\\
		\caption*{Traceplot}  
		\label{TraceplotSigma_ASIP}
	\end{subfigure}
	\hfill
	\begin{subfigure}[b]{0.495\textwidth}  
		\centering 
		\includegraphics[width=\textwidth]{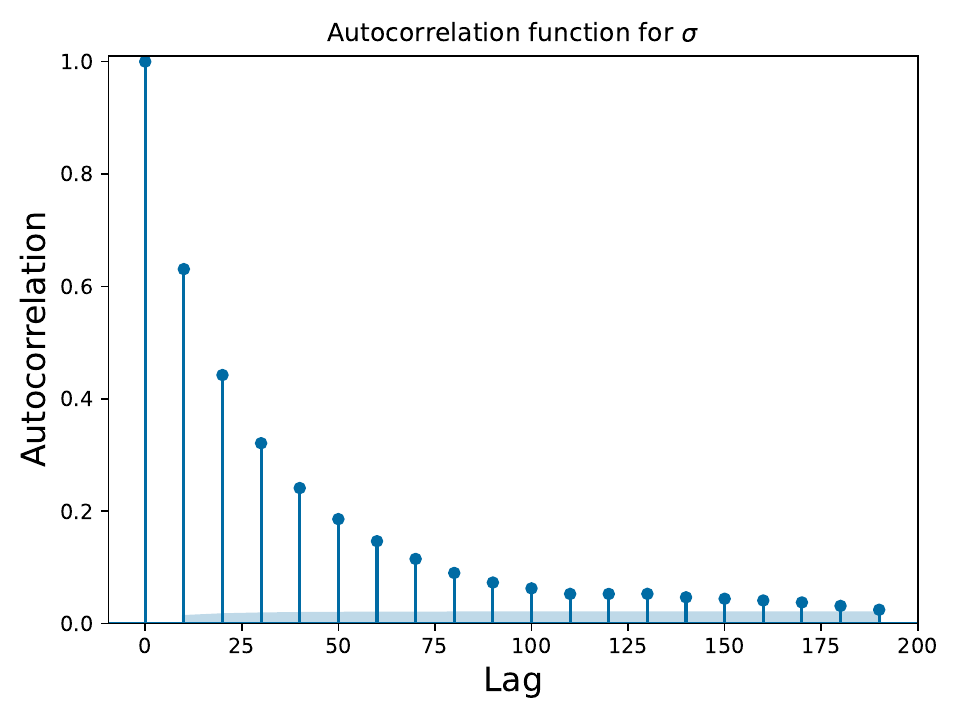}\\
		\caption*{Autocorrelation function}  
		\label{AutoCorrSigma_ASIP}
	\end{subfigure}\\
	\caption[Traceplot and autocorrelation function for the selection coefficient]{Traceplot and autocorrelation plot for $\sigma$ for the ASIP gene} 
	\label{TraceplotAutoCorrSigma_ASIP}
\end{figure}

\begin{figure}[H]
	\centering
	\begin{subfigure}[b]{0.495\textwidth}
		\centering
		\includegraphics[width=\textwidth]{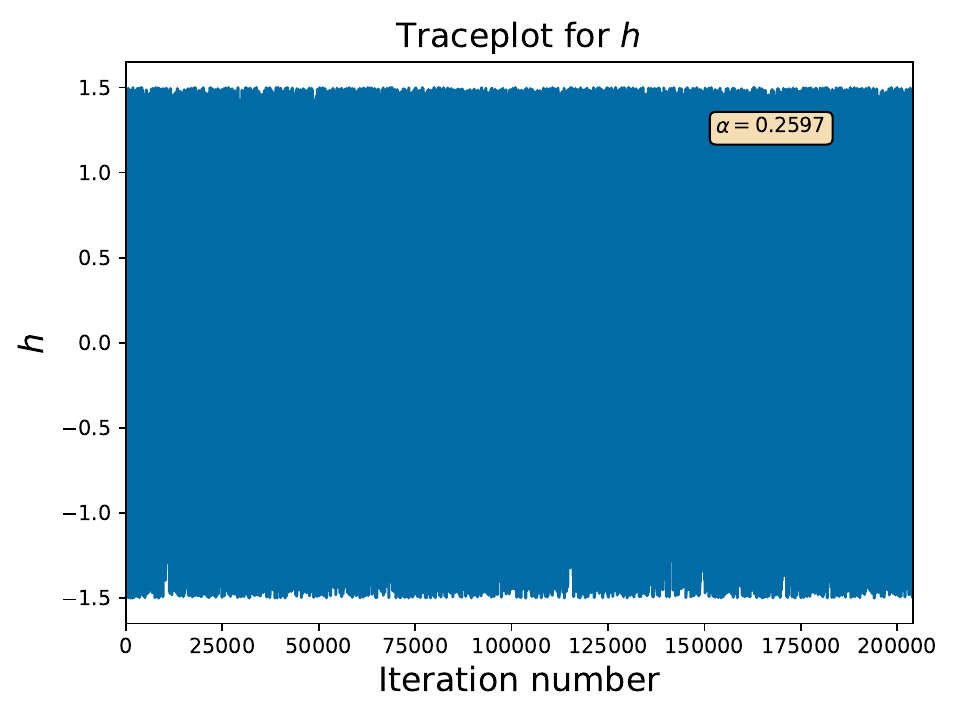}\\
		\caption*{Traceplot}  
		\label{TraceplotH_ASIP}
	\end{subfigure}
	\begin{subfigure}[b]{0.495\textwidth}  
		\centering 
		\includegraphics[width=\textwidth]{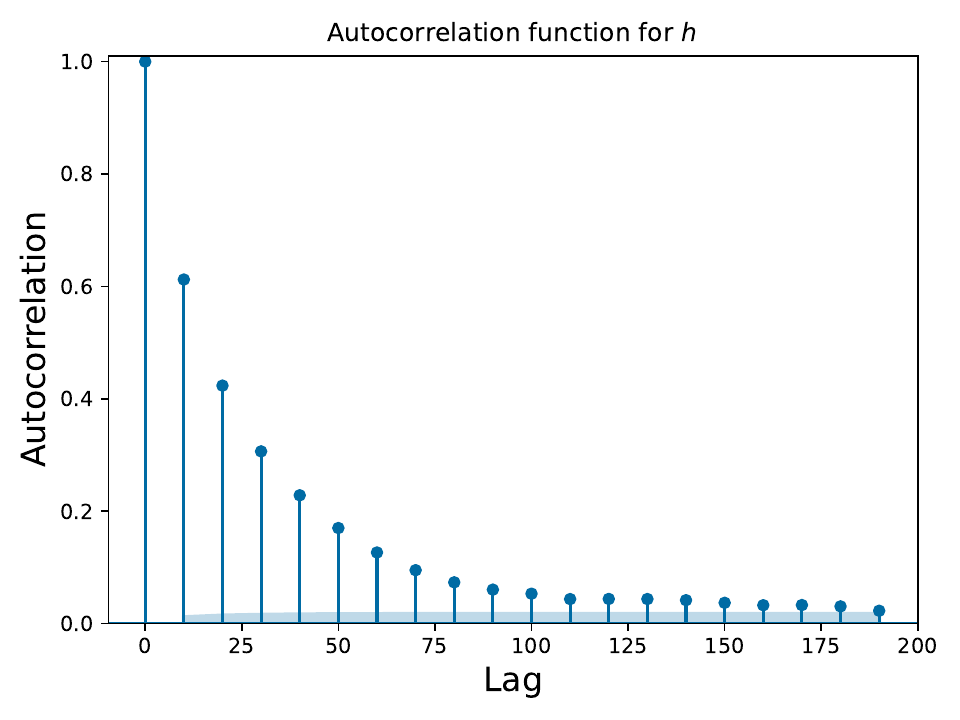}\\
		\caption*{Autocorrelation function}  
		\label{AutoCorrH_ASIP}
	\end{subfigure}\\
	\caption[Traceplot and autocorrelation function for the dominance parameter]{Traceplot and autocorrelation plot for $h$ for the ASIP gene} 
	\label{TraceplotAutoCorrH_ASIP}
\end{figure}

\begin{figure}[H]
	\centering
	\begin{subfigure}[b]{0.495\textwidth}
		\centering
		\includegraphics[width=\textwidth]{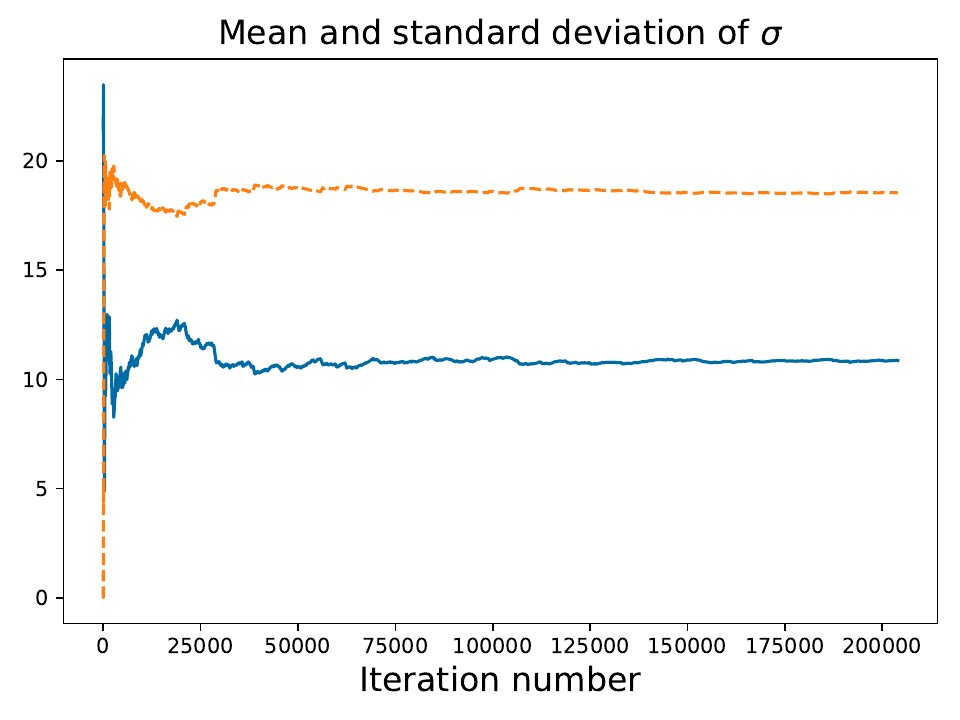}\\
		\caption*{$\bar{\sigma}$}  
		\label{SigmaMean_ASIP}
	\end{subfigure}
	\hfill
	\begin{subfigure}[b]{0.495\textwidth}  
		\centering 
		\includegraphics[width=\textwidth]{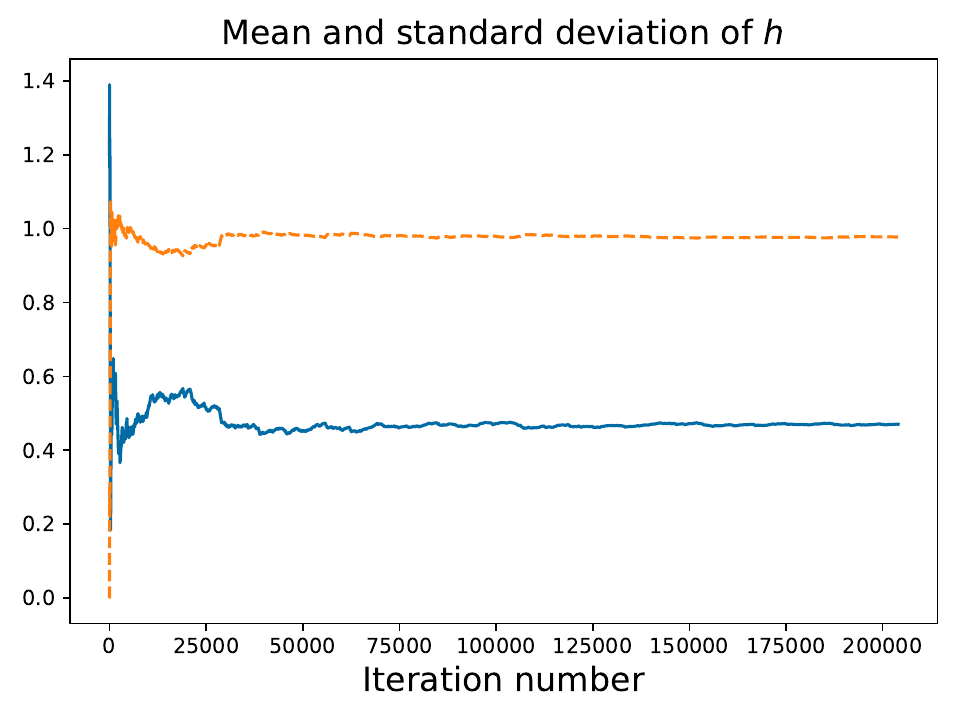}\\
		\caption*{$\hat{\sigma}$}  
		\label{SigmaStdev_ASIP}
	\end{subfigure}\\
	\caption[Plots of the mean and standard deviation of the selection coefficient and dominance parameter $h$]{Plots of the mean and standard deviation of $\sigma$ and $h$ for the ASIP gene} 
	\label{SigmaMeanStdev_ASIP}
\end{figure}

\begin{figure}[H]
	\centering
	\begin{subfigure}[b]{0.495\textwidth}
		\centering
		\includegraphics[width=\textwidth]{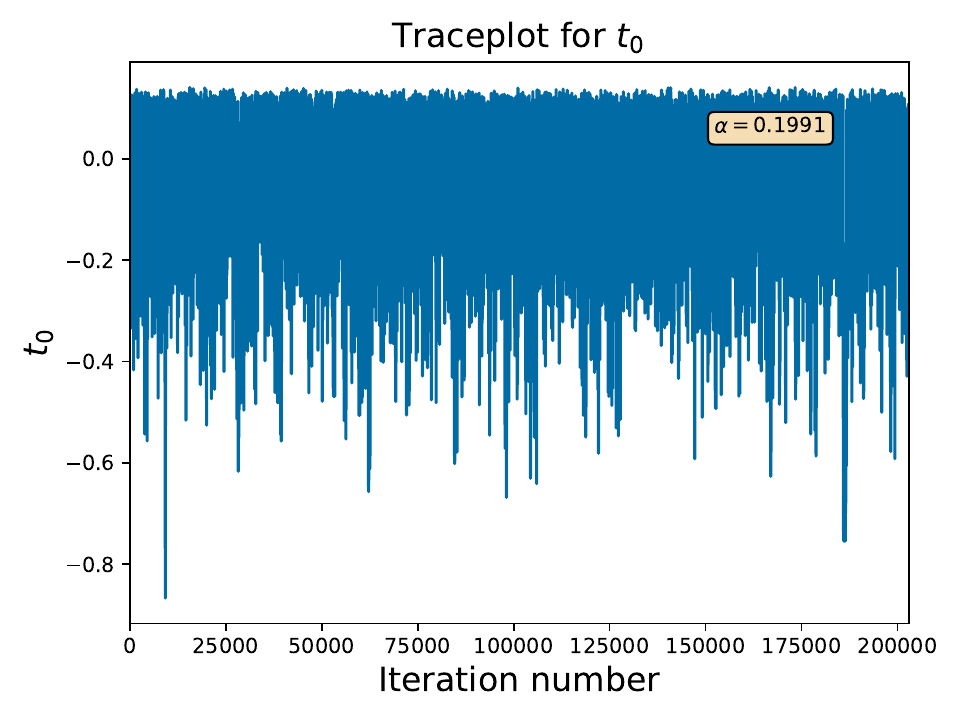}\\
		\caption*{Traceplot}  
		\label{TraceplotT0_ASIP}
	\end{subfigure}
	\hfill
	\begin{subfigure}[b]{0.495\textwidth}  
		\centering 
		\includegraphics[width=\textwidth]{simulations/plots/2024-08-20-09-46T0AutoCorr.pdf}\\
		\caption*{Autocorrelation function}  
		\label{AutoCorrT0_ASIP}
	\end{subfigure}\\
	\caption[Traceplot and autocorrelation function of the allele age]{Traceplot and autocorrelation plot for $t_0$ for the ASIP gene} 
	\label{TraceplotAutoCorrT0_ASIP}
\end{figure}

\begin{figure}[H]
	\centering
	\includegraphics[width=0.49\textwidth]{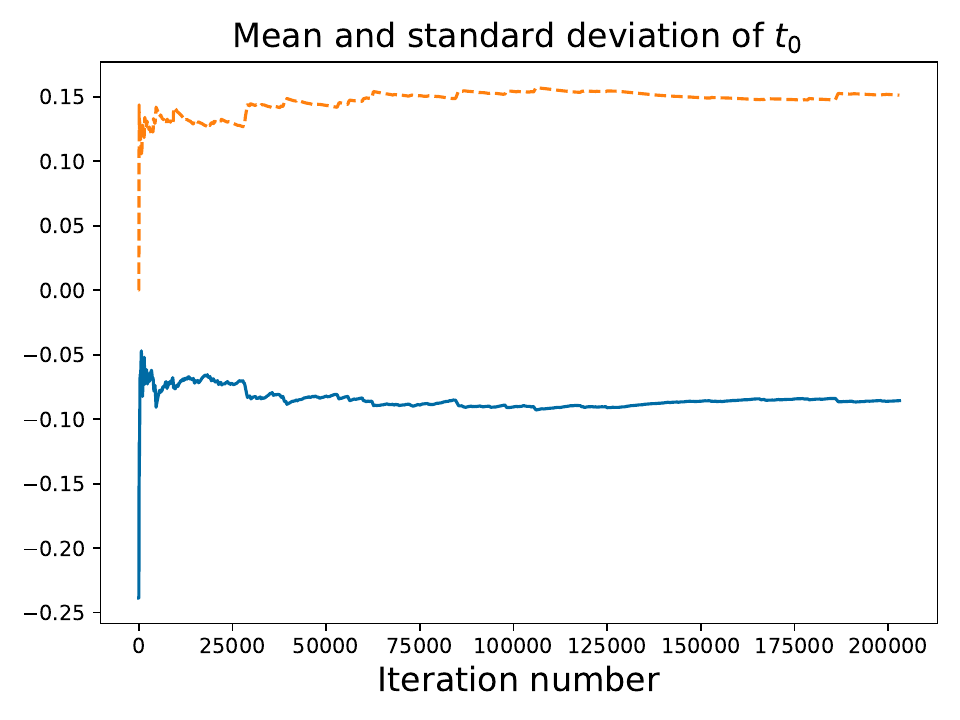}\\
	\label{T0Mean_ASIP}
	\caption[Plots of the mean and standard deviation of the allele age]{Plots of the mean and standard deviation of $t_0$ for the ASIP gene} 
	\label{T0MeanStdev_ASIP}
\end{figure}

\begin{figure}[H]
	\centering
	\includegraphics[width=\textwidth]{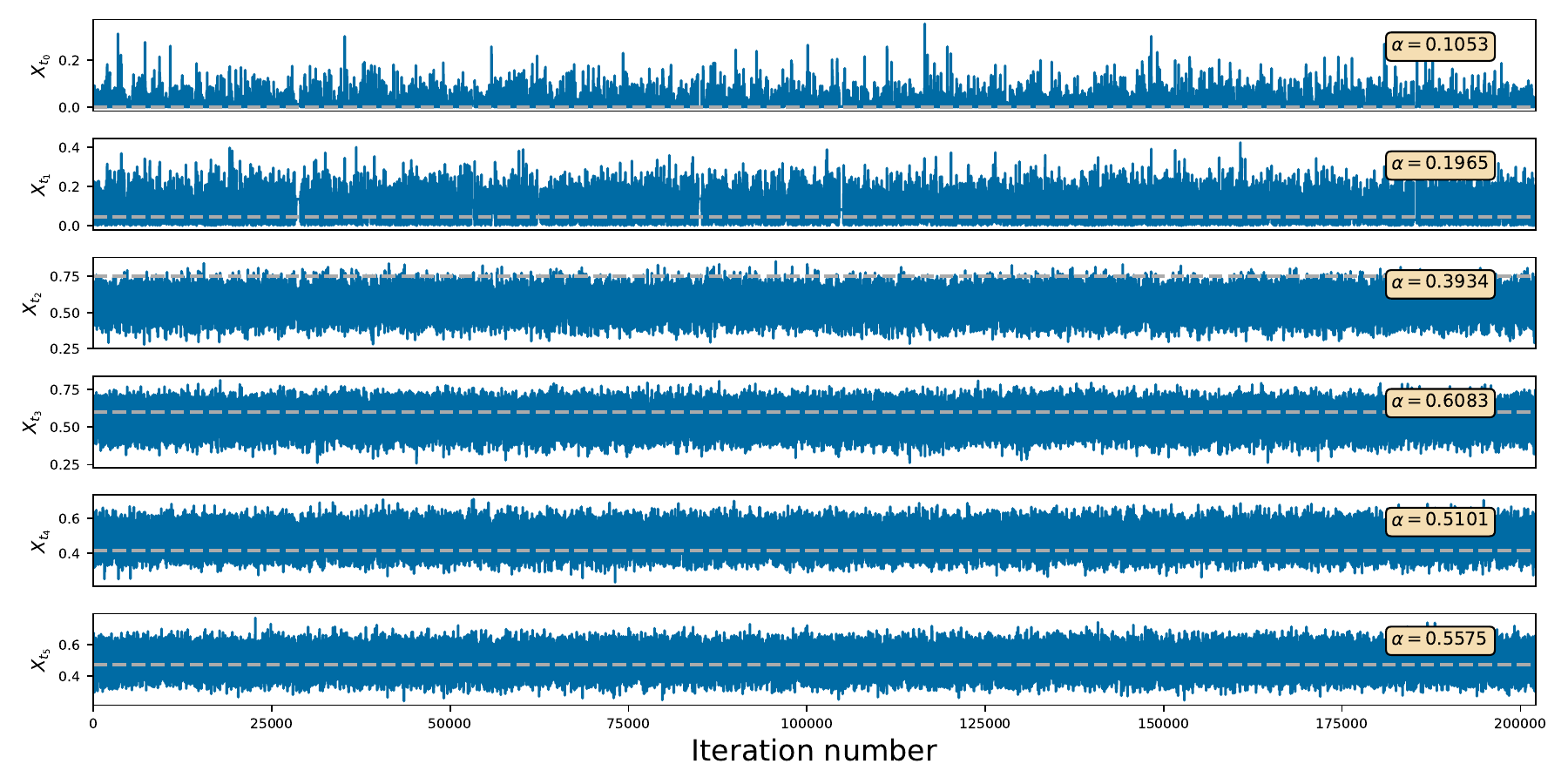}\\
	\caption[Mixing of the latent paths]{Mixing of the latent paths for the ASIP gene} 
	\label{MixingX_ASIP}
\end{figure}

\subsection{Output for the MC1R gene}

\begin{figure}[H]
	\centering
	\begin{subfigure}[b]{0.495\textwidth}
		\centering
		\includegraphics[width=\textwidth]{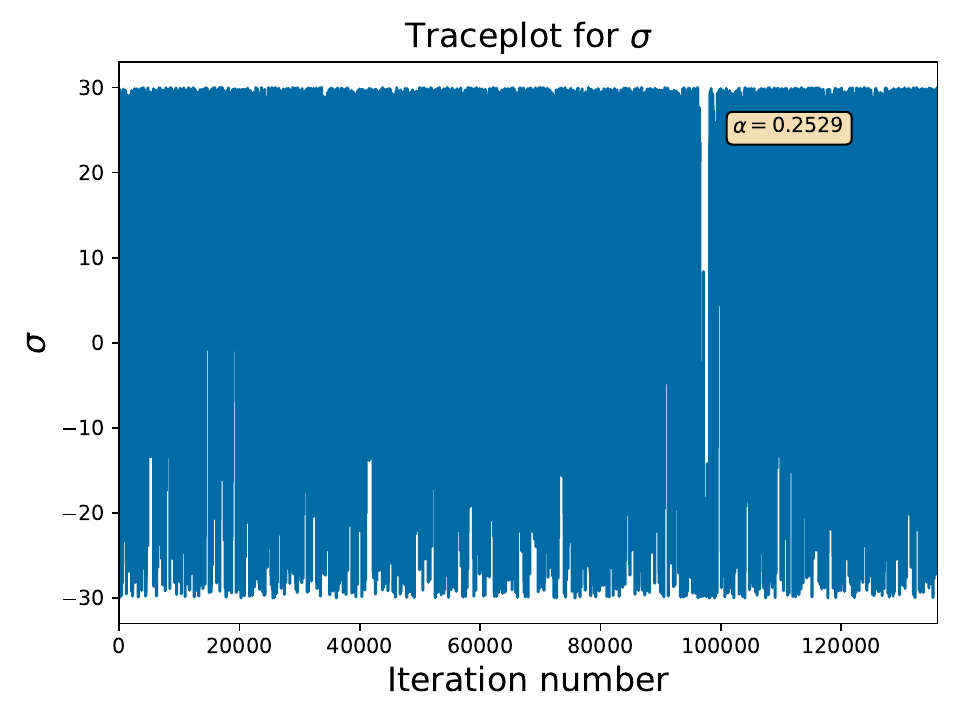}\\
		\caption*{Traceplot}  
		\label{TraceplotSigma_MC1R}
	\end{subfigure}
	\hfill
	\begin{subfigure}[b]{0.495\textwidth}  
		\centering 
		\includegraphics[width=\textwidth]{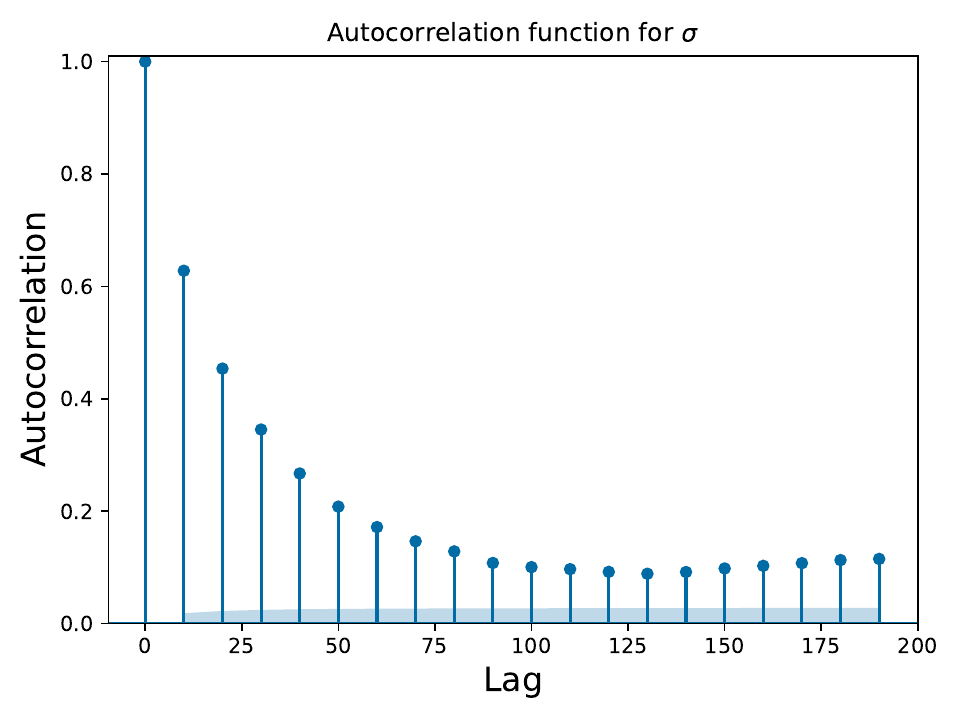}\\
		\caption*{Autocorrelation function}  
		\label{AutoCorrSigma_MC1R}
	\end{subfigure}\\
	\caption[Traceplot and autocorrelation function for the selection coefficient]{Traceplot and autocorrelation plot for $\sigma$ for the MC1R gene} 
	\label{TraceplotAutoCorrSigma_MC1R}
\end{figure}

\begin{figure}[H]
	\centering
	\begin{subfigure}[b]{0.495\textwidth}
		\centering
		\includegraphics[width=\textwidth]{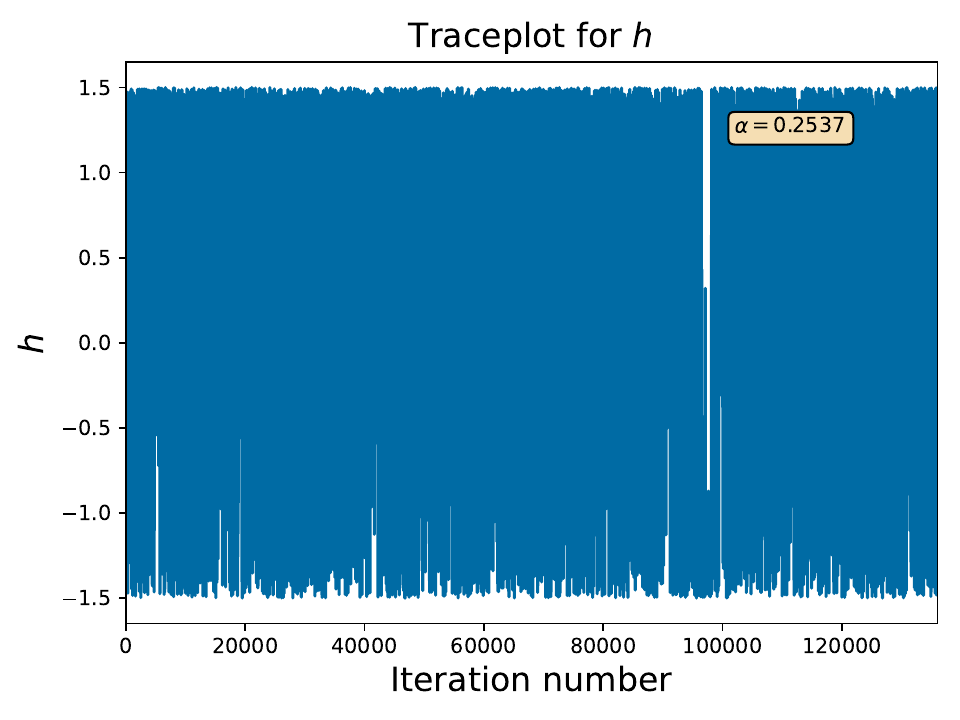}\\
		\caption*{Traceplot}  
		\label{TraceplotH_MC1R}
	\end{subfigure}
	\begin{subfigure}[b]{0.495\textwidth}  
		\centering 
		\includegraphics[width=\textwidth]{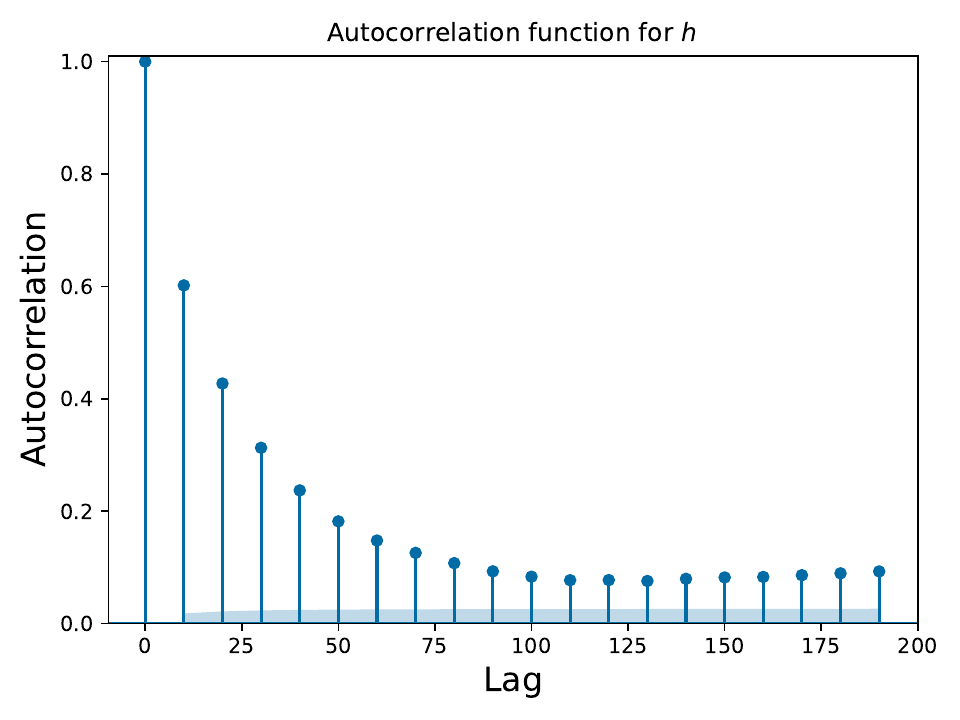}\\
		\caption*{Autocorrelation function}  
		\label{AutoCorrH_MC1R}
	\end{subfigure}\\
	\caption[Traceplot and autocorrelation function for the dominance parameter]{Traceplot and autocorrelation plot for $h$ for the MC1R gene} 
	\label{TraceplotAutoCorrH_MC1R}
\end{figure}

\begin{figure}[H]
	\centering
	\begin{subfigure}[b]{0.495\textwidth}
		\centering
		\includegraphics[width=\textwidth]{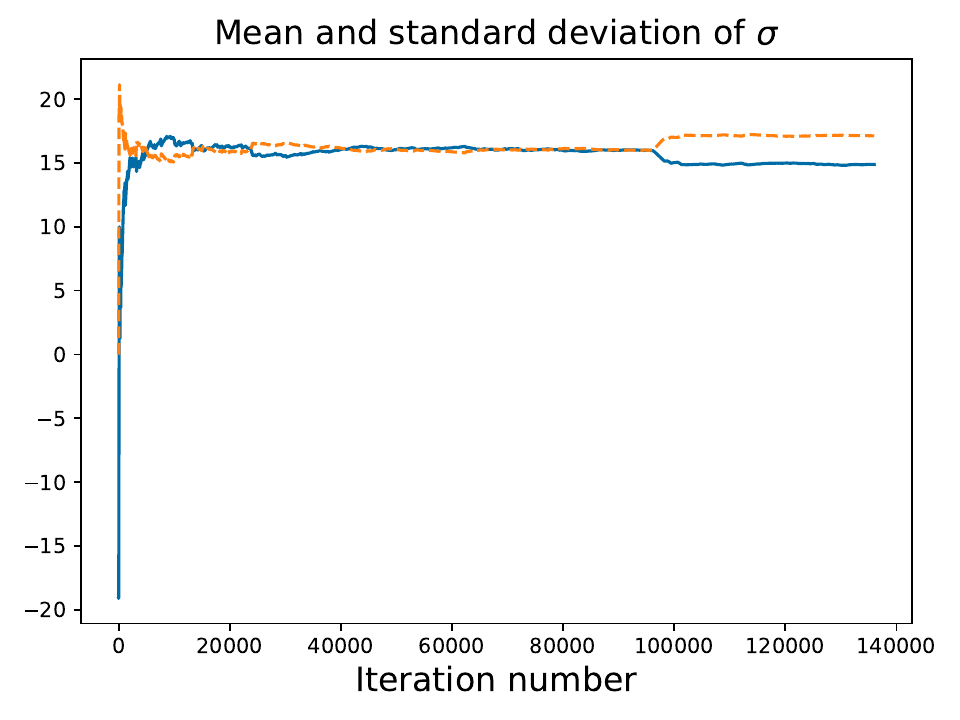}\\
		\caption*{$\bar{\sigma}$}  
		\label{SigmaMean_MC1R}
	\end{subfigure}
	\hfill
	\begin{subfigure}[b]{0.495\textwidth}  
		\centering 
		\includegraphics[width=\textwidth]{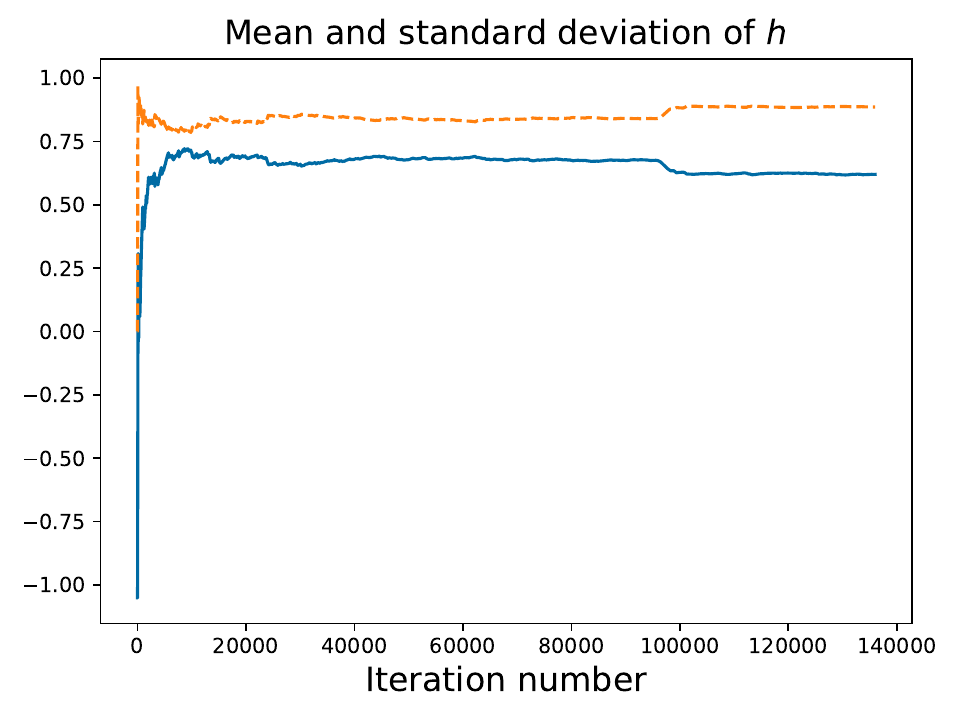}\\
		\caption*{$\hat{\sigma}$}  
		\label{SigmaStdev_MC1R}
	\end{subfigure}\\
	\caption[Plots of the mean and standard deviation of the selection coefficient and the dominance parameter]{Plots of the mean and standard deviation of $\sigma$ and $h$ for the MC1R gene} 
	\label{SigmaMeanStdev_MC1R}
\end{figure}

\begin{figure}[H]
	\centering
	\begin{subfigure}[b]{0.495\textwidth}
		\centering
		\includegraphics[width=\textwidth]{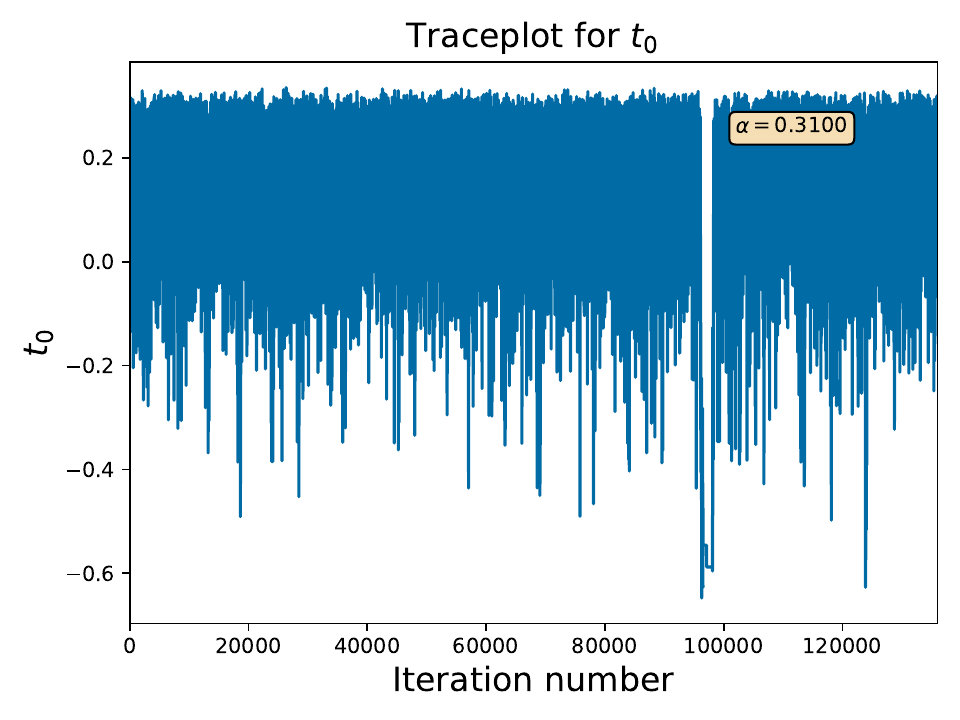}\\
		\caption*{Traceplot}  
		\label{TraceplotT0_MC1R}
	\end{subfigure}
	\hfill
	\begin{subfigure}[b]{0.495\textwidth}  
		\centering 
		\includegraphics[width=\textwidth]{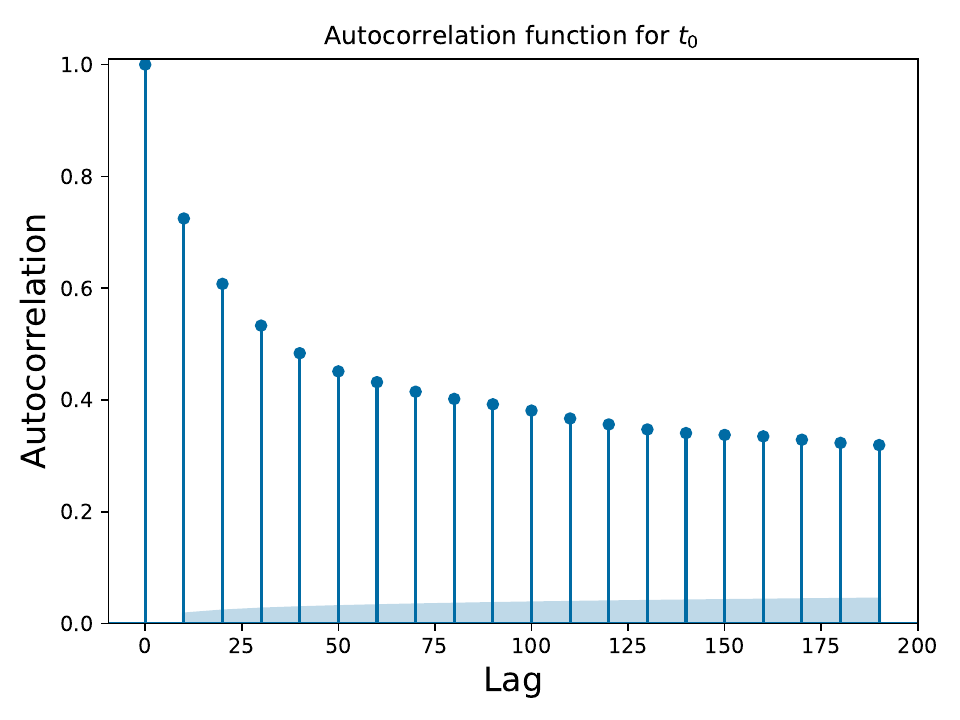}\\
		\caption*{Autocorrelation function}  
		\label{AutoCorrT0_MC1R}
	\end{subfigure}\\
	\caption[Traceplot and autocorrelation function of the allele age]{Traceplot and autocorrelation plot for $t_0$ for the MC1R gene} 
	\label{TraceplotAutoCorrT0_MC1R}
\end{figure}

\begin{figure}[H]
	\centering
	\includegraphics[width=0.49\textwidth]{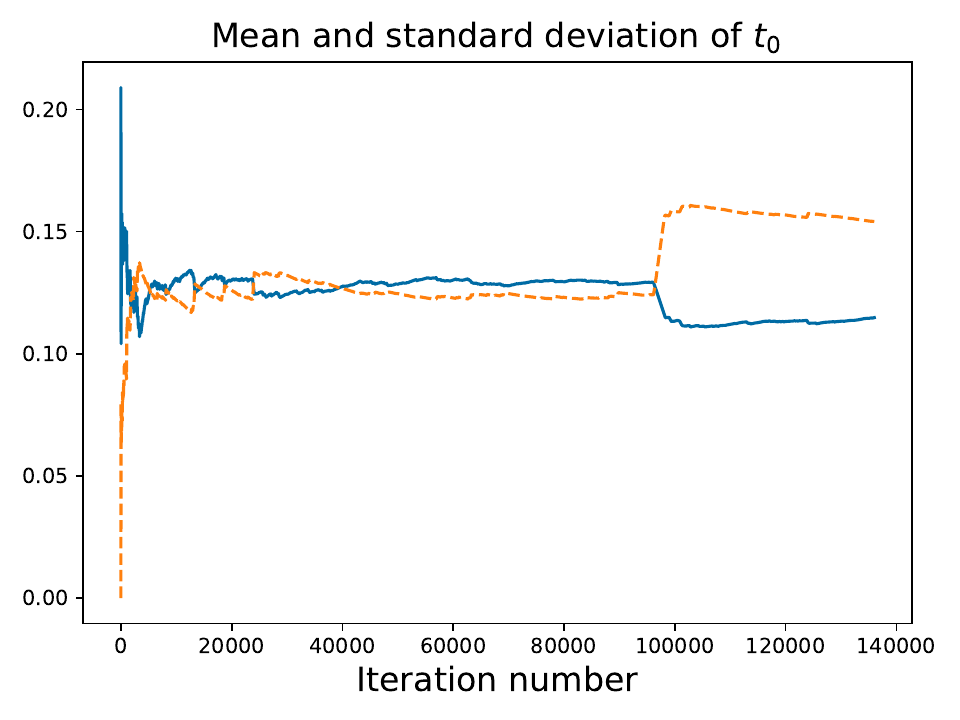}\\
	\caption*{$\bar{t_0}$}  
	\label{T0Mean_MC1R}
	\caption[Plots of the mean and standard deviation of the allele age]{Plots of the mean and standard deviation of $t_0$ for the MC1R gene} 
	\label{T0MeanStdev_MC1R}
\end{figure}

\begin{figure}[H]
	\centering
	\includegraphics[width=\textwidth]{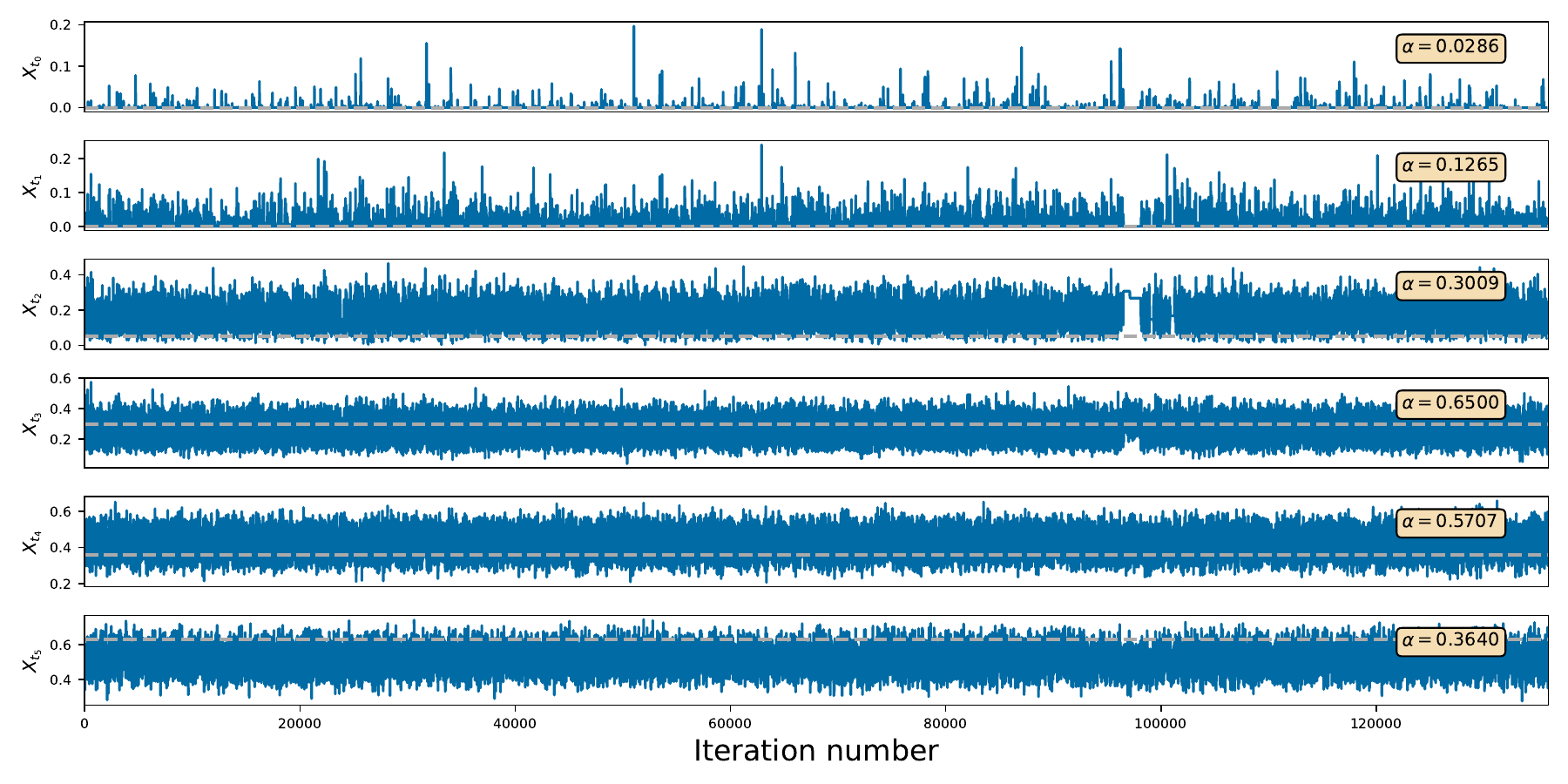}\\
	\caption[Mixing of the latent paths]{Mixing of the latent paths for the MC1R gene} 
	\label{MixingX_MC1R}
\end{figure}

\end{document}